\documentclass[structabstract]{aa}  
\usepackage{graphicx}
\usepackage{txfonts}
\usepackage[T1]{fontenc}
\usepackage{natbib}
\usepackage{hyperref}
\usepackage{siunitx}  
\usepackage{xcolor}
\usepackage{booktabs}   
\usepackage{placeins}   

\newcommand{\mci}[1]{\multicolumn{1}{c}{#1}}
\newcommand{\mcii}[1]{\multicolumn{2}{c}{#1}}
\newcommand{\mciii}[1]{\multicolumn{3}{c}{#1}}
\newcommand{\mciv}[1]{\multicolumn{4}{c}{#1}}

\newcommand{\mcvii}[1]{\multicolumn{7}{c}{#1}}

\newcommand{\mcx}[1]{\multicolumn{10}{c}{#1}}

\newcommand{\Teff}{\mbox{$T_{\rm eff}$}}
\newcommand{\logg}{\mbox{$\log g$}}
\newcommand{\logQ}{\mbox{$\log Q$}}

\newcommand{\vsini}{\mbox{$v\,\sin i$}}

\newcommand{\Msun}{\mbox{M$_\odot$}}

\newcommand{\MTW}{M$^3$W}

\newcommand{\cre}{\color{red}}
\newcommand{\cgr}{\color{green}}
\newcommand{\cbl}{\color{blue}}

\newcommand{\lili}{LiLiMaRlin}

\begin{document}

   \title{Multiplicity of Massive stars in the Milky Way (\MTW). \linebreak 
          II. The brightest single O stars: variability and systemic velocities}
   \titlerunning{\MTW~II. The brightest single O stars}
   \authorrunning{Ma\'{\i}z Apell\'aniz et al.}


   \author{J. Ma\'{\i}z Apell\'aniz\inst{1}
          \and
          E. Madero Fuentes\inst{1,2}
          \and
          G. Holgado\inst{3,4}
          \and
          M. Pantaleoni Gonz\'alez\inst{5}
          \and
          A. Cifuentes Santos\inst{1,2,6}
          \and \\
          F. A. Paredes Montalv\'an\inst{1,2}
          \and
          S. Sim\'on-D\'{\i}az\inst{3,4}
          \and 
          S. Rosu\inst{7,8}
          \and 
          J. A. Molina-Calzada\inst{1,2}
          \and 
          L. Sciarini\inst{7}
          \and \linebreak
          M. Abdul-Masih\inst{3,4}
          \and 
          C. Mart\'{\i}nez Sebasti\'an\inst{3,4}
          }

   \institute{Centro de Astrobiolog\'{\i}a, CSIC-INTA. Campus ESAC. 
              C. bajo del castillo s/n. 
              E-\num[detect-all]{28692} Villanueva de la Ca\~nada, Madrid, Spain.\\
              \email{jmaiz@cab.inta-csic.es}
         \and
              Departamento de Astrof{\'\i}sica y F{\'\i}sica de la Atm\'osfera. 
              Universidad Complutense de Madrid. 
              E-\num[detect-all]{28040} Madrid, Spain. 
         \and
              Instituto de Astrof\'isica de Canarias. 
              E-\num[detect-all]{38200} La Laguna, Tenerife, Spain.
         \and
              Departamento de Astrof\'isica, Universidad de La Laguna.
              E-\num[detect-all]{38205} La Laguna, Tenerife, Spain.
         \and
              Department of Astrophysics, University of Vienna.  
              T\"urkenschanzstrasse 17, 1180 Wien, Austria.
         \and
              Centro de Investigaciones Energ\'eticas, Medioambientales y Tecnol\'ogicas.
              E-\num[detect-all]{28040} Madrid, Spain.
         \and
              D\'epartement d'Astronomie, Universit\'e de Gen\`eve.
              Chemin Pegasi 51.
              CH-1290, Versoix, Switzerland.
         \and 
              Institut d'Astronomie et d'Astrophysique, Universit\'e Libre de Bruxelles.
              CP 226, B-1050 Brussels, Belgium.
             }

   \date{Received 10 September 2026; accepted 25 October 2026}

 
  \abstract
   {Multiplicity of Massive stars in the Milky Way (hereafter, \MTW) is a project designed to augment our knowledge of the dynamical properties 
    of massive stars within several kpc of the Sun. An often overlooked aspect of such properties is the determination of accurate
    systemic velocities, as they are affected by variability effects such as pulsations, line infilling by wind and disks, and incomplete
    knowledge about the central wavelengths. Such lack of accuracy impairs studies on the origin of runaway/walkaway stars and on the 
    velocity dispersions of stellar clusters.}
   {In this second paper of the \MTW\ series we want to increase the accuracy of the O-stars systemic velocities in 
    both isolated and multiple systems. To that end, we establish how different wavelength regions are affected by
    variability, determine the thresholds for the detection of spectroscopic binarity, and obtain accurate central wavelengths including
    the effects of contamination, forbidden lines, and wind infilling.}
   {Our sample consists of the 41 brightest O stars ($B \le 7.0$~mag) thought to be single at the beginning of the project (summer 
    2025). We used multi-epoch high-resolution spectroscopy from the \lili\ project cleaned with the UNWIND code and stellar models computed
    with FASTWIND. The epoch spectra were cross-correlated to study their variability in the form of radial velocity (RV)
    dispersions. The RVs were computed from (a) cross-correlation of H and He lines with FASTWIND models, (b) profile fitting of metallic 
    lines, and (c) (for very fast rotators) cross-correlation of the full spectrum with a template star.}
   {A comparison between the RV dispersions for wavelength regions respectively dominated by He lines and by H$\beta$ reveals different groups
    of stars in terms of variability: (a) non-supergiant slow rotators, (b) fast rotators, (c) supergiants, and (d) Oe stars.
    This is likely a manifestation of the different role of pulsations, winds, and disks in RV variability. We use that analysis to determine
    RV dispersion thresholds for the detection of new spectroscopic binaries (SBs).
    We also use a FASTWIND grid to establish the effects of (a) the presence of \ion{He}{ii} lines close to Balmer lines, (b) the existence of 
    forbidden components in two \ion{He}{i} lines, and (c) wind infilling in the determination of accurate RVs. With all of that, we
    determined RVs for our 41 stars, checked their accuracy comparing results from our three methods, and in some cases discovered 
    significant differences with literature values. Our detailed analysis has further
    revealed four SB1 systems unknown in the summer of 2025 (68~Cyg, HD~\num{152424}, HD~\num{209339}~A, and HD~\num{93521}), one new SB2 
    (HD~\num{46150}), two new Oe stars (HD~\num{57682} and HD~\num{93521}), and three new Onfp stars ($\xi$~Per, 68~Cyg, and HD~\num{175876}).}
   {}


   \maketitle
%

\section{Introduction}

$\,\!$\indent The main objective of the Multiplicity of Massive stars in the Milky Way (\MTW) project is to determine the kinematical 
and dynamical properties (systemic or centre-of-mass velocity, orbital parameters, mass, and internal structure constant or $k_2$) of as many 
Galactic massive stars (those with $M \gtrsim 8$~\Msun) as possible. In \citet{Maizetal26b}, from now on Paper~I, we presented the project, 
data, and methods and applied them to analyse the very massive twin system GLS~\num{11448}~Aa,Ab and generate the most complete DIB library to date. 
In this Paper~II, as part of a systematic analysis of those kinematical and dynamical properties, we study the determination of systemic velocities 
for the most simple case for O stars, namely bright and isolated systems for which we have many high-resolution spectroscopic epochs. We address the 
different uncertainty sources in order to obtain accurate and precise values.

Radial velocities (RVs) can be determined in cool stars with high precision (m/s to cm/s) because their spectra are composed of a large number
of very narrow, mostly metallic lines for which cross-correlation with a template spectrum can yield results with very low uncertainties. For 
hot stars, RV uncertainties are orders of magnitude larger due to several reasons. The first and most important one is the much larger 
intrinsic width of the lines, which are broadened by micro- and macro-turbulence, rotation, and (for H and He) the Stark effect \citep{SimDetal14}.
Second, the number of available lines is much lower, especially as we go to higher values of \Teff. Third, line shapes can be modified by 
non-photospheric effects such as wind infilling and disks, hence difficulting the determination of the real stellar RV. Fourth, some commonly used
H and He lines have multiple unresolved levels and/or contaminating lines that make their effective central wavelengths change from star to star, 
a fact compounded with uncertainties derived from atomic physics. Fifth, massive stars are known for their high degree of multiplicity and there
are certainly many hidden spectroscopic binaries (SBs) waiting to be discovered. Finally, even single stars have additional sources of 
variability such as pulsations that can shift the measured RVs from epoch to epoch \citep{SimDetal24}.

The effects presented in the previous paragraph complicate the determination of spectroscopic orbits for binaries and, in particular, have three
consequences. First, the non-orbital RV variability introduces noise in the orbits (see \citealt{Barbetal26} for several examples). Second, 
some previous works have confounded orbital and non-orbital variability, leading to the publication of pseudo-orbits where the RV variations 
were actually caused by pulsations (see \citealt{Trigetal21} for some examples). The third one is the existence of systematic effects in the
determination of the $\gamma$ values (systemic velocities) of the orbit. Those are especially obvious for SB2 systems where $\gamma$ is
determined independently for the two components and they differ substantially. Two examples with differences in the range 10-20 km/s 
(HD~\num{169727}, an O+O system, and HD~\num{1810}, an O+B system) are shown in \citet{Holgetal25a} but for WR+O systems the differences can 
be much larger (see \citealt{Marcetal98,Hirvetal06} for examples).

Of those three consequences for SB systems, the third one is commonly ignored because the main purpose of binary 
studies is to determine the orbital properties that lead to the calculation of masses and those are (almost) independent of the systemic 
velocity\footnote{One small dependence is usually ignored because the effect is too small (except for hypothetical hyper-velocity binaries). 
Periods used to determine masses should 
be calculated in the rest frame of the system as $\gamma$ produces a correction factor of (1+$\gamma/c$).}. However, an accurate determination 
of the systemic velocities is needed for two purposes. First, for the determination of velocity dispersions in clusters with OB stars, which
require identifying SBs as they can be the main source of scatter when using just a single epoch \citep{HenBetal12b}. Second, for the 
retropropagation or backtracing of the trajectories of runaway/walkaway stars \citep{Hoogetal01,Renzetal19b}, for which systematic effects 
in $\gamma$ can significantly alter the identification of the original point of the ejection. 

The possible biases in the SB systemic velocities described in the previous paragraph for cluster velocity dispersions and trajectories of 
ejected stars are also relevant for single stars. Such stars are better targets to analyse the effects of non-orbital variability, 
line-profile modifications due to non-photospheric effects, and changes in the central wavelength due to atomic physics, as for them the most 
important reason for RV variability in O stars (orbital motion) is absent. That is the
reasoning behind this paper: to \textit{select a sample of bright single O stars and measure the biases in the systemic velocities and 
the effects of non-orbital variability to derive the best possible RVs with the available data.}

A second reason behind this paper is the need for a multiplicity study to \textit{determine which stars are really single to avoid biases in
the multiplicity fraction.} Ultimately, we want to address whether all O stars are formed in multiple systems or not. An affirmative answer 
would require that all stars that are currently single are the result of a merger (to be identified by a high \vsini\ or by being a blue straggler, 
\citealt{deMietal13,Schnetal19}) or of an ejection event (to be identified by its peculiar motion and, eventually, by retropropagating its 
trajectory to a point of origin, \citealt{Hoogetal01}). This should have important implications on our understanding of the IMF and the
formation of massive stars.

This paper is structured as follows.
In Section~\ref{section2}  we present the sample, the spectroscopic data we use, and the methods for their analysis. The 
\hyperref[section3]{main results} section consists of eight parts. 
\hyperref[subsection31]{First}, we present the initial UNWIND run used to generate clean spectra and produce the spectral classifications of the sample. 
\hyperref[subsection32]{Second}, we discuss the auto-cross-correlation of the spectra and the associated variability analysis. 
Then, we present how our grid of FASTWIND models is used to analyse the effects in the RVs of the 
\hyperref[subsection33]{[a]} contamination of Balmer lines by \ion{He}{ii} lines, 
\hyperref[subsection34]{[b]} presence of forbidden components in some \ion{He}{i} lines, and 
\hyperref[subsection35]{[c]} wind infilling in H and He lines. We end the section presenting the determination of RVs by 
\hyperref[subsection36]{[a]} cross-correlating FASTWIND models with H and He lines, 
\hyperref[subsection37]{[b]} profile fitting for metallic lines, and 
\hyperref[subsection38]{[c]} (for very fast rotators) cross-correlating a template star with the full spectra. 
The main body of the paper ends with an 
\hyperref[section4]{analysis and future work section}. The Appendices provide the 
\hyperref[sample]{sample information, spectral classifications, and number of epochs}; the
\hyperref[appendixb]{spectral classification plots}; a 
\hyperref[appendixc]{description of the FASTWIND grid used}; and a
\hyperref[appendixd]{glossary of terms}.

\section{Data and methods}\label{section2}

\begin{table*}
\caption{Results from the FASTWIND modelling and RV dispersion analysis.}
\centerline{
\addtolength{\tabcolsep}{-1mm}
\begin{tabular}{l@{\hspace{-3mm}}rr@{$\pm$}lr@{$\pm$}lr@{$\pm$}lrrrrrrrrrr}
\midrule
Name                  & \mcvii{\rule[0.5ex]{21mm}{0.5pt} Modelling \rule[0.5ex]{21mm}{0.5pt}} & \mcx{\rule[0.5ex]{32mm}{0.5pt} RV dispersions \rule[0.5ex]{32mm}{0.5pt}}        \\
                      & \mci{\vsini} & \mcii{\Teff} & \mcii{\logg} & \mcii{\logQ}           & H$\gamma$ & H$\beta$ & H$\alpha$ & 4200 & 4500 & 4650 & 4686 & 4980 & 5412 & 5876 \\
                      & \mci{(km/s)} & \mcii{(kK)}  & \mcii{(cgs)} & \mcii{(solar)}         & \mcx{(km/s)}                                                                      \\
\midrule
$\zeta$ Pup           & \cbl 199 & 41.0&0.9   & 3.69&0.08  & $-$12.11&0.40  &      3.5 &      3.9 &      8.9 &      5.8 &      6.5 &     13.0 &      9.1 &     11.6 &      6.2 &      5.7 \\
$\zeta$ Oph           & \cre 385 & 32.0&0.3   & 3.59&0.04  & $-$13.14&1.05  &     10.5 &      3.7 &      6.3 &      5.0 &      9.2 &     19.6 &     12.5 &      8.4 &     11.0 &      6.7 \\
$\lambda$ Ori A       &       51 & 35.4&0.6   & 3.54&0.06  & $-$13.03&0.75  &      1.2 &      1.4 &      2.6 &      1.5 &      0.9 &      1.3 &      1.9 &      1.2 &      1.1 &      1.0 \\
$\xi$ Per             & \cbl 193 & 36.0&0.4   & 3.58&0.06  & $-$12.69&1.15  &      6.1 &      6.5 &     13.4 &      6.9 &      5.3 &      7.8 &      6.8 &      5.6 &      5.2 &      7.3 \\
$\alpha$ Cam          &       95 & 29.7&0.6   & 2.97&0.06  & $-$12.31&0.40  &      6.6 &      9.9 &     21.2 &      6.4 &      5.2 &      7.1 &     12.4 &      5.2 &      5.4 &      6.8 \\
$\upsilon$ Ori        &       13 & 33.5&0.7   & 4.11&0.13  & $-$13.50&0.05  &      0.4 &      0.6 &      0.9 &      0.2 &      0.2 &      0.1 &      0.3 &      0.2 &      2.4 &      0.2 \\
10 Lac                &       13 & 35.4&0.7   & 3.93&0.09  & $-$14.00&0.51  &      0.6 &      1.0 &      1.7 &      0.3 &      0.4 &      0.3 &      0.3 &      0.3 &      1.0 &      0.3 \\
$\mu$ Col             &      110 & 33.1&0.4   & 3.81&0.05  & $-$13.50&0.09  &      1.4 &      0.9 &      1.7 &      1.2 &      0.8 &      1.2 &      1.1 &      1.5 &      2.5 &      0.9 \\
68 Cyg                & \cgr 312 & 35.8&0.4   & 3.65&0.08  & $-$12.71&1.15  &      9.8 &     12.1 &     23.1 &     14.8 &     10.1 &     15.1 &     17.7 &      9.3 &     14.8 &     12.5 \\
                      &          & \mcii{   } & \mcii{   } &  \mcii{      } & \bf  7.8 & \bf  9.0 & \bf 15.6 & \bf 14.3 & \bf  9.0 & \bf 11.3 & \bf 12.1 & \bf 12.0 & \bf  6.8 & \bf  8.1 \\
$\mu$ Nor             &       51 & 29.4&0.7   & 3.14&0.07  & $-$12.71&0.85  &      4.1 &      6.0 &     19.1 &      4.2 &      4.2 &      3.7 &      5.0 &      2.9 &      2.1 &      4.9 \\
19 Cep                &       50 & 32.1&0.3   & 3.31&0.05  & $-$12.69&0.85  &      3.7 &      4.4 &     11.1 &      3.4 &      2.3 &      2.6 &      4.4 &      2.0 &      1.8 &      3.5 \\
HD \num{151804}       &       72 & 30.9&0.9   & 3.04&0.14  & $-$12.12&0.30  &      7.8 &      9.9 &      5.0 &      8.9 &      5.8 &      5.2 &     10.6 &      7.9 &      7.2 &      6.7 \\
$\lambda$ Cep         & \cbl 200 & 36.3&0.6   & 3.51&0.09  & $-$12.31&0.65  &      4.3 &      5.5 &     18.1 &      6.8 &      4.8 &     11.7 &     10.2 &      6.2 &      5.6 &      5.8 \\
HD \num{135591}       &       60 & 35.1&0.3   & 3.55&0.08  & $-$13.05&0.75  &      1.1 &      0.6 &      1.1 &      0.8 &      0.4 &      0.8 &      1.3 &      0.4 &      0.8 &      0.3 \\
HD \num{188209}       &       54 & 30.8&0.4   & 3.15&0.07  & $-$12.51&0.95  &      5.6 &      7.7 &     22.3 &      4.1 &      3.8 &      5.4 &      8.5 &      4.2 &      3.6 &      6.4 \\
HD \num{55879}        &       28 & 31.5&0.7   & 3.54&0.09  & $-$13.50&0.06  &      1.0 &      1.2 &      0.7 &      0.7 &      0.7 &      0.6 &      1.0 &      0.4 &      0.7 &      0.6 \\
HD \num{152408}       &      --- & \mcii{---} & \mcii{---} & \mcii{---}     &      9.0 &      8.7 &      7.5 &      4.0 &      3.2 &      4.0 &      3.9 &      4.7 &      5.0 &      4.5 \\
1 Cam A               & \cgr 298 & 29.7&0.7   & 3.49&0.09  & $-$13.50&0.08  &      4.8 &      4.0 &      6.2 &      9.9 &      4.3 &      7.1 &     10.9 &      6.7 &     15.8 &      6.1 \\
HD \num{155806} A,B   &       49 & 36.1&0.8   & 3.69&0.09  & $-$13.00&0.13  &      4.8 &     28.8 &     21.9 &      2.9 &      2.3 &      1.7 &      2.9 &      4.1 &      5.5 &      8.5 \\
63 Oph                &       53 & 35.0&0.3   & 3.51&0.03  & $-$12.51&0.85  &      2.9 &      3.8 &      9.1 &      5.3 &      1.7 &      4.0 &      6.2 &      3.2 &      2.6 &      2.7 \\
9 Sge                 &       68 & 32.2&0.4   & 3.12&0.05  & $-$12.11&0.40  &      6.6 &      7.9 &     12.5 &      6.1 &      4.3 &      5.1 &     26.9 &      5.9 &      5.3 &      6.9 \\
AE Aur                &       13 & 34.8&0.9   & 4.11&0.12  & $-$13.50&0.06  &      0.7 &      1.2 &      1.0 &      1.0 &      0.5 &      0.3 &      0.6 &      0.5 &      2.5 &      0.4 \\
HD \num{57682}        &        6 & 35.5&1.0   & 4.17&0.12  & $-$12.73&0.75  &      0.9 &      4.0 &     23.3 &      0.5 &      0.6 &      0.5 &      0.5 &      0.7 &      1.2 &      1.6 \\
HD \num{207198}       &       51 & 33.1&0.3   & 3.32&0.05  & $-$12.71&0.85  &      3.2 &      3.9 &      9.8 &      3.2 &      2.2 &      2.7 &      5.1 &      2.2 &      2.3 &      2.8 \\
HD \num{47432}        &       97 & 29.4&0.6   & 3.04&0.06  & $-$12.51&0.95  &      6.1 &      9.5 &     25.3 &      4.6 &      4.0 &      4.5 &      4.1 &      4.8 &      3.4 &      5.6 \\
HD \num{68450}        &       38 & 30.9&0.4   & 3.39&0.04  & $-$13.06&0.65  &      1.2 &      2.0 &      3.0 &      1.2 &      1.2 &      0.9 &      1.3 &      1.1 &      2.1 &      1.1 \\
HD \num{152249}       &       71 & 31.6&0.6   & 3.26&0.05  & $-$12.50&0.95  &      5.7 &      8.1 &      9.8 &      4.7 &      4.3 &      5.3 &      8.8 &      4.2 &      3.9 &      5.8 \\
HD \num{154368}       &       65 & 30.4&0.6   & 3.09&0.11  & $-$12.52&0.95  &      4.9 &      6.5 &      7.6 &      3.9 &      3.1 &      4.7 &     10.4 &      6.2 &      7.1 &      5.3 \\
HD \num{202214} Aa,Ab &       25 & 32.1&1.0   & 3.86&0.15  & $-$13.50&0.12  &      0.9 &      0.8 &      3.1 &      1.1 &      0.9 &      1.0 &      0.6 &      0.7 &      1.9 &      0.3 \\
BN Gem                & \cbl 136 & 31.9&1.6   & 3.48&0.27  & $-$13.00&0.31  &      7.8 &      4.8 &      3.4 &     11.2 &      7.3 &      6.3 &      5.3 &      9.5 &     11.0 &      5.6 \\
HD \num{152424}       &       59 & 30.1&0.6   & 3.14&0.10  & $-$12.31&0.40  &     17.6 &     20.2 &     16.9 &     16.5 &     15.7 &     16.8 &     22.6 &     17.5 &     15.4 &     19.1 \\
                      &          & \mcii{   } & \mcii{   } &  \mcii{      } & \bf  9.6 & \bf 12.9 & \bf 15.2 & \bf  7.9 & \bf  7.2 & \bf  8.0 & \bf 14.2 & \bf  9.7 & \bf  7.0 & \bf 10.0 \\
HD \num{41161}        & \cgr 330 & 35.9&0.5   & 3.68&0.07  & $-$13.13&1.05  &      4.6 &      3.7 &      4.0 &      8.6 &      7.4 &     15.3 &      5.7 &      9.4 &      6.6 &      3.0 \\
PZ Gem                & \cgr 272 & 30.0&0.9   & 3.63&0.15  & $-$14.00&1.53  &    >99.9 &    >99.9 &    >99.9 &    >99.9 &     25.3 &     14.7 &     14.3 &     41.2 &     19.3 &    >99.9 \\
HD \num{175876}       & \cgr 282 & 36.2&0.4   & 3.52&0.05  & $-$12.72&1.15  &      5.2 &      4.4 &      8.0 &      9.8 &      5.5 &     10.1 &      8.3 &     12.1 &      4.9 &      4.7 \\
HD \num{209339} A     &       84 & 31.9&0.7   & 3.79&0.07  & $-$14.00&0.54  &      6.2 &      5.9 &      6.9 &      6.7 &      7.2 &      8.2 &      6.8 &      7.4 &      7.9 &      6.5 \\
                      &          & \mcii{   } & \mcii{   } &  \mcii{      } & \bf  3.7 & \bf  3.6 & \bf  4.4 & \bf  4.0 & \bf  4.8 & \bf  4.8 & \bf  3.6 & \bf  3.7 & \bf  5.0 & \bf  3.6 \\
HD \num{93521}        & \cre 386 & 32.1&0.4   & 3.53&0.08  & $-$12.73&1.15  &     13.3 &     12.2 &     12.9 &     18.2 &     14.4 &     28.5 &     15.9 &     15.9 &     21.5 &     13.9 \\
                      &          & \mcii{   } & \mcii{   } &  \mcii{      } & \bf  8.2 & \bf  6.8 & \bf  5.5 & \bf 17.1 & \bf  8.7 & \bf 25.1 & \bf 12.5 & \bf 11.1 & \bf 21.7 & \bf  6.1 \\
HD \num{161807}       & \cre 400 & 30.1&1.6   & 3.49&0.25  & $-$13.00&0.16  &      8.8 &      9.9 &     10.5 &     31.7 &     15.3 &     22.3 &     19.4 &     19.5 &     38.6 &     11.7 \\
HD \num{175754}       & \cbl 182 & 33.7&0.5   & 3.27&0.05  & $-$12.53&0.95  &      5.0 &      7.5 &     18.6 &      9.2 &      4.4 &      8.9 &      8.5 &      6.6 &      4.9 &      6.6 \\
HD \num{34656} Aa     &       66 & 36.0&0.3   & 3.50&0.03  & $-$12.71&0.85  &      3.0 &      3.2 &      4.7 &      4.0 &      2.6 &      4.1 &     12.8 &      2.7 &      2.6 &      2.7 \\
HD \num{46150}        &       59 & 40.9&1.1   & 3.74&0.07  & $-$13.34&0.65  &      5.9 &      6.3 &      5.3 &      9.4 &      6.5 &     15.4 &      6.1 &     23.0 &      9.2 &      3.6 \\
                      &          & \mcii{   } & \mcii{   } &     \mcii{   } & \bf  6.2 & \bf  6.0 & \bf  6.5 & \bf  7.1 & \bf  5.9 & \bf  8.5 & \bf  6.0 & \bf 31.9 & \bf  4.9 & \bf  7.4 \\
HD \num{69106}        & \cgr 277 & 29.3&0.5   & 3.42&0.06  & $-$13.46&0.95  &      5.5 &      7.1 &      7.1 &     14.5 &      8.8 &     10.4 &     11.2 &      7.6 &     14.3 &      5.4 \\
\midrule
\multicolumn{18}{l}{Notes. n index colour code for \vsini: none, {\cbl (n)}, {\cgr n}, {\cre nn}. Values in boldface values are calculated using the orbital solution for the system.} \\
\end{tabular}
\addtolength{\tabcolsep}{1mm}
}
\label{results1}
\end{table*}

\begin{table*}
\caption{RV results.}
\centerline{
\addtolength{\tabcolsep}{-1mm}
\begin{tabular}{l@{\hspace{5mm}}rr@{$\pm$}r@{\hspace{5mm}}rrr@{$\pm$}r@{\hspace{5mm}}r@{$\pm$}rccr@{$\pm$}rcr@{$\pm$}r}
\midrule
Name                  & \mciii{FASTWIND\hphantom{kkkk}} & \mciv{\rule[0.5ex]{5mm}{0.5pt} Line fitting \rule[0.5ex]{5mm}{0.5pt}} & \mciv{Literature}             & \mciii{Recomm.}    & \mcii{Diff.}           \\
                      & $n_l$ & \mcii{(km/s)}           & $n_r$ & $n_l$ & \mcii{(km/s)}                                         & \mcii{(km/s)} & $n_u$ & $n_d$ & \mcii{(km/s)} & ty & \mcii{(km/s)}          \\
\midrule
$\zeta$ Pup           &     7 &     $-$1.6&    6.0 &     3 &     15 &     $-$0.7&    2.5 &  $-$24.8& 3.4 &    11 &     0 &   $-$0.8& 2.3 &  3 & \cre $-$24.0&\cre  4.1 \\
$\zeta$ Oph           &    13 &     $-$4.6&   11.8 &    -- &    --- &       \mcii{---}   &   $-$5.7&16.1 &    17 &     0 &      6.7& 7.0 &  5 & \cre $-$12.4&\cre 17.6 \\
$\lambda$ Ori A       &    16 &       33.6&    2.4 &    11 &    349 &       31.1&    2.4 &     33.1& 2.1 &    21 &     0 &     33.6& 2.4 &  1 & \cre  $-$0.5&\cre  3.2 \\
$\xi$ Per             &    11 &       61.0&    6.4 &     6 &     56 &       63.7&    3.4 &     64.0& 5.7 &    16 &     3 &     61.0& 6.4 &  1 & \cbl     3.0&\cbl  8.6 \\
$\alpha$ Cam          &    12 &       16.8&    4.0 &    11 &    288 &       13.7&    5.2 &     10.1& 5.6 &    22 &     0 &     16.8& 4.0 &  1 & \cre  $-$6.7&\cre  6.9 \\
$\upsilon$ Ori        &    16 &       17.7&    2.2 &    11 &    417 &       16.8&    0.4 &     17.0& 0.6 &     9 &     1 &     16.8& 0.4 &  2 & \cbl     0.1&\cbl  0.8 \\
10 Lac                &    14 &     $-$8.6&    2.1 &    11 &    399 &    $-$10.8&    0.4 &   $-$9.0& 1.7 &    21 &     1 &  $-$10.7& 0.4 &  2 & \cbl     1.7&\cbl  1.8 \\
$\mu$ Col             &    14 &      107.6&    3.8 &    11 &    311 &      107.8&    0.3 &    109.1& 1.4 &    15 &     1 &    107.8& 0.3 &  2 & \cbl     1.3&\cbl  1.4 \\
68 Cyg                &    10 &       23.8&    6.7 &     4 &     59 &       20.8&    0.9 &   $-$0.4&15.2 &    17 &     1 &     23.8& 6.7 &  1 & \cre $-$24.2&\cre 16.6 \\
$\mu$ Nor             &    11 &       12.6&    2.8 &    11 &    247 &        6.7&    3.3 &      6.3& 1.4 &    14 &     1 &     12.6& 2.8 &  1 & \cre  $-$6.3&\cre  3.1 \\
19 Cep                &    11 &     $-$9.6&    2.0 &    11 &    251 &    $-$14.2&    2.1 &  $-$11.2& 3.9 &    16 &     1 &   $-$9.6& 2.0 &  1 & \cre  $-$1.6&\cre  4.4 \\
HD \num{151804}       &    10 &    $-$42.8&    4.8 &     4 &     56 &    $-$44.1&    3.3 &  $-$56.2&10.9 &    13 &     2 &  $-$42.8& 4.8 &  1 & \cre $-$13.4&\cre 11.9 \\
$\lambda$ Cep         &    10 &    $-$65.3&    5.7 &     6 &     37 &    $-$61.3&    2.4 &  $-$73.8& 2.0 &    17 &     2 &  $-$65.3& 5.7 &  1 & \cre  $-$8.5&\cre  6.0 \\
HD \num{135591}       &    16 &        2.0&    2.8 &    11 &    196 &        1.1&    1.4 &   $-$1.0& 3.9 &    11 &     0 &      1.3& 1.3 &  2 & \cre  $-$2.3&\cre  4.1 \\
HD \num{188209}       &    13 &     $-$6.1&    2.3 &    11 &    275 &    $-$11.3&    3.1 &   $-$6.7& 5.5 &    18 &     0 &   $-$6.1& 2.3 &  1 & \cre  $-$0.6&\cre  6.0 \\
HD \num{55879}        &    17 &       28.6&    2.1 &    11 &    396 &       26.0&    0.8 &     32.6& 0.8 &     8 &     0 &     28.6& 2.1 &  1 & \cbl     4.0&\cbl  2.3 \\
HD \num{152408}       &    -- &       \mcii{---}   &     3 &      6 &    $-$30.8&   11.3 &  $-$96.2&68.9 &    13 &     0 &  $-$30.8&11.3 &  4 & \cre $-$65.4&\cre 69.9 \\
1 Cam A               &    11 &     $-$0.7&    7.5 &    10 &    128 &     $-$2.0&    2.3 &   $-$0.3& 9.2 &     6 &     0 &   $-$0.7& 7.5 &  1 & \cbl     0.4&\cbl 11.9 \\
HD \num{155806} A,B   &    12 &        5.5&    3.8 &     9 &    172 &        4.7&    2.3 &      8.0& 5.8 &     9 &     1 &      4.9& 2.0 &  2 & \cbl     3.1&\cbl  6.1 \\
63 Oph                &    11 &     $-$3.6&    2.9 &    10 &    195 &     $-$6.4&    3.5 &  $-$11.9& 2.2 &     8 &     1 &   $-$3.6& 2.9 &  1 & \cre  $-$8.4&\cre  3.6 \\
9 Sge                 &    11 &       15.6&    4.2 &     9 &    160 &       14.6&    5.7 &     11.7& 5.8 &    22 &     1 &     15.6& 4.2 &  1 & \cre  $-$3.9&\cre  7.2 \\
AE Aur                &    16 &       54.0&    1.7 &    11 &    421 &       53.5&    0.3 &     57.1& 2.5 &    25 &     1 &     53.5& 0.3 &  2 & \cbl     3.5&\cbl  2.5 \\
HD \num{57682}        &    10 &       23.0&    1.7 &    11 &    410 &       22.9&    0.5 &     24.2& 1.8 &    13 &     1 &     22.9& 0.5 &  2 & \cbl     1.3&\cbl  1.9 \\
HD \num{207198}       &    16 &    $-$19.4&    3.0 &    11 &    261 &    $-$22.4&    3.2 &  $-$18.3& 2.1 &    13 &     0 &  $-$19.4& 3.0 &  1 & \cbl     1.1&\cbl  3.7 \\
HD \num{47432}        &    14 &       64.6&    4.3 &    11 &    161 &       64.2&    1.4 &     58.6& 4.0 &    10 &     0 &     64.6& 4.3 &  1 & \cre  $-$6.0&\cre  5.9 \\
HD \num{68450}        &    14 &       35.0&    1.9 &    11 &    363 &       31.8&    2.0 &     36.4& 2.5 &     8 &     1 &     35.0& 1.9 &  1 & \cbl     1.4&\cbl  3.1 \\
HD \num{152249}       &    13 &    $-$19.3&    2.2 &    11 &    242 &    $-$23.9&    3.6 &  $-$25.7&13.2 &    15 &     0 &  $-$19.3& 2.2 &  1 & \cre  $-$6.4&\cre 13.4 \\
HD \num{154368}       &    12 &       17.8&    2.3 &    11 &    213 &       14.7&    3.5 &      8.7& 6.4 &    10 &     0 &     17.8& 2.3 &  1 & \cre  $-$9.1&\cre  6.8 \\
HD \num{202214} Aa,Ab &    16 &    $-$17.0&    1.7 &    11 &    414 &    $-$20.6&    1.0 &  $-$16.5& 1.0 &     7 &     0 &  $-$19.7& 0.9 &  2 & \cbl     3.2&\cbl  1.4 \\
BN Gem                &     8 &       19.6&    8.4 &     7 &     75 &       20.0&    7.8 &     14.9&10.8 &     9 &     2 &     19.8& 5.7 &  2 & \cre  $-$4.9&\cre 12.2 \\
HD \num{152424}       &    12 &    $-$28.9&    3.6 &    10 &    220 &    $-$29.8&    3.1 &  $-$22.5&16.5 &     9 &     0 &  $-$28.9& 3.6 &  1 & \cbl     6.4&\cbl 16.9 \\
HD \num{41161}        &    13 &    $-$20.1&    7.3 &     7 &     57 &    $-$20.8&    2.8 &   $-$7.9&19.4 &     7 &     0 &  $-$20.7& 2.6 &  2 & \cbl    12.9&\cbl 19.5 \\
PZ Gem                &     7 &        9.0&    8.9 &     7 &    105 &        8.7&    5.1 &      1.0&11.1 &     9 &     0 &      9.0& 8.9 &  1 & \cre  $-$8.0&\cre 14.2 \\
HD \num{175876}       &    10 &        7.2&    5.7 &     7 &     76 &        8.3&    2.8 &      8.2&10.6 &    13 &     0 &      7.2& 5.7 &  1 & \cbl     1.0&\cbl 12.0 \\
HD \num{209339} A     &    13 &    $-$13.6&    2.8 &    11 &    283 &    $-$13.0&    2.2 &  $-$11.7&15.4 &     8 &     0 &  $-$13.2& 1.7 &  2 & \cbl     1.5&\cbl 15.5 \\
HD \num{93521}        &     9 &        4.6&    7.8 &    -- &    --- &       \mcii{---}   &  $-$13.6& 3.1 &     8 &     0 &   $-$1.4& 2.8 &  5 & \cre $-$12.1&\cre  4.1 \\
HD \num{161807}       &    10 &    $-$24.0&    6.8 &    -- &    --- &       \mcii{---}   &  $-$24.8& 8.1 &     3 &     0 &  $-$24.5& 6.2 &  5 & \cre  $-$0.3&\cre 10.2 \\
HD \num{175754}       &    12 &        4.6&    4.9 &     9 &    109 &        3.5&    5.7 &   $-$8.0& 5.8 &     7 &     1 &      4.6& 4.9 &  1 & \cre $-$12.6&\cre  7.6 \\
HD \num{34656} Aa     &    11 &     $-$7.1&    3.2 &    10 &    240 &     $-$7.9&    2.2 &   $-$4.7& 5.2 &    12 &     2 &   $-$7.1& 3.2 &  1 & \cbl     2.4&\cbl  6.1 \\
HD \num{46150}        &    12 &       35.7&    2.0 &     5 &     44 &       30.3&    3.4 &     37.4& 3.2 &    17 &     1 &     34.3& 1.7 &  2 & \cbl     3.1&\cbl  3.6 \\
HD \num{69106}        &    13 &       18.1&    8.4 &    10 &    123 &       15.5&    8.7 &      9.3& 5.4 &     8 &     0 &     18.1& 8.4 &  1 & \cre  $-$8.8&\cre 10.0 \\
\midrule
\multicolumn{17}{l}{$n_l$: \# of lines used. $n_r$: \# of regions used. $n_u$: \# of values used. $n_d$: \# of values discarded. ty: See subsection~\ref{subsection41}.}
\end{tabular}
\addtolength{\tabcolsep}{1mm}
}
\label{results2}
\end{table*}

\subsection{Data}

$\,\!$\indent As presented in Paper I, the primary source of data for \MTW\ is \lili\ (Library of Libraries of Massive-star high-Resolution
spectra, \citealt{Maizetal19a}), a collection of $\sim 10^5$ ground-based high-resolution spectroscopic epochs of hot (mostly massive)
stars from database searches (such as ESO and OHP) and our own programs at ORM, CAHA, and OLS. The number of epochs in \lili\ for 
approximately half of our sample was already sufficient for our needs at the beginning of our project (2025). For the rest, we added more
epochs with HERMES@Mercator, FIES@NOT, and FEROS@OLS2.2 in 2025-26 to reach the desired number. In total, we used data from eight telescopes
listed in Tables~\ref{sample}~and~\ref{spectrographs} (see the \hyperref[appendixd]{Glossary} for the meaning of the acronyms) for the 
analysis of the blue-violet spectral region and we added CARMENES@CAHA3.5 spectra for the analysis of the yellow-red spectral region.

The goal for this paper is to analyse \textit{a complete magnitude-limited sample of single spectroscopic O stars.} We settled on a limit of 
$B = 7.0$ as a compromise between the need of having a sample large enough and the limitations of data availability in \lili. The sample 
was selected from the O stars in our GOSC \citep{Maizetal04b} and GOSSS \citep{Maizetal11} projects, using OWN and MONOS 
\citep{Maizetal19b,Trigetal21,Holgetal25a} to eliminate the spectroscopic binaries. The result from the selection process is that the 
79 Galactic O-type systems with $B \le 7.0$ and contained within a maximum separation of 0\farcs5 can be divided into: 

\begin{itemize}
 \item Spectroscopic binaries: 37 systems have 1-3 calculated orbits each, mostly spectroscopic, some also astrometric. The record-holder of 
       three orbits in this group is $\tau$~CMa~A, to which a future paper of the \MTW\ series will be dedicated \citep{Rosuetal25}, 
       and includes two O stars with $B \le 7.0$ (counted as one system). $\delta$~Ori~Aa,Ab also includes two O stars with $B \le 7.0$
       within 0\farcs5 and is also counted as one system. For $\zeta$~Ori~Aa,Ab (the brightest Galactic O-type system in the
       $B$ band) and its visual companion $\zeta$~Ori~B (the sixth brightest) we make the exception of counting them as one system, as they
       are separated by 2\farcs4 but are one of the closest O-type systems and are expected to follow an orbit with a period longer than 
       1000~a \citep{Masoetal01}. We plan to dedicate a separate analysis to the little-studied $\zeta$~Ori~B, as with good seeing it is
       possible to obtain ground-based \lili\ spectra uncontaminated from the brighter Aa,Ab contribution.
 \item Single spectroscopic stars, the sample in this paper: 41 objects with no spectroscopic orbit or clear sign of spectroscopic
       motion at the beginning of this study, listed in Table~\ref{sample}. Some stars have visual companions, either 
       included in the spectrograph aperture (e.g. HD~\num{155806}) or outside of it (e.g. $\lambda$~Ori~A) but 
       show no signs of orbital motion.
 \item FN~CMa~B is an O star with a brighter B0.7~Ib companion 0\farcs57 away unresolved in ground-based high-resolution spectroscopy, 
       the companion being an SB1 \citep{Rivietal11b}. We exclude it from our sample because we cannot spatially resolve it with \lili\ 
       data (even though we did it in \citealt{Maizetal21b} with lucky spectroscopy) but we note it is at least a three-body system. 
\end{itemize}

\subsection{Methods}

$\,\!$\indent The spectra were rectified, ISM-cleaned, and telluric-line corrected using UNWIND, the disentangling code
presented in Paper I, which can be applied to single stars for those purposes. The details of the UNWIND runs are given in the next
section. The \ion{Ca}{ii}~H+K ISM lines were analysed epoch by epoch to determine which had wavelength calibration errors; those were
discarded. The number of epochs selected for each star and spectrograph is given in Table~\ref{sample}. RVs were determined
by either cross correlation with a template or by profile fitting, depending on the different parts of the analysis in Section~\ref{section3}. We
used different routines written in IDL for the \MTW\ project. In particular, multiple-line profile fitting was done with the LINESFIT code,
which uses the non-linear least squares package MPFIT \citep{Mark09} to fit Gaussian and Stark profiles broadened by rotation and spectrograph
resolution and with the possibility of adding constraints to the different fitted parameters.

As described below, some of the UNWIND runs were degraded to $R=2500$ for the purpose of spectral classification. The classification itself was
done using MGB \citep{Maizetal12}, a visual interactive tool that compares the spectrum to be classified with a grid of standards that allows
for the possibility of using different rotation broadening indices and that follows the general criteria of spectral classification
\citep{Maizetal26a}. The grid of standards is an update of the one presented in the GOSSS project in \citet{Maizetal16} and will appear in a
separate paper (Ma\'{\i}z Apell\'aniz et al. in prep.). 

The effective temperature (\Teff) and surface gravity (\logg) were derived through quantitative spectroscopic 
analysis using a grid of FASTWIND stellar atmosphere models \citep{Santetal97,Pulsetal05,RivGetal11} and the IACOB-GBAT fitting tool 
\citep{SimDetal11d}, while projected rotational velocities (\vsini) were determined with IACOB-BROAD following the Fourier-transform 
and goodness-of-fit approach described by \citet{SimDHerr07,SimDHerr14}. Results are shown in Table~\ref{results1}. 
The values for some of the stars here had already been presented 
elsewhere \citep{Holg19} but some have been done for the first time using the UNWIND output and taking advantage of observations of strong
Oe stars obtained at low-emission phases. Only for one O~Iafpe star, HD~\num{152408}, we were unable to complete a FASTWIND analysis due 
to its strong wind, as discussed in Sect.~\ref{section3}. In addition, 
Appendix~\ref{appendixc} presents a second grid of FASTWIND models derived for the purposes of analysing the 
\hyperref[subsection33]{contamination of Balmer lines by \ion{He}{ii}}, the 
\hyperref[subsection34]{presence of forbidden components in two \ion{He}{i} lines}, and the 
\hyperref[subsection35]{effect of wind infilling for H+He lines}.

\section{Results}\label{section3}

\subsection{UNWIND runs and spectral classification}\label{subsection31}

\begin{table}
\caption{Spectrograph configurations used for the blue-violet spectral region.}
\centerline{
\begin{tabular}{ccrr@{-}l}
\midrule
Telescope & Spectrograph & \mci{Resolution} & \mcii{Wav. range}      \\
          &              &                  & \mcii{(\AA)}           \\
\midrule
Mercator  & HERMES       & \num{85000}      & \num{3785}&\num{8995}  \\
OHP1.9    & SOPHIE       & \num{40000}      & \num{3875}&\num{6930}  \\
          &              & \num{75000}      & \num{3875}&\num{6930}  \\
OLS2.2    & FEROS        & \num{48000}      & \num{3570}&\num{9210}  \\
NOT       & FIES         & \num{25000}      & \num{3700}&\num{9000}  \\
TNG       & HARPS-N      & \num{115000}     & \num{3875}&\num{6905}  \\
OLS3.6    & HARPS        & \num{120000}     & \num{3785}&\num{6905}  \\
VLT       & UVES         & \num{59000}      & \num{3760}&\num{4975}  \\
SALT      & HRS          & \num{65000}      & \num{3850}&\num{5560}  \\
\midrule
\end{tabular}
}              
\label{spectrographs}   
\end{table}

$\,\!$\indent To obtain our results, we performed different UNWIND runs. First, we assumed zero RV for all epochs and covered in different runs
the line-rich blue-violet region plus three lines in the yellow-red spectral region: \ion{He}{ii}~$\lambda$5411.53, \ion{He}{i}~$\lambda$5875.70, and
H$\alpha$.
Assuming zero RV allowed us to directly process all epochs and obtain clean spectra (rectified and with ISM 
lines eliminated) that could be used to determine velocities and other line characteristics for a large number of epochs. For five stars, 68~Cyg, 
HD~\num{152424}, HD~\num{209339}~A, HD~\num{93521}, and HD~\num{46150}, we calculated new SB orbits (see below) and for those we did independent 
runs with the orbit applied but with a zero systemic velocity. 
The numbers of spectra used in the blue-violet region are given in Table~\ref{sample}.
For the second runs, we used the systemic velocities calculated in this paper (and for the five stars mentioned above, the full new SB orbits)
to derive their spectral classifications, as a first step in our analysis. Doing an UNWIND run with the star
systemic velocity shifts the output to the stellar reference frame, which is needed for spectral classification.

The spectra resulting from the second runs are shown in Figs.~\ref{ALS_spectra_1}~to~\ref{ALS_spectra_Oslow3} degraded to $R=2500$, as
appropriate for spectral classification \citep{Maizetal26a},
and without ISM lines (though for some stars a residual \ion{Na}{i}~D doublet absorption is seen).
The spectral classifications are given in Table~\ref{sample}, along with their distances $d$ from \citet{Pantetal25b}. 
Here we briefly discuss the spectral classifications dividing them in four groups (also used for the figures in Appendix~C): seven Oe stars, 
six Onfp stars, one O~Iafpe star, and the rest of the sample. 

Classical Oe stars are the extension of classical Be stars to hotter \Teff, and are non-supergiant fast-rotating stars with emission lines 
originating from a disk that is tied up to pulsations in the star \citep{Neguetal04,Maizetal26a}. The emission-line profile is 
double-peaked when seen at high inclinations and single-peaked when seen close to pole-on and can experience drastic variations as the disk
grows and shrinks. Emission is seen mostly in Balmer lines (stronger in H$\alpha$) and \ion{He}{i} lines, with huge differences among 
them (e.g. compare the three lines in the 4900-5080~\AA\ range in Fig.~\ref{ALS_spectra_Oe_strong}).
Three stars in our sample (PZ~Gem, BN~Gem, and HD~\num{155806}~A,B) are well-known strong Oe stars with a high degree
of variability, as shown in Fig.~\ref{ALS_spectra_Oe_strong}, where for each star we show two average spectra, one derived from epochs with
the system in a high state and the other in a low state. Another four stars in our sample (HD~\num{57682}, $\zeta$~Oph, HD~\num{93521}, and
HD~\num{161807}, Fig.~\ref{ALS_spectra_1}) are weak Oe stars, with emission lines not visible or barely so in the blue-violet region and
only clearly visible in H$\alpha$ (and even there sometimes almost disappearing, as it is the case for $\zeta$~Oph). Of the seven Oe stars, 
two are close to pole-on (single-peaked, HD~\num{155806}~A,B and HD~\num{57682}), hence showing little rotational broadening in the stellar
absorption profiles (HD~\num{57682} has the lowest measured \vsini\ of the whole sample, so it must have an extremely low inclination).
The other five have higher inclinations, which is evidenced in their rotation indices of (n), n, or nn. To our knowledge, two of the weak 
Oe stars, HD~\num{57682} and HD~\num{93521}, had never been reported as Oe stars before. The first one is highly variable, with H$\alpha$ 
emission becoming weak at times in the time span of our spectra (which starts in 2002). The second one
is more steady in the data we have starting in 1994, but the detection of the weak ``horns'' (emission peaks at both sides of an absorption
line) around \ion{He}{i}~$\lambda$5015.6776 or 
H$\alpha$ is difficult. $\zeta$~Oph, HD~\num{93521}, and HD~\num{161807} are ONnn stars of luminosity class
II to IV, that is, rapidly-rotating, evolved, and nitrogen enriched O stars, possibly the result of a merger
(\citealt{Walbetal11,Walbetal16}; see also \citealt{Sciaetal26}). 
They receive the ON classification for the first time here, primarily due to the higher S/N of the spectra compared to previous efforts and to 
their strong \ion{N}{iii}~$\lambda$4513 triplet in absorption. To our knowledge, this is the first time that ONnn stars have been also 
identified as Oe stars, connecting two spectral classification phenomena linked by high values of \vsini.

\begin{figure*}
 \centerline{\includegraphics*[width=0.49\linewidth]{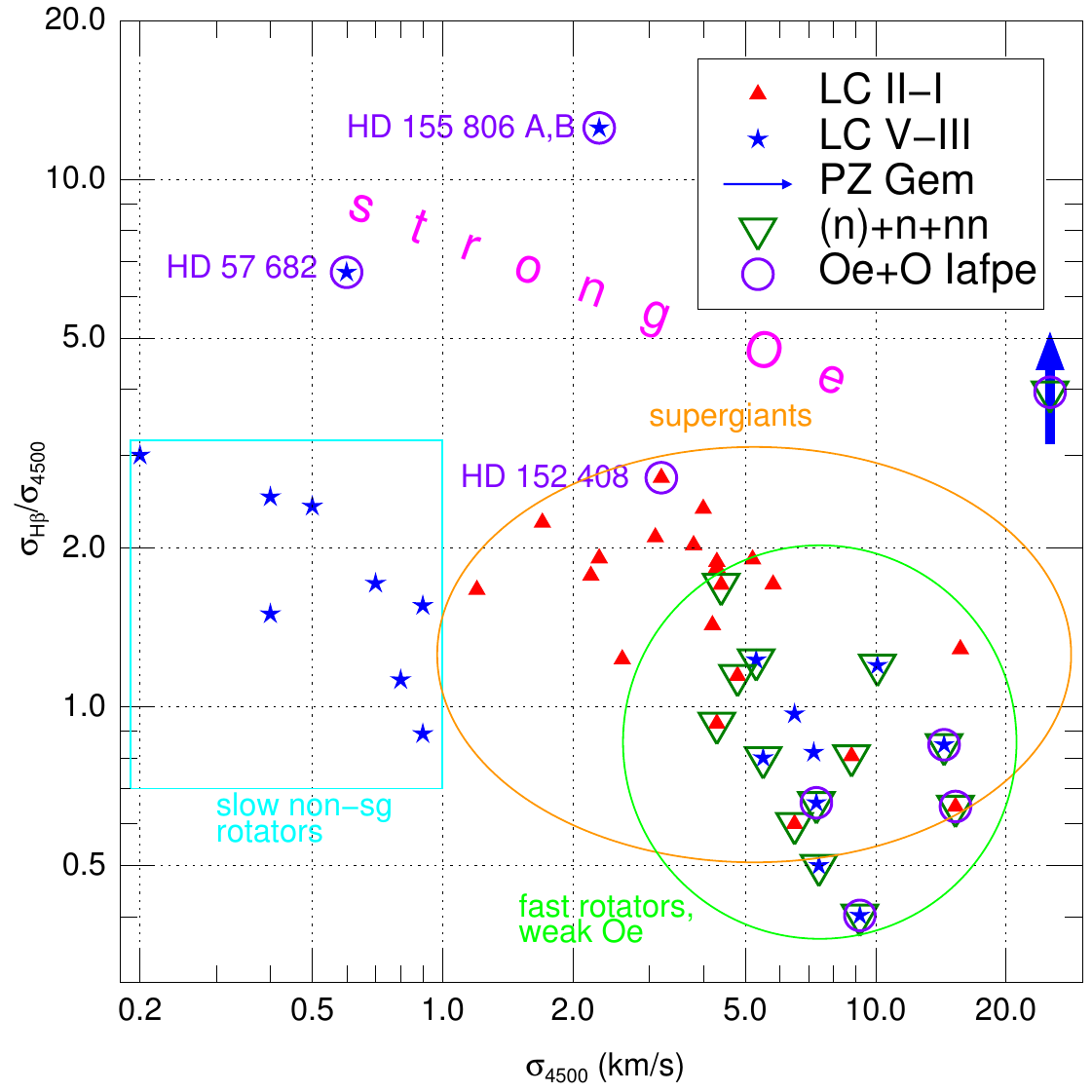} \
             \includegraphics*[width=0.49\linewidth]{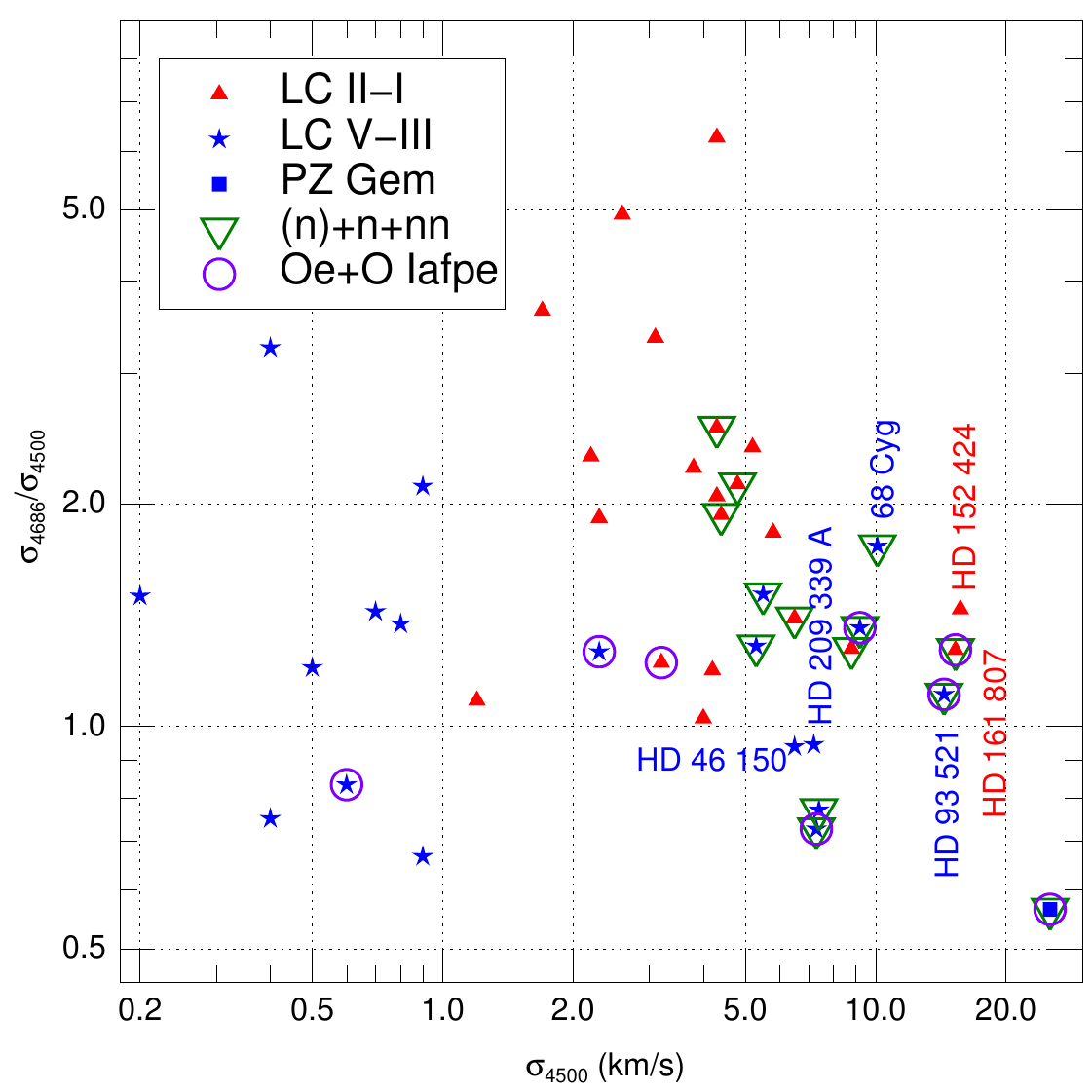}}
 \caption{RV dispersion plots comparing $\sigma_{4500}$ with $\sigma_{\rm H\beta}$ (left) and $\sigma_{4686}$ (right). The sample is divided
          into three sets: supergiants (LC II and I), non-supergiants (LC V to III), and PZ Gem (the strongest 
          Oe star in the sample, for which the calculated values are of little use and only a lower limit for the ratios in the vertical axes
          can be derived). In addition, fast rotators and Oe stars are marked with additional symbols and different groups (or individual
          stars) are identified in each panel (see the text).}
 \label{dispersion}      
\end{figure*}

\begin{figure}
 \centerline{\includegraphics*[width=\linewidth]{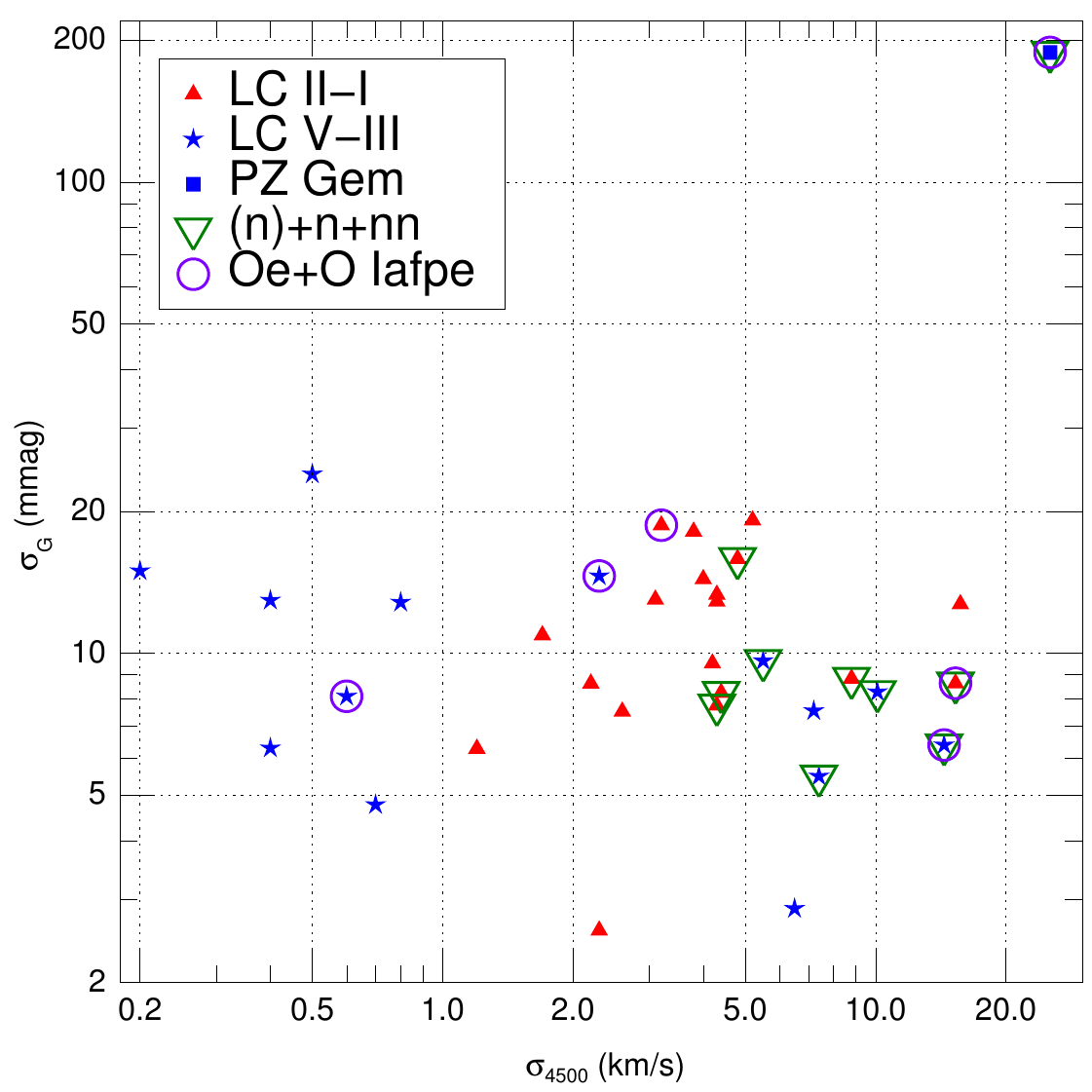}}
 \caption{Comparison between the RV dispersion ($\sigma_{4500}$) from this paper and the photometric dispersion ($\sigma_G$) from
          \citet{Maizetal23} for the 34 stars with values in the latter.}
 \label{RVphot}      
\end{figure}

Onfp stars are Of stars where \ion{He}{ii}~$\lambda$4685.71 shows an emission profile with an absorption reversal, similar to those of
high-inclination Oe stars (for Balmer and \ion{He}{i} lines) described in the previous paragraph, 
and most of them are fast rotators \citep{Walb72,Sotaetal11a}. Six of the 
stars in our sample display such characteristics: $\zeta$~Pup, $\lambda$~Cep, HD~\num{175754}, HD~\num{175876}, $\xi$~Per, and 68~Cyg
(Fig.~\ref{ALS_spectra_Onfp}). To our knowledge, the last three had never been reported as Onfp, likely because the weakness of their
features allows them to be seen only in high S/N clean spectra such as ours, though 68~Cyg was at one point reported as an Oe
\citep{Neguetal04}. The likely reason in the case of 68~Cyg is its high H$\alpha$ variability (Fig.~\ref{ALS_spectra_Onfp}), with the line
switching between full absorption to filling most of it with a double-peaked profile. We note, however, the lack of classical Oe profiles 
in the blue-violet spectra and the constancy of the \ion{He}{ii}~$\lambda$4685.71 profile. 

HD~\num{152408} is an O~Iafpe or ``cool slash'' star, not a classical Oe star despite the e suffix in the spectral classification
\citep{Sotaetal14}. These objects are supergiants with very high mass loss rates (Fig.~\ref{ALS_spectra_1}). Hence, we were unable to 
derive a FASTWIND model for this star and some results are missing for it in Table~\ref{results1} and the analysis below. 
Despite the high intensity of the emission lines in HD~\num{152408}, their profiles are quite stable, especially when compared to
those of classical Oe stars.

The remaining 27 objects in our sample are normal O stars and their spectra are shown in Fig~\ref{ALS_spectra_On} for the three fast
rotators and in Figs.~\ref{ALS_spectra_Oslow1}~to~\ref{ALS_spectra_Oslow3} for the 24 slow rotators. The normal O stars span from dwarfs to
supergiants, with the sample heavily biased towards late types (two thirds of the 27 objects are of subtypes O9 to O9.7).
We have discovered that the earliest star, HD~\num{46150}, is an SB2, and we 
disentangled the contribution of the secondary (Ab), resulting in a spectral classification of O4~V((fc))z for the primary (Aa), one subtype
earlier than the (contaminated) result of \citet{Sotaetal11a}. 

\subsection{Multi-epoch cross-correlation with data}\label{subsection32}

$\,\!$\indent In this subsection we present our work on the RV variability of the sample. We analysed the RV stability of
different wavelength regions to determine how appropriate they are for the measurement of the true stellar RV, what are the detectability 
limits of spectroscopic binaries, and the origins of the variability in different wavelength regions: pulsations, winds, and/or disks.

For each star we selected a representative epoch that has a good S/N, we cross-correlate it with all epochs, and we measure the dispersion in the RVs. 
For this procedure, we selected ten different wavelength regions, each producing a dispersion measurement. Three are centred on Balmer lines:
$\sigma_{\rm H\gamma}$ (4320-4360~\AA), $\sigma_{\rm H\beta}$ (4840-4860~\AA), and $\sigma_{\rm H\alpha}$ (4840-4860~\AA). Another three are centred 
on single strong He lines: $\sigma_{4686}$ (\ion{He}{ii}~$\lambda$4685.71, 4670-4695~\AA), $\sigma_{5412}$ ( \ion{He}{ii}~$\lambda$5411.53, 5400-5424~\AA),
and $\sigma_{5876}$ (\ion{He}{i}~$\lambda$5875.70, 5861-5888~\AA). The last four are larger regions without Balmer lines and with He and metallic lines: 
$\sigma_{4200}$ (4110-4320~\AA), $\sigma_{4500}$ (4360-4600~\AA), $\sigma_{4650}$ (4600-4670~\AA), and $\sigma_{4980}$ (4880-5080~\AA).
Results are given in Table~\ref{results1} and plotted in Fig.~\ref{dispersion}. In addition,
Fig.~\ref{RVphot} shows a comparison between $\sigma_{4650}$ and $\sigma_G$, the photometric variability in the \textit{Gaia} $G$ band from
\citet{Maizetal23}, for the 34 stars in common between both samples.

Before analysing the results, we mention the PZ~Gem case. As seen in Fig.~\ref{ALS_spectra_Oe_strong}, it is an extreme Oe star 
for which the spectrum differs completely between the high and low states among our epochs due to changes in its disk. Therefore, the 
cross-correlation results for PZ~Gem are mostly meaningless other than to say its spectrum is highly variable. The only exception is the 
relatively low value of $\sigma_{4650}$, 
which indicates that region (dominated by C and N absorption lines and without H and He lines) is the only one among the six that (at least 
partially) reflects the photospheric RV variability of the star. The location of PZ~Gem in Fig.~\ref{dispersion} is marked with an arrow due to
the issues regarding H$\beta$ but it can be considered to be in the upper right quadrant of the plot, as the H$\beta$ variability is extreme
for the star.

We find the two most useful dispersions to be $\sigma_{4500}$ and $\sigma_{\rm H\beta}$. The first one is dominated by 
\ion{He}{i}~$\lambda$4471.55 and \ion{He}{ii}~$\lambda$4541.591, the two main lines used for horizontal spectral classification in O
stars. Its variability is expected to originate mostly in pulsations and, possibly, by hidden spectroscopic binaries
\citep{SimDetal24}. H$\beta$, on the 
other hand, should also include the effects of disk activity in Oe stars and of winds in general. A comparison between those two 
quantities is shown in Fig.~\ref{dispersion}, which also includes a comparison with $\sigma_{4686}$, the \ion{He}{ii} line on 
the blue-violet range most affected by stellar winds for O stars. For discussion purposes, we divide the sample in different groups: normal
(or non Oe) non-supergiants, supergiants, and Oe stars.

\textit{Normal non-supergiant stars.} Normal stars with LC between V and III have $\sigma_{\rm H\beta} \sim \sigma_{4500}$. Slow rotators 
have ratios typically slightly above 1.0 and very low RV dispersions in general. The two exceptions to that rule, HD~\num{209339}~A and
HD~\num{46150}, turned out to be SB whose nature we discovered during our analysis and whose
orbital motion is the dominant source of their RV dispersions.
On the other hand, fast rotators have typical ratios of $\sim 1$ or lower and significantly higher (by an order of magnitude)
RV dispersions. Indeed, there is a gap between the two groups for $\sigma_{4500}$ in the range 1-5~km/s. As the main source of RV variability 
for non-supergiant non-Oe stars is expected to be pulsations, their amplitude appears to increase with \vsini. 

\textit{Supergiants.} All O supergiants in the sample have $\sigma_{4500} > 1.0$~km/s, so pulsations are in general more intense than for 
stars of lower LCs, a correlation already detected by \citet{SimDetal24}. For slow rotators, $\sigma_{\rm H\beta} > \sigma_{4500}$, indicating 
that H$\beta$ is likely affected by the wind variability characteristic of supergiants, but for fast rotators the effect is diminished, in the
same direction we see for lower luminosity classes. The effect of wind variability on supergiants is stronger than for dwarfs and
giants for $\sigma_{4686}$, as expected for that \ion{He}{ii} line, which is infilled by wind as luminosity increases at constant spectral
subtype \citep{Walb71a}. However, the range of values for $\sigma_{4686}/\sigma_{4500}$ is quite large. Of the supergiants in the sample, the 
one with the largest $\sigma_{\rm H\beta}/\sigma_{4500}$ is the O~Iafpe star HD~\num{152408}, not surprising given its strong H$\beta$ 
emission (Fig.~\ref{ALS_spectra_1}). Somewhat surprisingly, the recently discovered magnetic nature of 63~Oph \citep{Barretal25}
does not lead to large values of $\sigma_{\rm H\beta}$ or $\sigma_{4686}$.

\textit{Oe stars.} Oe stars have a diverse behaviour in Fig.~\ref{dispersion}. Fast-rotating weak Oe stars in general behave like their
fast-rotating normal counterparts, with $\sigma_{4500} \sim 10$~km/s and $\sigma_{\rm H\beta} \lesssim \sigma_{4500}$. Therefore, once
strong disk activity is removed, pulsations affect H$\beta$ for such stars even less than other lines. For the strong Oe
stars, HD~\num{155806}~A,B is located close to the top of the left panel of Fig.~\ref{dispersion}, also the likely real location of PZ~Gem.
However, BN~Gem, the third strong Oe, has $\sigma_{\rm H\beta} \sim \sigma_{4500}$ because in all of our epochs H$\beta$ is observed as a
double-peaked emission profile on top of the photospheric absorption profile. An interesting case is that of HD~\num{57682}, which until now
had not been identified as an Oe star but shows a clear variable H$\alpha$ emission and a weaker variability in H$\beta$ 
(Fig.~\ref{ALS_spectra_1}) that is sufficient to place the object in the top part of the left panel of Fig.~\ref{dispersion}. Yet,
HD~\num{57682} has the lowest \vsini\ of our sample (6~km/s) so it is an extreme (apparent) slow rotator. Those characteristics can be 
reconciled by a viewing angle that places the star (nearly) pole-on, which agrees with the single-peaked H$\alpha$ emission in
Fig.~\ref{ALS_spectra_1}. HD~\num{57682} and HD~\num{93521} (the other Oe star in our sample not previously identified as such) have similar
spectral classifications with the difference of the nn rotation index for HD~\num{93521}. Ignoring the latitudinal variation
in \Teff\ in rapidly rotating stars, this means that we are seeing two very similar stars from near-perpendicular viewing angles. If that is the
case, the very different position of the two stars in the left panel of Fig.~\ref{dispersion} suggests that the observed $\sigma_{4500}$ is a
strong function of viewing angle for fast rotators: pulsations would cause significant changes in the broad profiles observed from 
directions close to the stellar equator but would introduce almost no shifts in the centroids of the narrow lines observed from directions
close to the poles.

For the other RV dispersions, $\sigma_{\rm H\gamma}$ and $\sigma_{\rm H\alpha}$ have a behaviour similar to that of $\sigma_{\rm H\beta}$, 
as expected for the Balmer lines, with H$\alpha$ being the most variable of the three in most cases. Most of the rest behave in a similar way to 
$\sigma_{4500}$, with some differences due to the added noise in early-O subtypes for \ion{He}{i} and in late-O subtypes for \ion{He}{ii}.
$\sigma_{4650}$ shows more differences due to the C and N lines in that range that are in absorption in some stars and in emission in others.

\begin{figure*}
 \centerline{\includegraphics*[height=22mm]{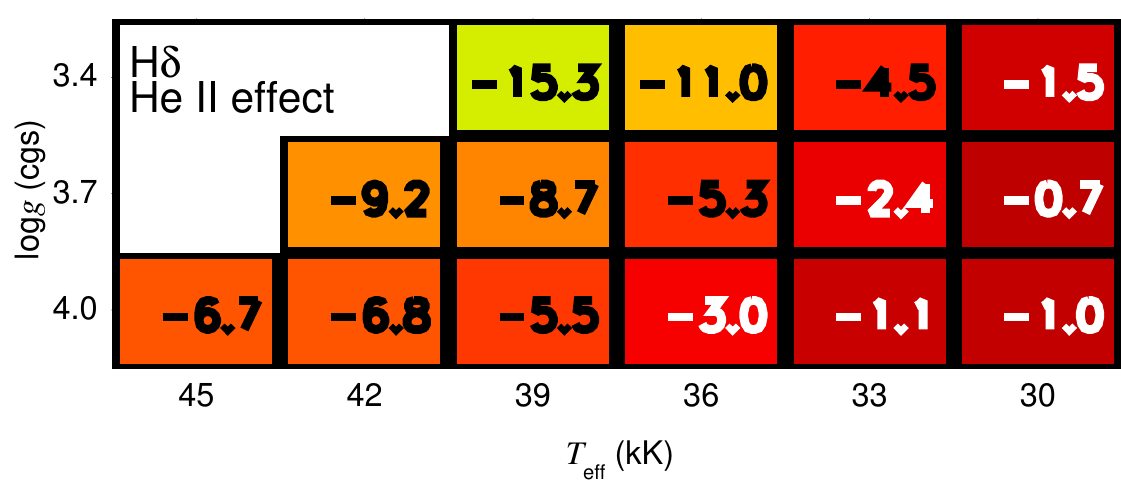}%
             \includegraphics*[height=22mm]{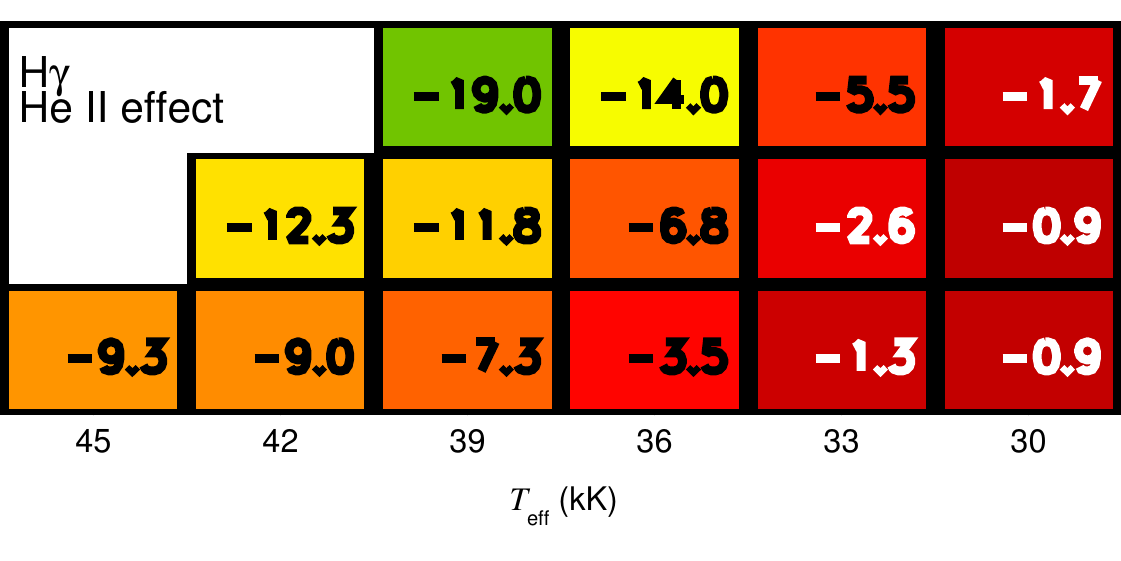}%
             \includegraphics*[height=22mm]{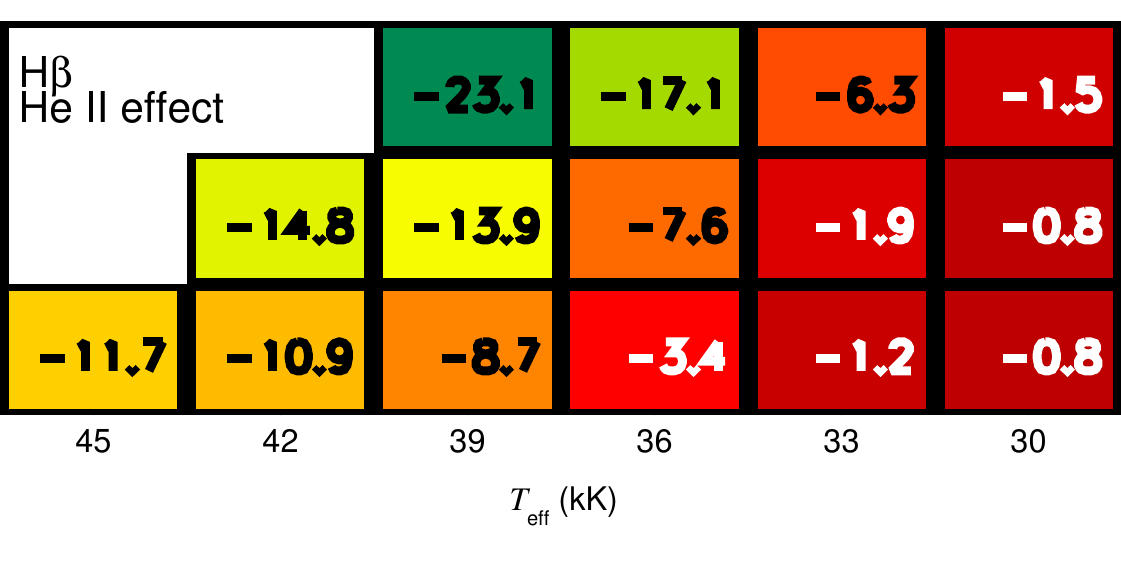}%
             \includegraphics*[height=22mm]{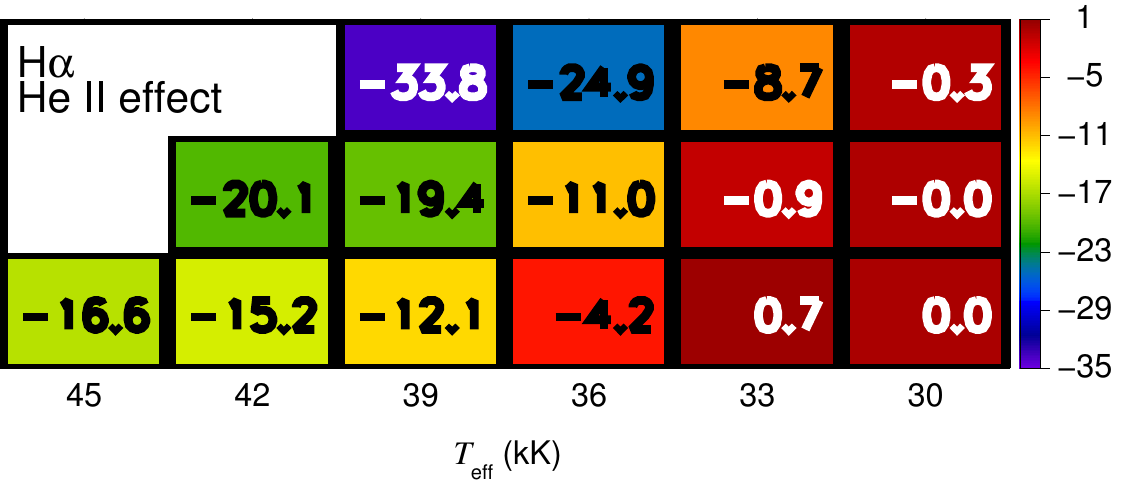}}
 \caption{RV displacement (in km/s) of the first four Balmer lines resulting from the \ion{He}{ii} effect for our FASTWIND grid 
          with $\logQ = -14.0$ in solar units.}
 \label{HeIIeffect}      
\end{figure*}

In addition to the extreme case of PZ~Gem, four objects stand out in the $\sigma_{4500}$ distribution as the only ones with values above
10~km/s, HD~\num{152424}, HD~\num{93521}, 68~Cyg, and HD~\num{161807}. Furthermore, the four objects have consistently high values of the ten RV 
dispersions. As it turns out, the first three 
objects have been independently discovered to be SB1 systems within the last year, the first one by Roberto Gamen \citep{Barbetal26}, the 
second one by Sergio Sim\'on-D\'{\i}az, and the third one during the analysis of this paper (their orbits will be published in the
\MTW\ series). For the fourth one, HD~\num{161807}, we suspect it may be a hidden SB1 but we need more epochs to confirm it with a periodogram
analysis. In summary, those four objects in our sample with $\sigma_{4500} > 10$~km/s include three 
new SB1 systems and one SB1 candidate, with another candidate just below that limit. Therefore, we propose that a consistent value of the RV 
dispersion in different bands above $\sim$10~km/s to be an indication of spectroscopic binarity that needs to be verified with a periodogram 
analysis and an orbit fit. A similar result with a single-band analysis was reached by \citet{SimDetal24}, note that paper uses peak-to-peak 
variations rather than dispersions. That proposed threshold of 10~km/s should not be taken as a strict limit, as low-inclination spectroscopic 
systems can still be hidden at lower RV dispersions 
(e.g. HD~\num{209339}~A and HD~\num{46150} above)
and some extreme pulsators may reach it. The threshold is significantly lower for slowly rotating
O dwarfs and giants with no Oe phenomenon, as in that case the eight stars that are not SBs have all RV dispersions $< 4$~km/s and 
$\sigma_{4500} < 1$~km/s.

To show the effect that the SB solutions have on the RV dispersions, we have calculated them with the orbital solutions for the five new
systems applied in separate UNWIND runs. Results are shown in boldface in Table~\ref{results1}. In most cases the values are 
reduced and, in particular, all the ones for $\sigma_{4500}$ now fall below the 10~km/s threshold, though some are still relatively high likely
due to pulsations. 
The difference between HD~\num{57682} and HD~\num{93521} discussed above is still significant after the correction.
The case of HD~\num{46150} is somewhat different, as in that case we uncovered and disentangled an early-B companion (Ab) that 
contributes $\sim10\%$ of the total flux. For such an SB2 system, most RV dispersions are also reduced to some degree, with the largest value 
happening for $\sigma_{4650}$, as there the primary has strong \ion{N}{iii} emission and the secondary has only weak lines. 
For the \ion{He}{i}-dominated $\sigma_{4650}$, however, the value increases due to the effect of the secondary counter-motion.

We end this subsection with a comparison between the RV dispersion ($\sigma_{4500}$) from this paper and 
$\sigma_G$ for the 34 stars with values in the latter (missing are $\zeta$~Pup, $\zeta$~Oph, $\lambda$~Ori~A, 
$\xi$~Per, HD~\num{151804}, HD~\num{202214}~Aa,Ab, and BN~Gem). Other than the strong photometric variability of PZ~Gem, there is no clear 
correlation between photometric variability and RV variability in Fig.~\ref{RVphot}. The strongest source of photometric variability for 
OB+WR stars are eclipses (not relevant for this sample), followed by strong wind variability (mostly in WR stars) and strong Oe events (only 
PZ~Gem in this sample).  AE~Aur, one of the most RV-stable stars in the sample (and second lowest in \vsini), is a curious case as it is the 
second most variable star in $\sigma_G$ after PZ~Gem. One possible source of that variability is dimming from the ISM for this runaway star 
\citep{BondKrel23}, which could be checked by comparing \textit{Gaia}~DR4 epoch photometry with our \lili\ spectroscopy.

\subsection{Contamination of Balmer lines by \ion{He}{ii}}\label{subsection33}

$\,\!$\indent In this subsection we determine the RV displacements in the first four hydrogen Balmer lines (H$\alpha$ at 6562.80~\AA, 
H$\beta$ at 4861.325~\AA, H$\gamma$ at 4340.464~\AA, and H$\delta$ at 4101.734~\AA) induced by the presence of the \ion{He}{ii} Brackett lines 
(Br$\beta$ at 6560.1~\AA, Br$\delta$ at 4859.320~\AA, Br$\zeta$ at 4338.671~\AA, and Br$\theta$ at 4100.041~\AA) as a function of \Teff\ and
\logg\ (as a proxy for luminosity). We do that using the FASTWIND grid in Appendix~\ref{appendixc} for the low-wind case of $\logQ = -14.0$ in solar units 
(wind infilling effects are treated in a separate subsection). For each model and line we fit a profile to the rectified flux of the form:

\begin{equation}
 f(\lambda) = 1 - \frac{I_0}{[((\lambda-\lambda_0)/\sigma)^2+1]^{1.2}},
\end{equation}

\noindent which is a good approximation for the profile caused by the Stark effect, convolved with a Gaussian + rotation profile. For each point in 
the \Teff+\logg\ grid we do the procedure for the two values of \vsini\ of 20~km/s and 300~km/s. We then calculate the RV displacements with
respect to the appropriate central wavelength of the hydrogen Balmer line and obtain the average of the two \vsini\ values. We use profile fitting 
instead of centroids (calculated by integration) because Balmer lines are broad, usually including one or several metallic lines in their wings, and
can have complex profiles when including wind effects (see subsection~\ref{subsection35}), making centroids impractical.

Results are shown in Fig.~\ref{HeIIeffect} but before discussing them we mention a caveat. We are fitting a profile that is
not exactly correct for two reasons: the combination of H and \ion{He}{ii} lines makes the real profile asymmetric (while our model is symmetric)
and the lines themselves are split due to their fine structure, further complicating the exact profile. However, under most observing conditions, the
effect of turbulence and rotation dominate and such asymmetries are hardly noticeable, so fitting an approximate profile to the FASTWIND models
should be a close approximation to reality. We evaluated such systematic effects by comparing the different results obtained for the two values of 
\vsini\ for a given grid point. The differences are of the order of 1~km/s, which can be taken as the precision for our results.

Figure~\ref{HeIIeffect} shows that the \ion{He}{ii} effect is negligible for late-O stars (especially so for dwarfs), as expected from the weakness
of their isolated \ion{He}{ii} lines. Starting from late-O dwarfs, the RV displacement increases not only towards higher values of \Teff\ 
(as expected from the strength of the isolated \ion{He}{ii} lines) but also towards lower gravities, where the effect is even larger. Indeed, the 
displacements for supergiants with \Teff~=~39~kK are about twice as large as for dwarfs with \Teff~=~45~kK. This is a manifestation of the decrease
in width of the Stark profiles for lower gravities. We note that, as the models in this subsection do not include the wind effects that are more
important for supergiants than for dwarfs, a comparison with real data must also take into account those effects (see subsection \ref{subsection35}).

The \ion{He}{ii} effect is largest for H$\alpha$ and decreases as we progress towards the upper Balmer series. Nevertheless, it is considerable
even for H$\delta$ for sufficiently large values of \Teff\ or small values of \logg. Therefore, it should be taken into account when using Balmer
lines to determine RVs of O stars.

\begin{figure}
 \centerline{\includegraphics*[height=21mm]{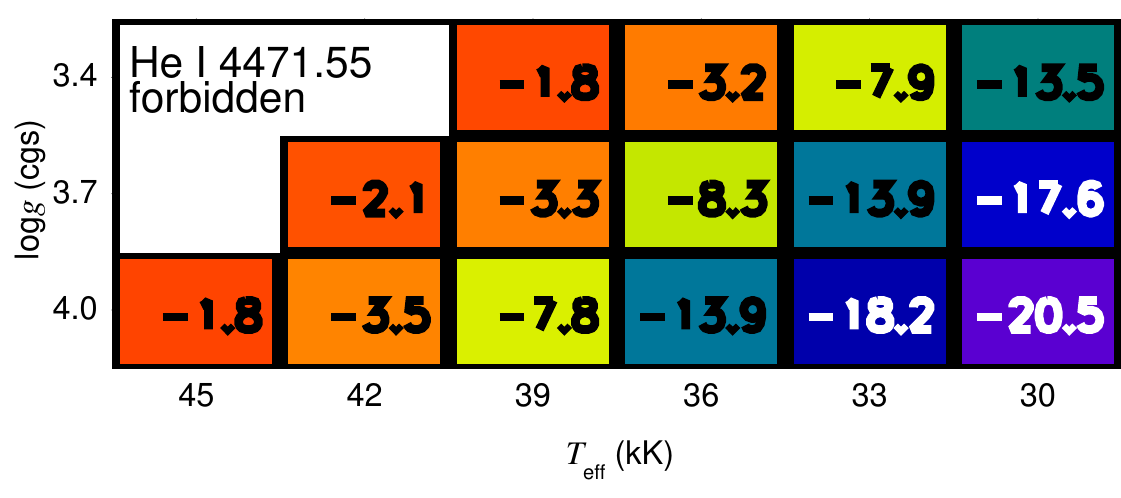}%
             \includegraphics*[height=21mm]{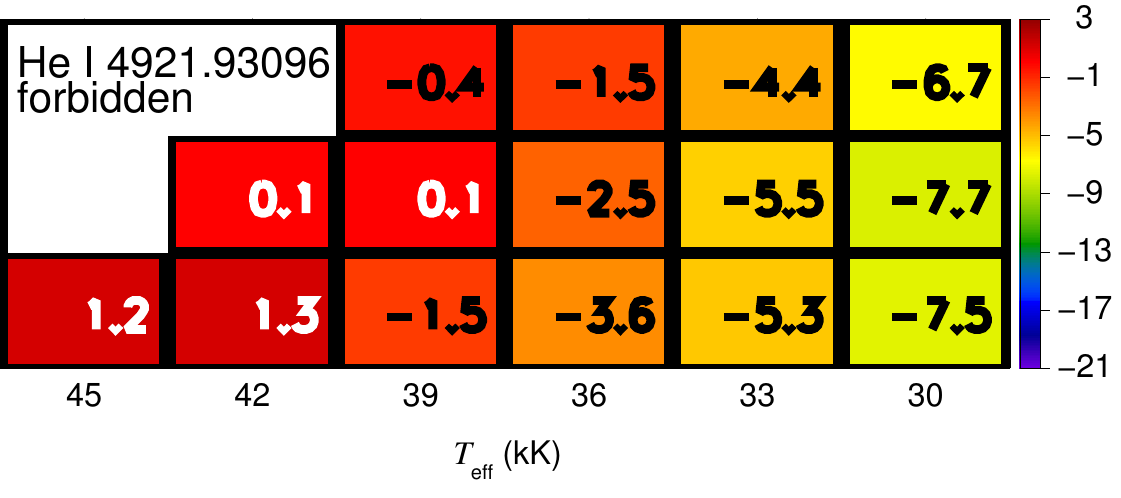}}
 \caption{RV displacement (in km/s) with respect to the reference wavelength for \ion{He}{i} lines with a forbidden component 
          for our FASTWIND grid with $\logQ = -14.0$ in solar units.}
 \label{forbidden}      
\end{figure}

\begin{figure*}
 \centerline{\includegraphics*[height=22mm]{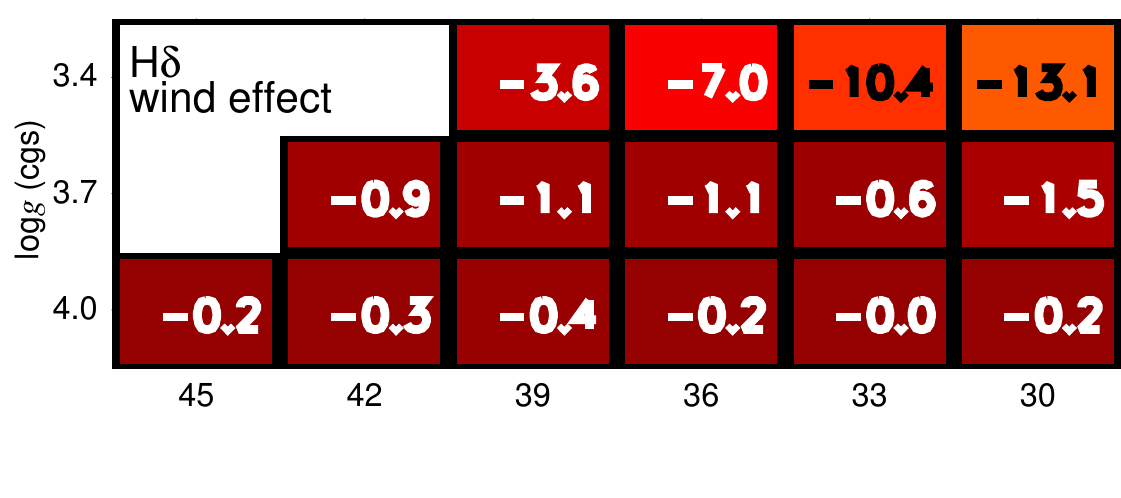}%
             \includegraphics*[height=22mm]{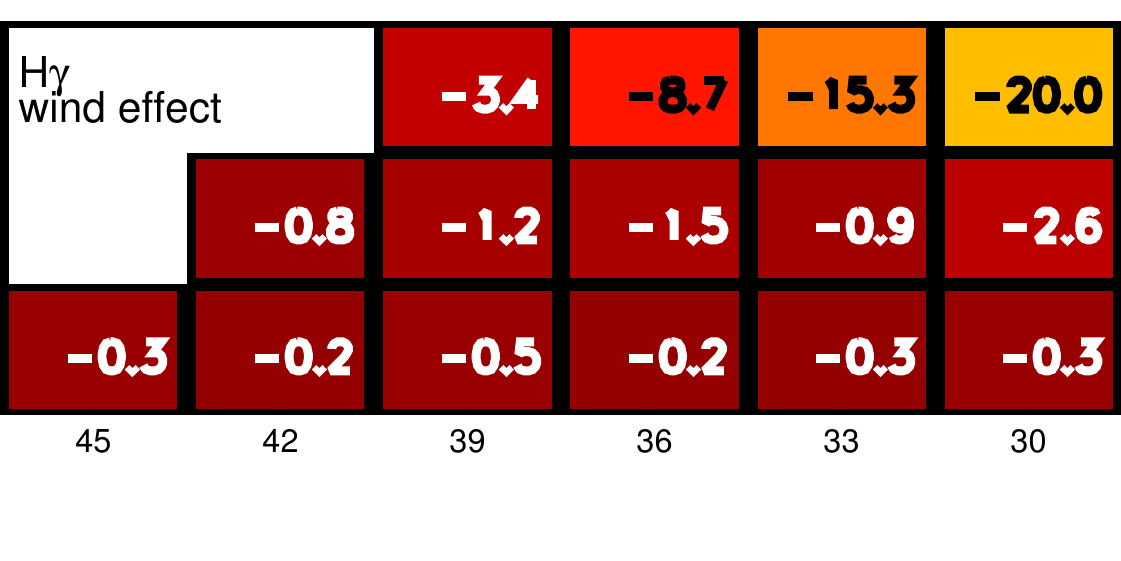}%
             \includegraphics*[height=22mm]{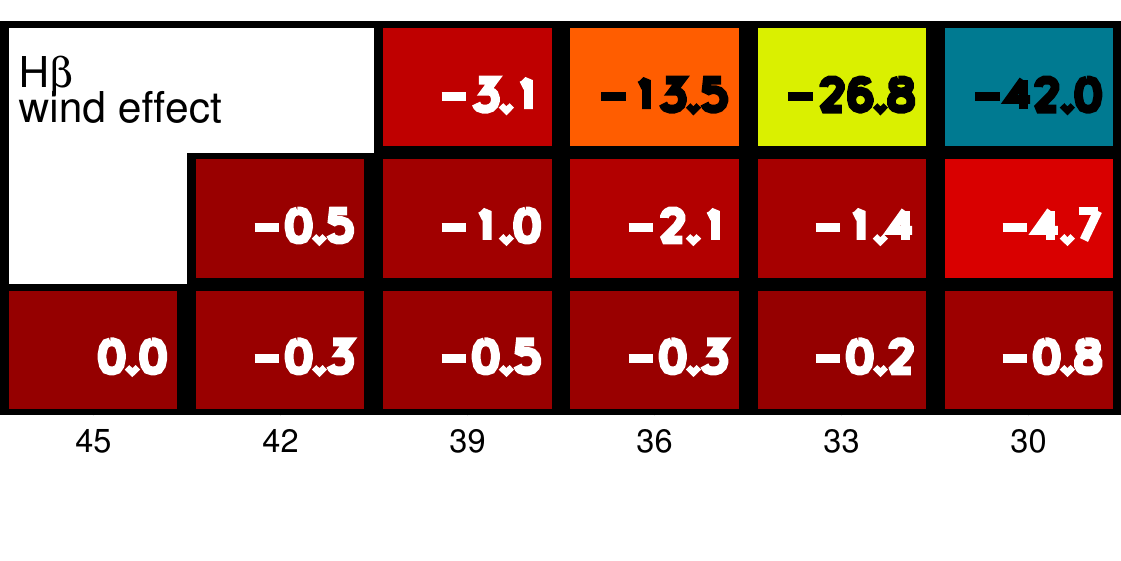}%
             \includegraphics*[height=22mm]{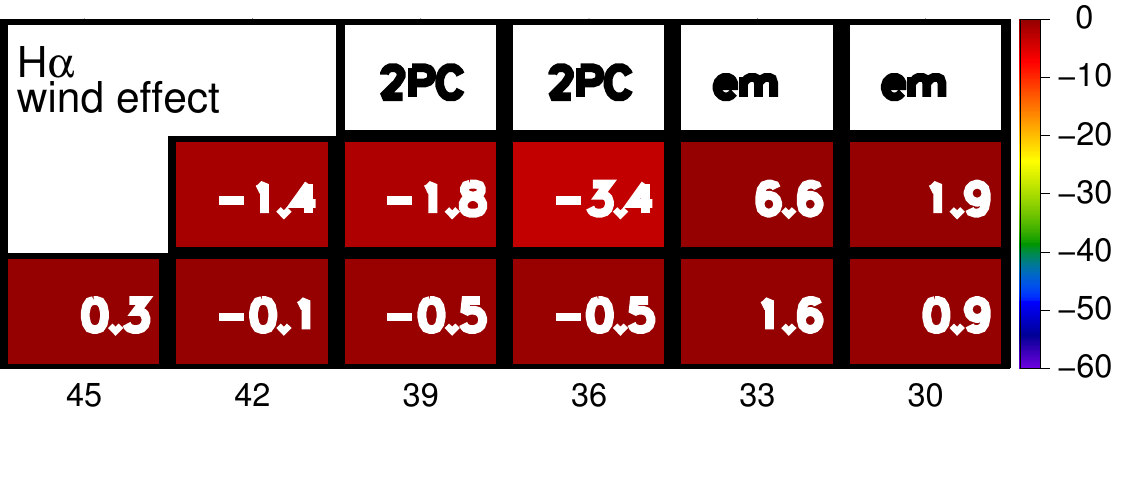}}
             \vspace{-4mm}
 \centerline{\includegraphics*[height=22mm]{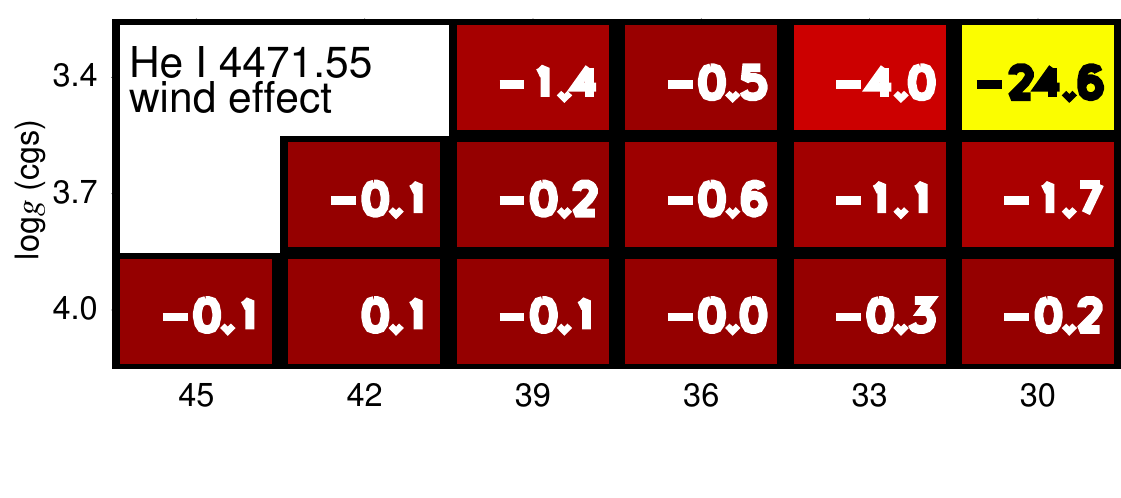}%
             \includegraphics*[height=22mm]{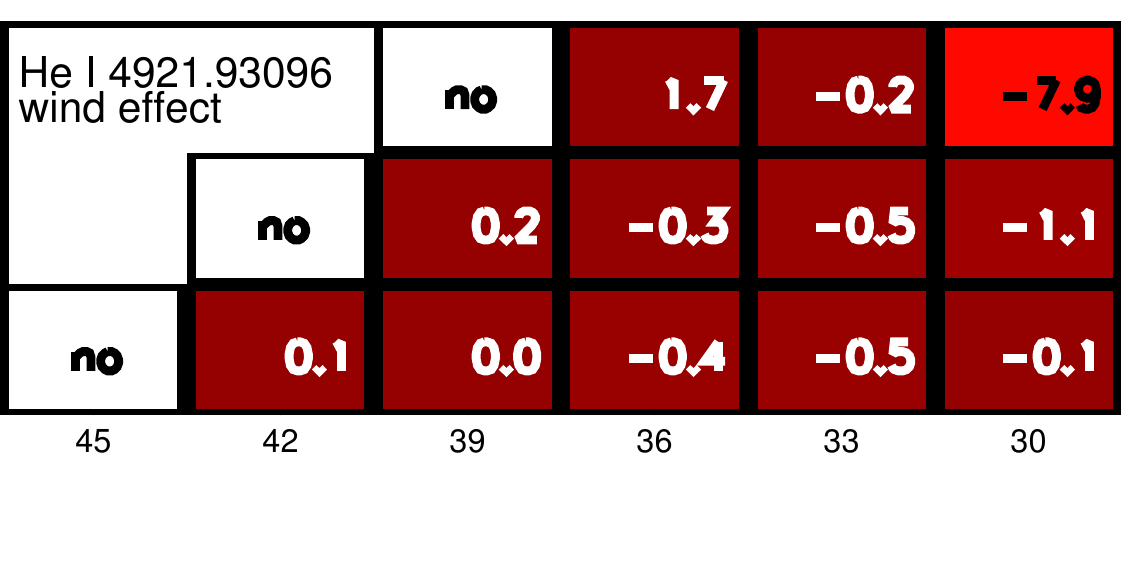}%
             \includegraphics*[height=22mm]{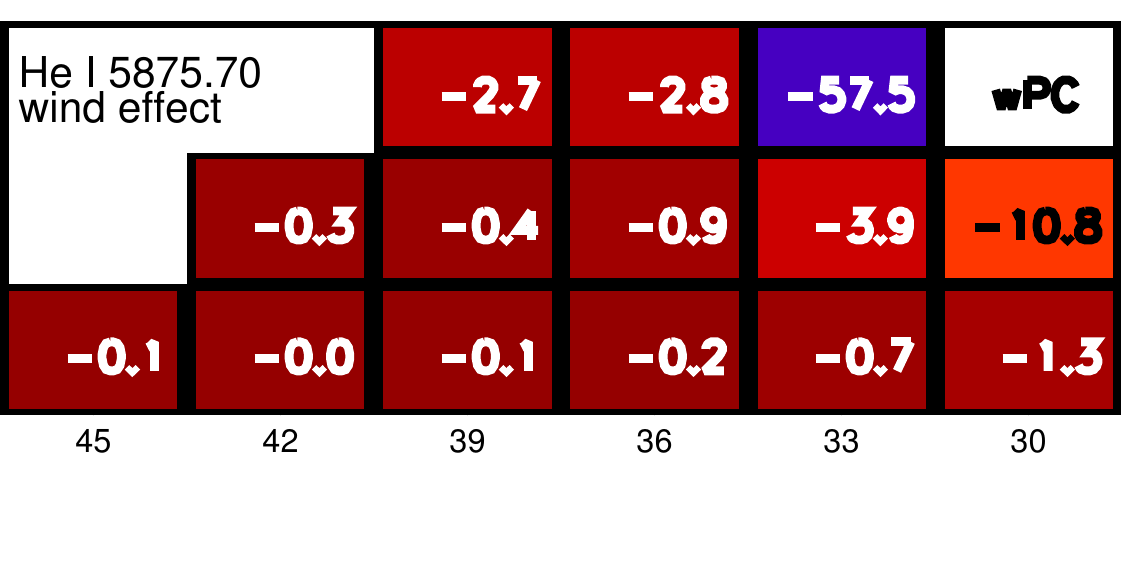}%
             \includegraphics*[height=22mm]{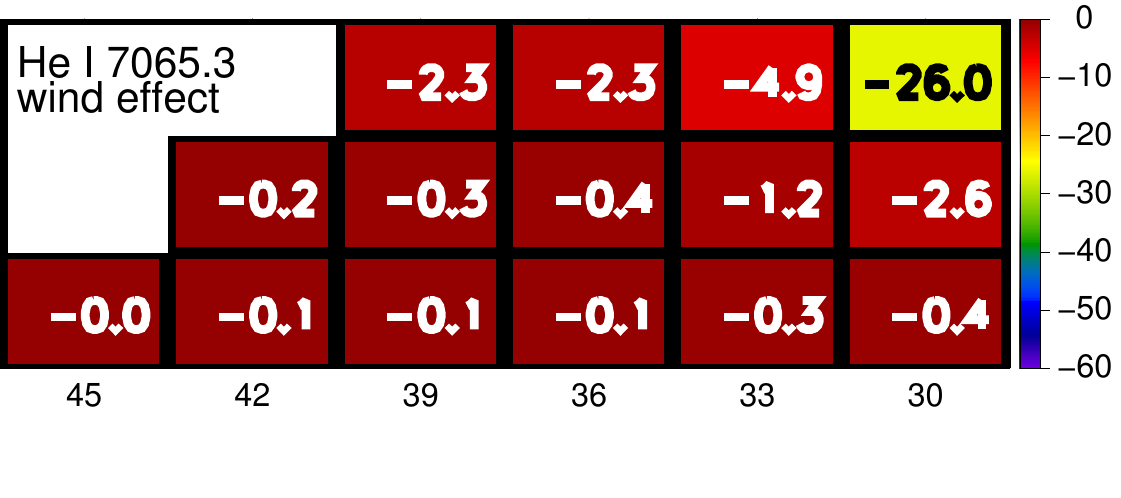}}
             \vspace{-4mm}
 \centerline{\includegraphics*[height=22mm]{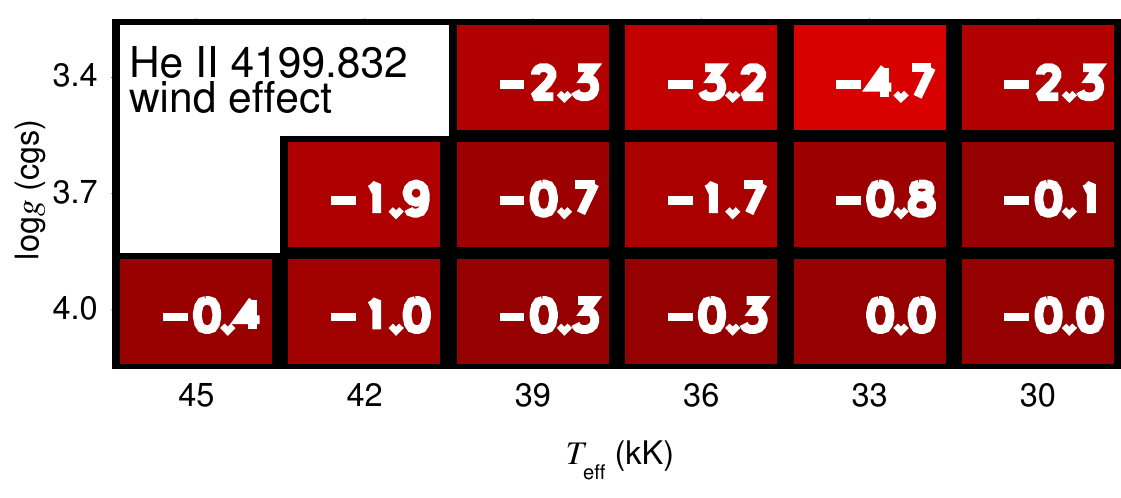}%
             \includegraphics*[height=22mm]{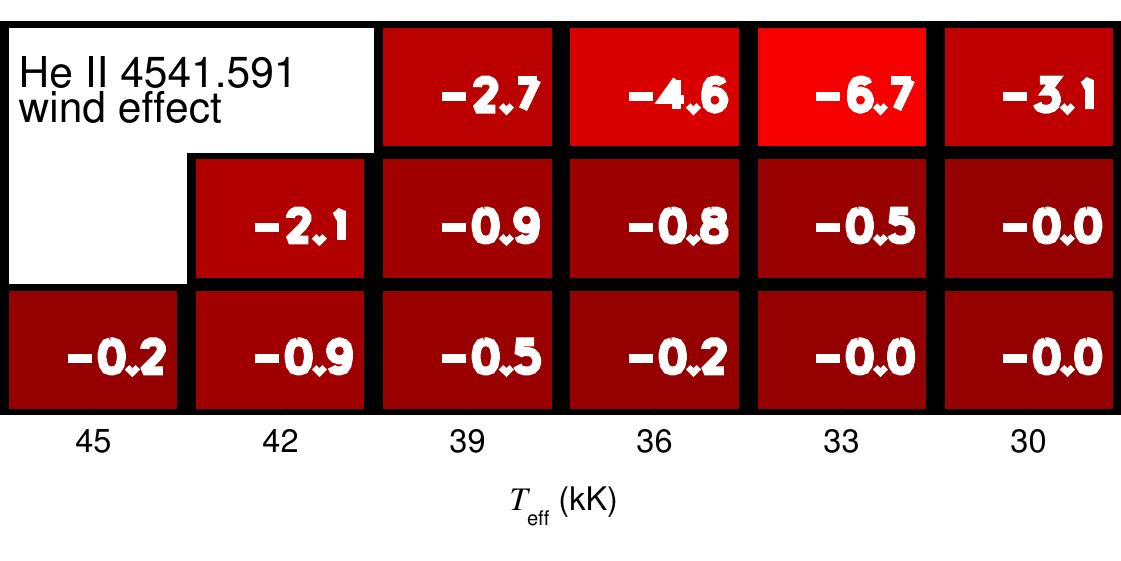}%
             \includegraphics*[height=22mm]{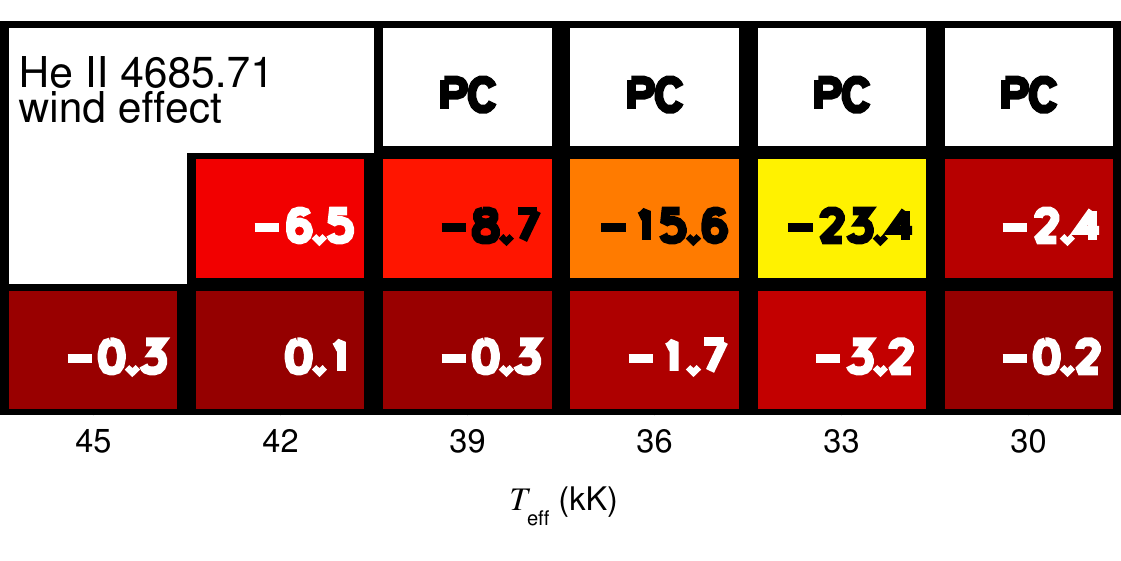}%
             \includegraphics*[height=22mm]{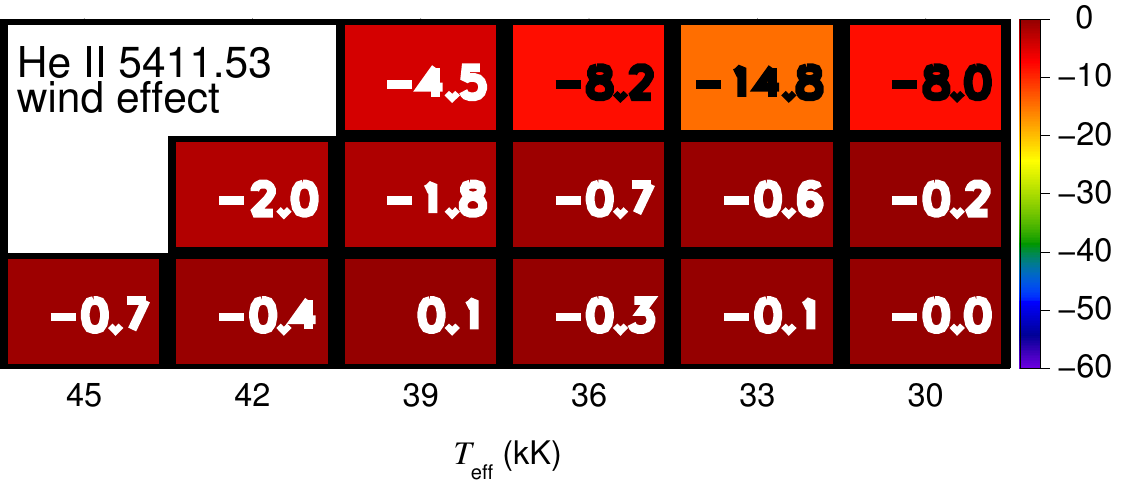}}
 \caption{Differences in RV displacement (in km/s) between the strong and weak wind cases for twelve lines using our FASTWIND grid. For the lines 
          shown in Figs.~\ref{HeIIeffect}~and~\ref{forbidden} the different effects should be added when comparing with data. Codes for missing 
          values: 2PC, double P-Cygni profile; em, emission; no, no line; wPC, weak P-Cygni profile; PC, P-Cygni profile.}
 \label{windinfilling}      
\end{figure*}

\subsection{The effect of forbidden components in \ion{He}{i} lines}\label{subsection34}

$\,\!$\indent Some \ion{He}{i} lines have forbidden components with relative strengths with respect to the main one that vary as a function
of \Teff\ and \logg\ \citep{Tremetal26}. For two lines commonly used to determine RVs, \ion{He}{i}~$\lambda$4471.55 and 
\ion{He}{i}~$\lambda$4921.93096, the forbidden component shifts the effective rest wavelength towards the violet.

Here we use the FASTWIND grid in Appendix~\ref{appendixc} for the low-wind case of $\logQ = -14.0$ in solar units to determine the RV displacement caused by 
the forbidden component in a similar way to what we did in the previous subsection for the \ion{He}{ii} effect. The main difference is that here we
determine the shift by measuring the velocity centroid, which yields more accurate determinations for narrower lines such as those of \ion{He}{i}. 
We also evaluated the systematic effects by comparing the different results obtained for the two values of \vsini\ for a given grid point. In this
case the differences are even lower, with typical values of 0.1~km/s for \ion{He}{i}~$\lambda$4471.55 and 0.3~km/s for 
\ion{He}{i}~$\lambda$4921.93096.

Figure~\ref{forbidden} shows that the forbidden-component effect runs in opposite directions to the \ion{He}{ii} effect. It increases with
decreasing \Teff\ and increasing \logg, thus being larger for the late-O dwarfs in our sample. The effect is larger for \ion{He}{i}~$\lambda$4471.55
than for \ion{He}{i}~$\lambda$4921.93096 and can be quite significant. Furthermore, the trends seen in Fig.~\ref{forbidden} suggest that the effect 
may be even larger for early-B stars. As those two lines are commonly used to determine orbits, they may be the main source of discrepancies in 
systemic velocities in SB2 systems, such as the $\sim$8~km/s seen for HD~1810~Aa,Ab by \citet{Holgetal25a}. Note that the Ab component there, a 
B1~V, has the more negative $\gamma$ value, which is the expected result of extrapolating the Fig.~\ref{forbidden} data into the early-B range, as 
the Aa component is of spectral type O9.5~V(n).

\subsection{Wind infilling of H+He lines}\label{subsection35}

$\,\!$\indent As a third effect that can be studied using the FASTWIND grid in Appendix~\ref{appendixc}, we analysed the effect of wind infilling in four 
Balmer lines, four \ion{He}{i} lines, and four \ion{He}{ii} lines. To do that we calculated the RV displacement in models with strong winds and weak 
winds (values in Table~\ref{gridfastwind}) and subtracted one from the other. Displacements were calculated by profile fitting for Balmer lines and 
by centroid determination for He lines, as above. Results are shown in Fig.~\ref{windinfilling}.

The overall result from Fig.~\ref{windinfilling} is that the RV displacement from wind infilling for the twelve lines analysed is negligible for 
dwarfs (\logg~=~4.0~cgs, \logQ~=~$-$13.5~solar) and significant to very large for supergiants (\logg~=~3.4~cgs, \logQ~=~$-$12.3~solar), with large 
differences among lines. For Balmer lines (fitted with Stark profiles), wind infilling is very large in supergiants, increasing with decreasing \Teff\ 
and order, to the point that early-O-type supergiants have double (caused by H and \ion{He}{ii}) P-Cygni profiles in H$\alpha$ and late-O-type 
supergiants have the line in emission (well known effects, see Appendix~\ref{appendixb} for examples). When compounded with the contamination from 
\ion{He}{ii} previously discussed, the overall conclusion is that Balmer lines yield poor measurements of RVs for O supergiants unless those effects are 
modelled. 

For the four \ion{He}{i} lines in Fig.~\ref{windinfilling}, the one that is more strongly affected by wind infilling is \ion{He}{i}~$\lambda$5875.70
and the one that is less affected is \ion{He}{i}~$\lambda$4921.93096. For all of them, the RV displacement caused by wind infilling increases with 
decreasing \Teff\ for giants and supergiants and for the two lines with forbidden components, that previously discussed effect dominates except for
the cooler O supergiants. Finally, among the four \ion{He}{ii} lines in Fig.~\ref{windinfilling} there is also one where wind infilling is stronger 
than in the others, \ion{He}{ii}~$\lambda$4865.71, another previously well-known effect. For the other three \ion{He}{ii} lines analysed, wind infilling 
effects are only significant for supergiants\footnote{In Paper~I we determined RV displacements for the two components in GLS~\num{11448} 
using five \ion{He}{ii} lines, including three in this analysis, and FASTWIND models, with displacements between $-$7~and~$-$11~km/s. Those two stars 
are not represented in the grid in this section because they have dwarf-like \logg\ and supergiant-like \logQ\ values in the upper left corner of the 
plots in Fig.~\ref{windinfilling}, and require specific modelling.}.

\subsection{Cross-correlation with FASTWIND models for H+He}\label{subsection36}

$\,\!$\indent As described in Section~\ref{section2}, we used the FASTWIND grid and methodology of \citet{Holg19} to fit the
stellar parameters to our sample. The results for \vsini, \Teff, \logg, and \logQ\ are shown in Table~\ref{results1}. For one star,
HD~\num{152408}, no FASTWIND analysis could be performed because its wind is too strong and such analysis would require a different code such
as CMFGEN \citep{HillMill98} or PoWR \citep{Hainetal19}.

The FASTWIND model for each star was cross-correlated with the equivalent UNWIND output. We note two important
characteristics of this technique: (a) using a FASTWIND model as reference takes into account the effects of the contamination of 
Balmer lines by \ion{He}{ii}, forbidden lines for \ion{He}{i}, and wind infilling of H+He lines previously described and (b) the previous cleaning with 
UNWIND eliminates the biases introduced by ISM lines, which may be especially important for H$\beta$ due to the existence of the adjacent DIB
(see Paper I). We selected the 19 wavelength regions in Table~\ref{FASTWIND_regions} and we cross-correlated the UNWIND output of each star in each
region, producing 19 velocity determinations for each star. In the case of the five stars with new SB orbits, we used the UNWIND runs with the 
orbit applied and with zero systemic velocity (as the calculated systemic velocity for the orbits will be actually the outcome of this analysis). 

\begin{table}
\caption{Wavelength regions used for the cross-correlation with FASTWIND models.}
\centerline{
\begin{tabular}{lr@{-}ll}
Name            & \mcii{Wav. range}     & H+He lines                                                              \\
                & \mcii{(\AA)}          &                                                                         \\
\midrule
H$\epsilon$     & \num{3950}&\num{3990} & H$\epsilon$, \ion{He}{ii}~$\lambda$3968.432                             \\
H$\delta$       & \num{4080}&\num{4132} & H$\delta$, \ion{He}{i}~$\lambda$4120.83, \ion{He}{ii}~$\lambda$4100.041 \\
H$\gamma$       & \num{4320}&\num{4360} & H$\gamma$, \ion{He}{ii}~$\lambda$4338.671                               \\
H$\beta$        & \num{4840}&\num{4880} & H$\beta$, \ion{He}{ii}~$\lambda$4859.320                                \\
H$\alpha$       & \num{6540}&\num{6585} & H$\alpha$, \ion{He}{ii}~$\lambda$6560.1                                 \\
\ion{He}{i+ii}  & \num{4010}&\num{4040} & \ion{He}{i}~$\lambda$4023.97981, \ion{He}{ii}~$\lambda$4025.601         \\
\ion{He}{i} 1   & \num{4132}&\num{4155} & \ion{He}{i}~$\lambda$4143.75900                                         \\
\ion{He}{i} 2   & \num{4373}&\num{4403} & \ion{He}{i}~$\lambda$4387.92910                                         \\
\ion{He}{i} 3   & \num{4462}&\num{4482} & \ion{He}{i}~$\lambda$4471.55                                            \\
\ion{He}{i} 4   & \num{4700}&\num{4725} & \ion{He}{i}~$\lambda$4713.24                                            \\
\ion{He}{i} 5   & \num{4907}&\num{4937} & \ion{He}{i}~$\lambda$4921.93096                                         \\
\ion{He}{i} 6   & \num{5006}&\num{5026} & \ion{He}{i}~$\lambda$5015.6776                                          \\
\ion{He}{i} 7   & \num{5038}&\num{5058} & \ion{He}{i}~$\lambda$5047.7384                                          \\
\ion{He}{i} 8   & \num{5861}&\num{5888} & \ion{He}{i}~$\lambda$5875.70                                            \\
\ion{He}{ii} 1  & \num{4190}&\num{4210} & \ion{He}{ii}~$\lambda$4199.832                                          \\
\ion{He}{ii} 2  & \num{4532}&\num{4552} & \ion{He}{ii}~$\lambda$4541.591                                          \\
\ion{He}{ii} 3  & \num{4670}&\num{4695} & \ion{He}{ii}~$\lambda$4685.71                                           \\
\ion{He}{ii} 4  & \num{5400}&\num{5424} & \ion{He}{ii}~$\lambda$5411.53                                           \\
\ion{He}{ii} 5  & \num{6515}&\num{6540} & \ion{He}{ii}~$\lambda$6527.099                                          \\
\midrule
\end{tabular}
}              
\label{FASTWIND_regions}   
\end{table}

For each star we then determined the final FASTWIND RV after eliminating some of the 19 velocity values based on the following 
criteria: (a) poor profile fitting (relatively common for the \ion{He}{ii} 3 region due to high wind strength), (b) contamination by strong 
neighbour metallic lines (common for the H$\delta$ region), and (c) weakness of the line (common for the \ion{He}{ii} regions for late-O 
stars). In Table~\ref{results2} we show the number of lines used $n_l$ and the final FASTWIND RVs+uncertainties obtained from the mean and
standard deviation. For the 40 stars we were able to derive a value using between 7 and 17 lines.

There is a clear correlation between the uncertainties for the FASTWIND RVs and the stellar properties. Slow rotators have lower
uncertainties than fast rotators. Oe stars have higher uncertainties than non-Oe stars but this is likely an effect of the observed \vsini,
not of the Oe phenomenon itself, as the two nearly pole-on Oe stars (HD~\num{155806}~A,B and HD~\num{57682}) have low uncertainties. 
Overall, the best stars (slow rotators) have typical uncertainties of 2-3~km/s. As in the subsection~\ref{subsection37} we derive 
alternative RVs using metallic lines, we defer the comparison with pre-existing RV values until we have computed the alternatives.

We point out that non-radial pulsations can make line profiles asymmetric and time-dependent, hence inducing RV variations on the centroids
and possibly questioning the validity of determining RVs from a time-average as we do in this and in the following subsections. We tested that
issue by also determining RVs from H+He lines using individual epochs (which can deviate more from a symmetric profile than the mean of a large 
number of epochs) and calculating the mean and dispersion of the resulting RVs. Comparing the results from the two methods, they turned out to be 
very similar and the possible bias can be neglected.

\subsection{Profile fitting of metallic lines}\label{subsection37}

\begin{table}
\caption{Wavelength regions used for the output line fitting with (mostly) metallic lines.}
\centerline{
\scriptsize
\begin{tabular}{r@{-}ll}
\mcii{Wav. range (\AA)} & Examples of relevant lines                                                                                          \\
\midrule
\num{4033}&\num{4092}   & \ion{Si}{iv}~$\lambda$4088.862, \ion{N}{iv}~$\lambda$4057.75, many \ion{C}{iii}+\ion{O}{ii}                         \\
\num{4108}&\num{4137}   & \ion{He}{i}~$\lambda$4120.83, \ion{Si}{iv}~$\lambda$4116.104                                                        \\
\num{4147}&\num{4194}   & \ion{He}{i}~$\lambda$4168.9716, \ion{C}{iii}~$\lambda$4186.900                                                      \\
\num{4209}&\num{4281}   & \ion{C}{ii}~$\lambda$4267 triplet, many \ion{O}{ii}                                                                 \\
\num{4281}&\num{4332}   & many \ion{O}{ii}                                                                                                    \\
\num{4342}&\num{4383}   & \ion{N}{iii}~$\lambda$4379 doublet                                                                                  \\
\num{4394}&\num{4462}   & \ion{He}{i}~$\lambda$4437.5537, \ion{O}{ii}~$\lambda$4416 doublet                                                   \\
\num{4477}&\num{4538}   & \ion{Mg}{ii}~$\lambda$4481 triplet, \ion{N}{iii}~$\lambda$4513 triplet, \ion{S}{iv}~$\lambda\lambda$4485.7,4504.2   \\
\num{4546}&\num{4624}   & \ion{Si}{iii}~$\lambda\lambda\lambda$4552.622,4567.841,4574.757, \ion{N}{v}~$\lambda\lambda$4603.74,4619.97         \\
\num{4624}&\num{4681}   & \ion{N}{iii}~$\lambda$4640 triplet, \ion{Si}{iv}~$\lambda\lambda$4631.273,4654.312, many \ion{C}{iii}+\ion{C}{iv}   \\
\num{4690}&\num{4712}   & many \ion{O}{ii}                                                                                                    \\
\midrule
\end{tabular}
\vspace{-1mm}
}              
\label{metallic_regions}   
\end{table}

$\,\!$\indent As an alternative measurement of RVs, we used the line fitting code LINESFIT, part of the UNWIND package. The code fits 
simultaneous multiple line profiles using different combinations. Here we use combined rotation+Gaussian (rotation+Stark for He lines) profiles in which 
(a) the central wavelengths,
\vsini, and spectral resolution (or mimimum gaussian FWHM) are fixed from external values and (b) the EW and Gaussian FWHM for each line separately
as well as a common RV are fitted. The central wavelengths are from the Atomic Line List v2.04\footnote{\url{https://linelist.pa.uky.edu/atomic/}.} with a
few exceptions (see below for an example). LINESFIT is used interactively to set the continuum rectification points (a major source of systematic errors,
especially for weak lines, if not done correctly) and to select/deselect different lines. 

We divided the metallic-rich parts of the blue-violet spectrum into eleven regions (Table~\ref{metallic_regions}), selected each to contain a significant 
number of metallic lines while avoiding as much as possible the strong H+He lines from the previous subsection. As this was not completely possible, three 
relatively weak \ion{He}{i} lines were included. For each star and wavelength region we did a separate fit using the combined UNWIND output.
The RV and uncertainty for each star was computed from the mean and
standard deviation of the regions measured (Table~\ref{results2}). Regions where there was not enough S/N or peculiar properties were observed were excluded,
with the number of regions used for each star, $n_r$, given in Table~\ref{results2}. The total number of lines fitted for each star, $n_l$ is also given
there, note that in the best cases (typically, late-O slowly rotating dwarfs) $\sim$400 lines were used. For $n_l$ the individual elements of a multiplet and
the individual components of the special $\zeta$~Pup case (see below) are counted separately. We now discuss the different cases and exceptions
that we found:

\begin{table}
\caption{RV differences for the three ON9.5nn stars determined by cross-correlation with the HD~\num{68450} template.}
\centerline{
\begin{tabular}{lr@{$\pm$}l}
Star            & \mcii{$\Delta$RV} \\
                & \mcii{(km/s)}     \\
\midrule
$\zeta$~Oph     & $-$28.2&7.0       \\
HD~\num{93521}  & $-$36.0&1.5       \\
HD~\num{161807} & $-$59.6&6.0       \\
\midrule
\end{tabular}
\vspace{-2mm}
}              
\label{ccresults}   
\end{table}

\textit{SB systems.} For these we calculated the RVs after applying the SB solution with $\gamma = 0$, as in the previous subsection.

\textit{Very fast rotators.} Three stars ($\zeta$~Oph, HD~\num{93521}, and HD~\num{161807}) have an nn index or $\vsini > 350$~km/s. For them,
it was not possible to obtain reliable RVs. In subsection~\ref{subsection38} we use a different method to calculate their RVs.

\textit{Onfp stars.} The Onfp phenomenon was initially assigned only to \ion{He}{ii}~$\lambda$4685.71 but here we detect a double-peaked structure in the
\ion{N}{iii}~$\lambda$4640 triplet in emission for all Onfp stars (Fig.~\ref{ALS_spectra_Onfp}) as well as hints of Oe-like ``horns'' in some other lines. 
To avoid biases, we eliminated the 4624-4681~\AA\ region for all of the Onfp results (except for $\zeta$~Pup, see next). 

\textit{$\zeta$~Pup.} This is the earliest and strongest Onfp in our sample, with few metallic lines (Fig.~\ref{ALS_spectra_Onfp}), typical for a 
fast-rotating early-O supergiant. Furthermore, several metallic lines show a quasi-symmetric double-peaked profile, some with a central absorption
component. Given that, for $\zeta$~Pup we ignored the absorption
lines and fitted the double-peaked lines \ion{N}{iv}~$\lambda$4057.75, \ion{Si}{iv}~$\lambda$4088.862, \ion{Si}{iv}~$\lambda$4116.104, 
\ion{N}{iii}~$\lambda\lambda$4378.99,79.20, and \ion{N}{iii}~$\lambda$4904.78 (the latter outside the ranges used for the rest of the stars) using multiple 
Gaussians in a symmetric disposition with respect to the central wavelength. Remarkably, the RV value we derived has a low dispersion and is consistent with
the FASTWIND one from the previous subsection. This is an indication that the FASTWIND treatment is accurate, 
as fitting such profiles should yield results close to the real systemic velocity.

\textit{HD~\num[detect-all]{152408}.} The strong wind in this star precluded us from calculating a FASTWIND model. Most metallic lines 
show P-Cyg profiles (Fig.~\ref{ALS_spectra_1}), again hampering an RV determination. For this object we searched for pure absorption lines in 
the whole optical spectrum which could be fitted and we found six of them: \ion{N}{iv}~$\lambda$4057.75, the \ion{N}{iv}~$\lambda$5204 triplet, 
\ion{C}{iv}~$\lambda$5801.34, and \ion{C}{iv}~$\lambda$5811.97. We did specific UNWIND runs to cover those lines, including a detailed fitting of the complex
DIB structure around the \ion{C}{iv} lines (see Paper~I), and fitted them with LINESFIT. The result has a significant dispersion and is likely affected by
wind infilling but is the best one we can produce.

Finally, we describe the case of the \ion{S}{iv}~$\lambda\lambda$4485.7,4504.2 emission lines in several of the stars in our sample.
They were discussed by \citet{Wolf63} but as of 2001 they remained unidentified, an ``embarrassment'' according to \citet{Walb01}, until soon afterwards they
were discovered to originate in \ion{S}{iv} by \citet{WernRauc01}. As these lines have been seldom studied, we decided to use the high-quality
spectrum of $\lambda$~Ori~A to determine their empirical central wavelengths by fitting other prominent lines to determine the RV and then 
fixing it to measure the central wavelengths of the two lines. The values we determine, 4485.7~\AA\ and 4504.2~\AA\ are close to the laboratory ones of 
4485.662~\AA\ and 4504.093~\AA\ from \citet{WernRauc01}.

\subsection{Cross-correlation analysis for very fast rotators}\label{subsection38}

\begin{figure}
 \centerline{\includegraphics*[width=\linewidth]{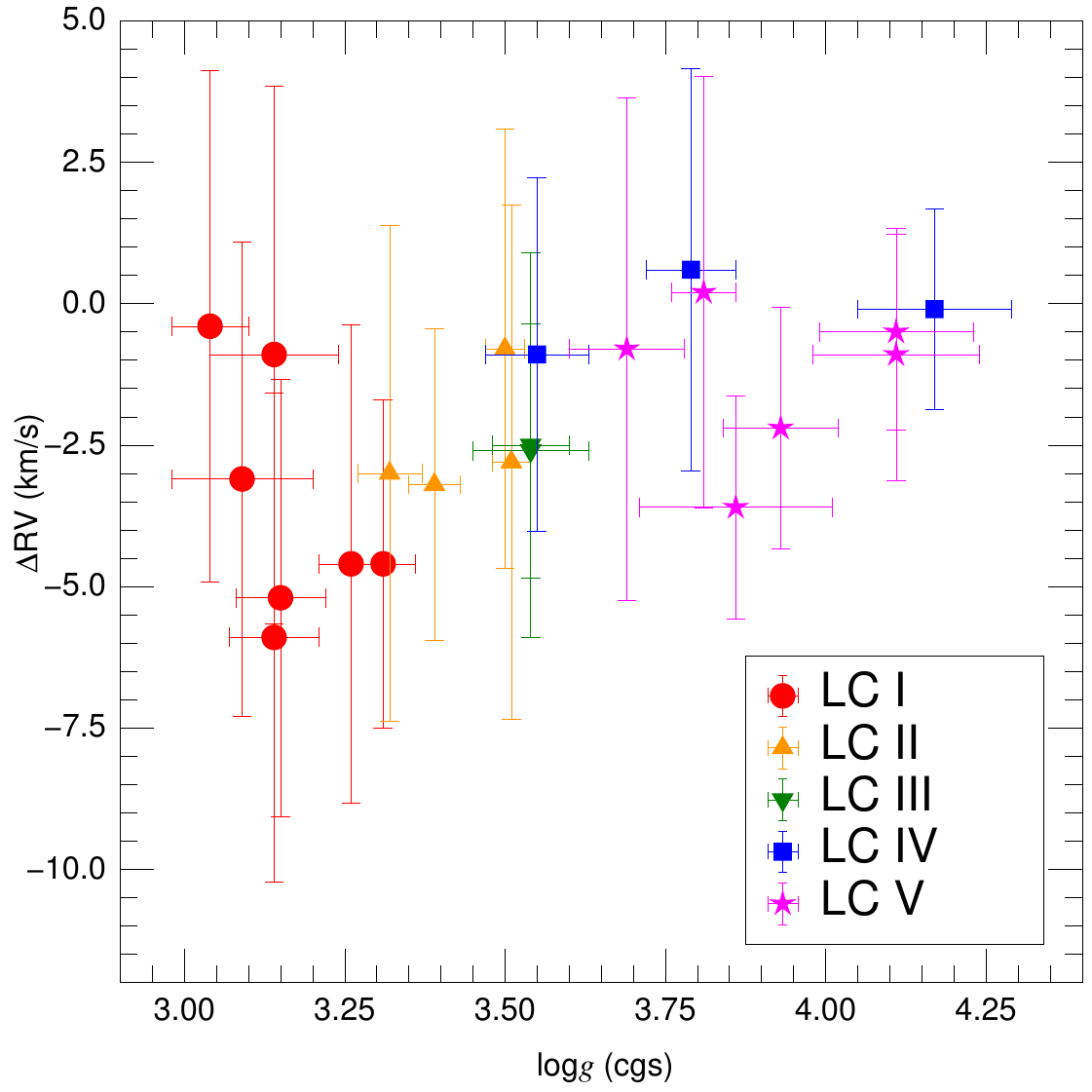}}
 \caption{RV differences between the metallic line and FASTWIND determinations for the 22 stars with the most precise values.}
 \label{logg_Drv}      
\end{figure}

$\,\!$\indent Given the impossibility of using the few broad metallic lines to obtain reliable RV measurements for the three ON9.5nn stars
($\zeta$~Oph, HD~\num{93521}, and HD~\num{161807}), we used another technique for them. Starting with the UNWIND output for HD~\num{68450}, a star with
a similar spectral classification with narrow lines and accurate RV determinations, we broadened it with the rotation profiles corresponding to the 
\vsini\ values for the three stars. We then divided the blue-violet range in three wavelength regions, cross-correlated each processed HD~\num{68450} UNWIND
output with the equivalent one for each star, and computed the RV difference between the pairs. Finally, for the three wavelength regions of each pair, we
computed the mean and standard deviation, with the results shown in Table~\ref{ccresults}. In Section~\ref{section4} we compare the results from the three RV
methods used in this paper.

\section{Analysis and future work}\label{section4}

\subsection{Comparison between different RV methods}\label{subsection41}

$\,\!$\indent We selected the 22 stars with the most precise RV determinations (combined uncertainty better than 5~km/s) and computed the difference between
the RV value from the metallic lines (subsection~\ref{subsection37}) and from the FASTWIND analysis (subsection~\ref{subsection36}), 
plotting the results in Fig.~\ref{logg_Drv}. Even
though the error bars are relatively large, the plot shows small differences between the two values for LCs IV and V (or high gravities) and significant ones
for LCs III to I (or low gravities). The negative signs (metallic-line values lower than FASTWIND ones) are consistent with the metallic lines being affected by
wind infilling (and uncorrected for in our determinations) for giants and supergiants, and the range of values is similar to the effects seen in He lines in
Fig.~\ref{windinfilling}. Therefore, we conclude that most likely direct RV determinations from metallic lines have little or no bias for O stars with LCs of 
IV and V but can be biased by several km/s for giants and supergiants.

What value should be used as the ``best'' RV for a given star? Based on the above, we propose the following algorithm. [1] For (most) giants and supergiants,
use the FASTWIND value, as the metallic-line result is biased. [2] For (most) dwarfs and subgiants, combine the two RV measurements considering them
independent and unbiased using the weighted mean and its uncertainty (all differences are less than two sigmas apart). [3] The two $\zeta$~Pup values can also
be combined, as the metallic-line determination comes from a symmetric configuration. [4] For HD~\num{152408} we only have the likely biased metallic-line 
result. [5] For the three ON9.5nn stars we use the analysis in subsection~3.8 combined with the FASTWIND value for HD~\num{68450}. Those recommended values are 
shown in Table~\ref{results2}.

\subsection{Comparison with previous velocity determinations}\label{subsection42}

$\,\!$\indent We have searched in the literature for previous RV determinations for the 41 stars in our sample, finding them in all cases and ranging in
number of measurements from 3 (HD~\num{161807}) to 26 (AE~Aur). For all the stars we have computed the robust mean and standard deviation from the
literature, discarding values more than 2.5 sigmas away from the mean. Results are shown in Table~\ref{results2}.

Most of our RV values are compatible with the literature ones and in most cases our precision is better than the one derived from the literature dispersion.
Overall, we observe a discrepancy: there are more negative differences with the literature (in red in the last column in Table~\ref{results2})
than positive ones (in blue) and they also have larger absolute values on average. Furthermore, the negative differences dominate among supergiants, as only 
HD~\num{152424} has a positive one (and is a new SB1). This is likely a consequence of wind infilling not beeing
taken into account in the past, with the two other effects discussed in this paper (Balmer line contamination by \ion{He}{ii} and forbidden components
in \ion{He}{i} lines) possibly playing a minor role.

The two most significant discrepancies with the literature are $\zeta$~Pup ($-$5.8 sigmas) and HD~\num{93521} ($-$3.2 sigmas), both runaway stars. We will
analyse the consequences on their trajectories in a future paper.

Two sources have particularly bad results regarding RVs for our sample. \textit{Gaia}~DR3 \citep{Valletal23} has RV measurements for four stars in our
sample and three of them (68~Cyg, $\mu$~Nor, are 63~Oph) are discarded when computing the literature value. SDSS~DR17 \citep{Abduetal22b} has RV measurements
for six stars in our sample and three of them (BN~Gem, HD~\num{175754}, and HD~\num{34656}~Aa) are similarly discarded. The situation is likely caused by 
the treatment of data by such large surveys when the wrong template is used to calculate RVs.

\subsection{Summary and future work}\label{subsection43}

$\,\!$\indent In this paper we have determined that different effects (presence of \ion{He}{ii} lines next to Balmer lines, forbidden 
components in \ion{He}{i} lines, and infilling by wind effects) can significantly bias the RV determinations of O stars. We have also established a
procedure that corrects for those biases while at the same time the application of the UNWIND code to the \lili\ data allows us to produce spectra of
unprecedented quality. In that way we have discovered four new SB1 and one new SB2 systems, proposing thresholds in the RV dispersions for the detection
of new SBs. We have also determined for the first time the Oe nature of two stars and the Onfp nature of another three. We plan to continue this line of 
work in 
four ways:

\begin{itemize}
 \item Analysing in detail the four new SB orbits (we exclude HD~\num{152424}, as we have already done it).
 \item Accurately measuring systemic velocities of single stars.
 \item Applying the analysis to the systemic velocities of spectroscopic binaries. 
 \item Using our newly determined RVs to trace back the trajectories of runaway/walkaway stars. 
\end{itemize}

\begin{acknowledgements}
J.M.A., E.M.F., M.P.G., S.R., and J.A.M.C. acknowledge support from the Spanish Government Ministerio de Ciencia, Innovaci\'on y
Universidades and Agencia Estatal de Investigaci\'on (AEI, \num{10.13039}/\num{501100011033}) through grant PID2022-\num{136640}~NB-C22.
G. H acknowledges support from the same entities and from the Fondo Europeo de Desarrollo Regional (FEDER) through grant PID2024-\num{159329}NB-C21.
The project leading to this application has received funding from the European Comission (EC) under Project OCEANS, HORIZON-MSCA-2023-SE-01, No G.A 
\num{101183150} funded by the European Union. G. H. also acknowledges financial support from the AEI through the Severo Ochoa Centre of Excellence 
accreditation awarded to the IAC, grant CEX2025-\num{001609}-S and from the ``La Caixa'' Foundation 
(ID \num{100010434}) under the fellowship code LCF/BQ/PI23/\num{11970035}.
M.P.G.  acknowledges funding from the European Research Council (ERC, ISM-FLOW, \num{101055318}) under the European Union's Horizon 2020 research and 
innovation programme (Grant agreement No. \num{851435}).
This research was supported by the Munich Institute for Astro-, Particle and BioPhysics (MIAPbP) which is funded by the Deutsche Forschungsgemeinschaft 
(DFG, German Research Foundation) under Germany's Excellence Strategy - EXC-2094 - \num{390783311}.
S.R. and L.S. acknowledge support from the Swiss National Foundation (SNF) project No \num{212143}.
S.R. is a Postdoctoral Researcher of the Fonds de la Recherche Scientifique - FNRS. 
Support provided to S.R. by the L'Or\'eal-UNESCO For Women In Sciences Programme.
This paper includes data obtained with:
the 1.2~m Mercator Telescope (MT),
the 2.6~m Nordic Optical Telescope (NOT),  and
the 3.6~m Telescopio Nazionale Galileo (TNG) at the Observatorio del Roque de los Muchachos (ORM), La Palma, Spain;
the 1.9~m telescope (OHP1.9) at the Observatoire de Haute-Provence (OHP), France;
the 3.5~m Telescope (CAHA3.5) at the Centro Astron\'omico Hispano-Alem\'an/Andaluz, Almer{\'\i}a, Spain;
the 2.2~m telescope (OLS2.2) and 
the 3.6~m telescope (OLS3.6) at the Observatorio~de~La~Silla (OLS), Chile; 
the 8.2~m Very Large Telescope (VLT) at Observatorio Paranal (OP), Chile; and
the 9.2~m Southern African large Telescope (SALT) at the South African Astronomical Observatory (SAAO), South Africa. 
We thank the staff at those observatories for their support.
\end{acknowledgements}

%
%
%
\bibliographystyle{aa} 
\bibliography{general}   

@Article{Abduetal22b,
 author        = {{Abdurro'uf} and others},
 title         = {The Seventeenth Data Release of the Sloan Digital Sky Surveys: Complete Release of MaNGA, MaStar, and APOGEE-2 Data},
 journal       = {ApJS},
 year          = 2022,
 month         = apr,
 volume        = {259},
 number        = {2},
 eid           = {35},
 pages         = {35},
 doi           = {10.3847/1538-4365/ac4414},
 archivePrefix = {arXiv},
 eprint        = {arXiv:2112.02026},
 primaryClass  = {astro-ph.GA},
 adsurl        = {https://ui.adsabs.harvard.edu/abs/2022ApJS..259...35A}}

@Article{Barbetal26,
 author        = {Barb{\'a}, R.~H. and Gamen, R. and Morrell, N.~I. and Arias, J.~I. and Ma{\'\i}z Apell{\'a}niz, J. and Higa, R. and Rodr{\'\i}guez, C. and Putkuri, C. and Ferrero, G.~A. and Ans{\'\i}n, T. and Moreno, J. and Molina-Calzada, J.~A. and Holgado, G. and Rosu, S.},
 title         = {The OWN Survey. A high-resolution spectroscopic survey of southern Galactic O and WN-type stars: I. Project description and single-lined spectroscopic orbits},
 journal       = {A\&A},
 year          = 2026,
 month         = apr,
 volume        = {708},
 eid           = {A98},
 pages         = {A98},
 doi           = {10.1051/0004-6361/202558665},
 adsurl        = {https://ui.adsabs.harvard.edu/abs/2026A&A...708A..98B}}

@Article{Barretal25,
 author        = {Barron, James A. and others},
 title         = {A Magnetic Field Detection in the Massive O-type Bright Giant 63 Oph},
 journal       = {ApJ},
 year          = 2025,
 month         = sep,
 volume        = {991},
 number        = {1},
 eid           = {79},
 pages         = {79},
 doi           = {10.3847/1538-4357/adf6d9},
 archivePrefix = {arXiv},
 eprint        = {arXiv:2505.11374},
 primaryClass  = {astro-ph.SR},
 adsurl        = {https://ui.adsabs.harvard.edu/abs/2025ApJ...991...79B}}

@Article{BondKrel23,
 author        = {Bondar, A. and Kre{\l}owski, J.},
 title         = {Spectral investigation of the interstellar cloud towards a runaway star AE Aur},
 journal       = {MNRAS},
 year          = 2023,
 month         = nov,
 volume        = {525},
 number        = {4},
 pages         = {5905-5914},
 doi           = {10.1093/mnras/stad2703},
 adsurl        = {https://adsabs.net/abs/2023MNRAS.525.5905B}}

@Article{deMietal13,
 author        = {de Mink, S.~E. and others},
 title         = {The Rotation Rates of Massive Stars: The Role of Binary Interaction through Tides, Mass Transfer, and Mergers},
 journal       = {ApJ},
 archivePrefix = "arXiv",
 eprint        = {arXiv:1211.3742},
 primaryClass  = "astro-ph.SR",
 year          = 2013,
 month         = feb,
 volume        = 764,
 eid           = {166},
 pages         = {166},
 doi           = {10.1088/0004-637X/764/2/166},
 adsurl        = {https://adsabs.net/abs/2013ApJ...764..166D}}

@Article{Hainetal19,
 author        = {Hainich, R. and Ramachandran, V. and Shenar, T. and Sander, A.~A.~C. and Todt, H. and Gruner, D. and Oskinova, L.~M. and Hamann, W. -R.},
 title         = {PoWR grids of non-LTE model atmospheres for OB-type stars of various metallicities},
 journal       = {A\&A},
 year          = 2019,
 month         = jan,
 volume        = {621},
 eid           = {A85},
 pages         = {A85},
 doi           = {10.1051/0004-6361/201833787},
 archivePrefix = {arXiv},
 eprint        = {arXiv:1811.06307},
 primaryClass  = {astro-ph.SR},
 adsurl        = {https://adsabs.net/abs/2019A&A...621A..85H}}

@Article{HenBetal12b,
 author        = {H{\'e}nault-Brunet, V. and Evans, C.~J. and Sana, H. and Gieles, M. and Bastian, N. and Ma{\'\i}z Apell{\'a}niz, J. and Markova, N. and Taylor, W.~D. and Bressert, E. and Crowther, P.~A. and van Loon, J.~T.},
 title         = {The VLT-FLAMES Tarantula Survey. VII. A low velocity dispersion for the young massive cluster R136},
 journal       = {A\&A},
 archivePrefix = "arXiv",
 eprint        = {arXiv:1208.0825},
 primaryClass  = "astro-ph.GA",
 year          = 2012,
 month         = oct,
 volume        = 546,
 eid           = {A73},
 pages         = {A73},
 doi           = {10.1051/0004-6361/201219471},
 adsurl        = {https://adsabs.net/abs/2012A&A...546A..73H}}

@Article{HillMill98,
 author        = {Hillier, D.~J. and Miller, D.~L.},
 title         = {The Treatment of Non-LTE Line Blanketing in Spherically Expanding Outflows},
 journal       = {ApJ},
 year          = {1998},
 month         = {mar},
 volume        = {496},
 firstpage     = {407},
 pages         = {407},
 adsurl        = {https://adsabs.net/abs/1998ApJ...496..407H},
 comment       = {CMFGEN(programa)}}

@Article{Hirvetal06,
 author        = {Hirv, A. and Annuk, K. and Eenm{\"a}e, T. and Liimets, T. and Pelt, J. and Puss, A. and Tempel, M.},
 title         = {Orbital Elements and Mass-Loss Rate of V 444 Cyg},
 journal       = {Baltic Astronomy},
 year          = 2006,
 month         = jan,
 volume        = {15},
 pages         = {405-412},
 adsurl        = {https://adsabs.net/abs/2006BaltA..15..405H}}

@Phdthesis{Holg19,
 author        = {Holgado, Gonzalo},
 title         = {Spectroscopic and physical characterization of the Galactic O-type stars targeted by the IACOB \& OWN surveys},
 school        = {University of La Laguna, Spain},
 year          = 2019,
 month         = jan,
 adsurl        = {https://ui.adsabs.harvard.edu/abs/2019PhDT.......144H}}

@Article{Holgetal18,
 author        = {Holgado, G. and Sim{\'o}n-D{\'\i}az, S. and Barb{\'a}, R.~H. and Puls, J. and Herrero, A. and Castro, N. and Garcia, M. and Ma{\'\i}z Apell{\'a}niz, J. and Negueruela, I. and Sab{\'\i}n-Sanjuli{\'a}n, C.},
 title         = {The IACOB project. V. Spectroscopic parameters of the O-type stars in the modern grid of standards for spectral classification},
 journal       = {A\&A},
 archivePrefix = "arXiv",
 eprint        = {arXiv:1711.10043},
 primaryClass  = "astro-ph.SR",
 year          = 2018,
 month         = jun,
 volume        = 613,
 eid           = {A65},
 pages         = {A65},
 doi           = {10.1051/0004-6361/201731543},
 adsurl        = {https://adsabs.net/abs/2018A&A...613A..65H}}

@Article{Holgetal25a,
 author        = {Holgado, G. and Ma{\'\i}z Apell{\'a}niz, J. and Gamen, R.~C.},
 title         = {MONOS: Multiplicity Of Northern O-type Spectroscopic systems: III. New orbits and Gaia{\textendash}TESS analysis for ten SB2E systems},
 journal       = {A\&A},
 year          = 2025,
 month         = sep,
 volume        = {701},
 eid           = {A246},
 pages         = {A246},
 doi           = {10.1051/0004-6361/202556068},
 archivePrefix = {arXiv},
 eprint        = {arXiv:2508.04272},
 primaryClass  = {astro-ph.SR},
 adsurl        = {https://ui.adsabs.harvard.edu/abs/2025A&A...701A.246H}}

@Article{Hoogetal01,
 author        = {Hoogerwerf, R. and de Bruijne, J.~H.~J. and de Zeeuw, P.~T.},
 title         = {On the origin of the O and B-type stars with high velocities. II. Runaway stars and pulsars ejected from the nearby young stellar groups},
 journal       = {A\&A},
 year          = {2001},
 month         = {jan},
 volume        = {365},
 firstpage     = {49},
 pages         = {49-77},
 doi           = {10.1051/0004-6361:20000014},
 adsurl        = {https://adsabs.net/abs/2001A&A...365...49H},
 comment       = {Lo mismo de arriba pero en art{\'\i}culo largo}}

@Article{Maizetal04b,
 author        = {Ma{\'\i}z Apell{\'a}niz, J. and others},
 title         = {A Galactic O Star Catalog},
 journal       = {ApJS},
 year          = {2004},
 month         = {mar},
 volume        = {151},
 firstpage     = {103},
 pages         = {103-148},
 adsurl        = {https://adsabs.net/abs/2004ApJS..151..103M},
 comment       = {Cat{\'a}logo O}}

@Inproceedings{Maizetal11,
 author        = {Ma{\'\i}z Apell{\'a}niz, J. and Sota, A. and Walborn, N.~R. and Alfaro, E.~J. and Barb{\'a}, R.~H. and Morrell, N.~I. and Gamen, R.~C. and Arias, J.~I.},
 title         = {The Galactic O-star spectroscopic survey (GOSSS)},
 booktitle     = {HSA 6},
 year          = 2011,
 archivePrefix = "arXiv",
 eprint        = {arXiv:1010.5680},
 primaryClass  = "astro-ph.SR",
 month         = nov,
 pages         = {467-472},
 adsurl        = {https://adsabs.net/abs/2011hsa6.conf..467M}}

@Inproceedings{Maizetal12,
 author        = {Ma{\'\i}z Apell{\'a}niz, J. and others},
 title         = {The Galactic O-Star Spectroscopic (GOSSS) and Northern Massive Dim Stars (NoMaDS) Surveys, the Galactic O-Star Catalog (GOSC), and Marxist Ghost Buster (MGB)},
 booktitle     = {ASPCS},
 year          = 2012,
 volume        = 465,
 archivePrefix = "arXiv",
 eprint        = {arXiv:1109.1492},
 primaryClass  = "astro-ph.SR",
 month         = dec,
 pages         = {484},
 adsurl        = {https://adsabs.net/abs/2012ASPC..465..484M}}

@Article{Maizetal16,
 author        = {Ma{\'\i}z Apell{\'a}niz, J. and Sota, A. and Arias, J.~I. and Barb{\'a}, R.~H. and Walborn, N.~R. and Sim{\'o}n-D{\'\i}az, S. and Negueruela, I. and Marco, A. and Le{\~a}o, J.~R.~S. and Herrero, A. and Gamen, R.~C. and Alfaro, E.~J.},
 title         = {The Galactic O-Star Spectroscopic Survey (GOSSS). III. 142 Additional O-type Systems.},
 journal       = {ApJS},
 archivePrefix = "arXiv",
 eprint        = {arXiv:1602.01336},
 primaryClass  = "astro-ph.SR",
 year          = 2016,
 month         = may,
 volume        = 224,
 eid           = {4},
 pages         = {4},
 doi           = {10.3847/0067-0049/224/1/4},
 adsurl        = {https://adsabs.net/abs/2016ApJS..224....4M}}

@Inproceedings{Maizetal19a,
 author        = {Ma{\'\i}z Apell{\'a}niz, J. and others},
 title         = {LiLiMaRlin, a Library of Libraries of Massive-Star High-Resolution Spectra with applications to OWN, MONOS, and CollDIBs},
 booktitle     = {HSA 10},
 year          = 2019,
 archivePrefix = "arXiv",
 eprint        = {arXiv:1810.10943},
 primaryClass  = "astro-ph.SR",
 month         = mar,
 pages         = {420},
 adsurl        = {https://adsabs.net/abs/2019hsax.conf..420M}}

@Article{Maizetal19b,
 author        = {Ma{\'\i}z Apell{\'a}niz, J. and others},
 title         = {MONOS: Multiplicity Of Northern O-type Spectroscopic systems. I. Project description and spectral classifications and visual multiplicity of previously known objects},
 journal       = {A\&A},
 archivePrefix = "arXiv",
 eprint        = {arXiv:1904.11385},
 primaryClass  = "astro-ph.SR",
 year          = 2019,
 month         = jun,
 volume        = 626,
 pages         = {A20}, 
 eid           = {A20},
 doi           = {10.1051/0004-6361/201935359},
 adsurl        = {https://adsabs.net/abs/2019A&A...626A..20M}}

@Article{Maizetal21b,
 author        = {Ma{\'\i}z Apell{\'a}niz, J. and Barb{\'a}, R.~H. and Fari{\~n}a, C. and Sota, A. and Pantaleoni Gonz{\'a}lez, M. and Holgado, G. and Negueruela, I. and Sim{\'o}n-D{\'\i}az, S.},
 title         = {Lucky spectroscopy, an equivalent technique to lucky imaging. II. Spatially resolved intermediate-resolution blue-violet spectroscopy of 19 close massive binaries using the William Herschel Telescope},
 journal       = {A\&A},
 year          = 2021,
 month         = feb,
 volume        = {646},
 eid           = {A11},
 pages         = {A11},
 doi           = {10.1051/0004-6361/202039479},
 archivePrefix = {arXiv},
 eprint        = {arXiv:2011.12250},
 primaryClass  = {astro-ph.SR},
 adsurl        = {https://adsabs.net/abs/2021A&A...646A..11M}}

@Article{Maizetal23,
 author        = {Ma{\'\i}z Apell{\'a}niz, J. and others},
 title         = {Stellar variability in Gaia DR3. I. Three-band photometric dispersions for 145 million sources},
 journal       = {A\&A},
 year          = 2023,
 month         = sep,
 volume        = {677},
 eid           = {A137},
 pages         = {A137},
 doi           = {10.1051/0004-6361/202346759},
 archivePrefix = {arXiv},
 eprint        = {arXiv:2304.14249},
 primaryClass  = {astro-ph.SR},
 adsurl        = {https://adsabs.net/abs/2023A&A...677A.137M}}

@Inproceedings{Maizetal26a,
 author        = {Ma{\'\i}z Apell{\'a}niz, Jes{\'u}s and others},
 title         = {Spectral classification},
 booktitle     = {Encyclopedia of Astrophysics},
 year          = 2026,
 volume        = {2},
 month         = jan,
 pages         = {43-84},
 doi           = {10.1016/B978-0-443-21439-4.00057-2},
 adsurl        = {https://ui.adsabs.harvard.edu/abs/2026enap....2...43M}}

@Article{Maizetal26b,
 author        = {Ma{\'\i}z Apell{\'a}niz, J. and others},
 title         = {Multiplicity of Massive stars in the Milky Way (M$^{3}$W): I. Project description, UNWIND, application to GLS 11 448 (=LS III +46 11), and DIB catalog},
 journal       = {A\&A},
 year          = 2026,
 month         = sep,
 volume        = {713},
 eid           = {A177},
 pages         = {A177 (Paper I)},
 doi           = {10.1051/0004-6361/202660211},
 archivePrefix = {arXiv},
 eprint        = {arXiv:2604.02111},
 primaryClass  = {astro-ph.SR},
 adsurl        = {https://ui.adsabs.harvard.edu/abs/2026A&A...713A.177M}}

@Article{Marcetal98,
 author        = {Marchenko, Sergey V. and Moffat, Anthony F.~J. and Eenens, Philippe R.~J.},
 title         = {The Wolf-Rayet Binary WR 141 (WN5O + O5 V-III) Revisited},
 journal       = {PASP},
 year          = 1998,
 month         = dec,
 volume        = {110},
 number        = {754},
 pages         = {1416-1422},
 doi           = {10.1086/316280},
 adsurl        = {https://adsabs.net/abs/1998PASP..110.1416M}}

@Inproceedings{Mark09,
 author        = {Markwardt, C.~B.},
 title         = {Non-linear Least-squares Fitting in IDL with MPFIT},
 year          = 2009,
 booktitle     = {ADASS XVIII},
 series        = {ASPCS},
 volume        = {411},
 month         = sep,
 pages         = {251},
 archivePrefix = {arXiv},
 eprint        = {arXiv:0902.2850},
 primaryClass  = {astro-ph.IM},
 adsurl        = {https://adsabs.net/abs/2009ASPC..411..251M}}

@Article{Masoetal01,
 author        = {Mason, B.~D. and others},
 title         = {The 2001 US Naval Observatory Double Star CD-ROM. I. The Washington Double Star Catalog},
 journal       = {AJ},
 year          = {2001},
 month         = {dec},
 volume        = {122},
 firstpage     = {3466},
 pages         = {3466-3471},
 doi           = {10.1086/323920},
 adsurl        = {https://adsabs.net/abs/2001AJ....122.3466M},
 comment       = {Cat{\'a}logo WDS}}

@Article{Neguetal04,
 author        = {Negueruela, I. and Steele, I.~A. and Bernab\'eu, G.},
 title         = {On the class of Oe stars},
 journal       = {AN},
 eprint        = {arXiv:astro-ph/0407113},
 year          = {2004},
 month         = {dec},
 volume        = {325},
 firstpage     = {749},
 pages         = {749-760},
 doi           = {10.1002/ansa.200310258},
 adsurl        = {https://adsabs.net/abs/2004AN....325..749N}}

@Article{Pantetal21,
 author        = {Pantaleoni Gonz{\'a}lez, M. and others},
 title         = {The Alma catalogue of OB stars - II. A cross-match with Gaia DR2 and an updated map of the solar neighbourhood},
 journal       = {MNRAS},
 year          = 2021,
 month         = jun,
 volume        = {504},
 number        = {2},
 pages         = {2968-2982},
 doi           = {10.1093/mnras/stab688},
 archivePrefix = {arXiv},
 eprint        = {arXiv:2103.02748},
 primaryClass  = {astro-ph.SR},
 adsurl        = {https://adsabs.net/abs/2021MNRAS.504.2968P}}

@Article{Pantetal25b,
 author        = {Pantaleoni Gonz{\'a}lez, M. and others},
 title         = {The Alma catalogue of OB stars {\textendash} III. A cross-match with Gaia DR3 and an extension based on new spectral classifications},
 journal       = {MNRAS},
 year          = 2025,
 month         = oct,
 volume        = {543},
 number        = {1},
 pages         = {63-82},
 doi           = {10.1093/mnras/staf1409},
 adsurl        = {https://ui.adsabs.harvard.edu/abs/2025MNRAS.543...63P}}

@Article{Pulsetal05,
 author        = {Puls, J. and Urbaneja, M.~A. and Venero, R. and Repolust, T. and Springmann, U. and Jokuthy, A. and Mokiem, M.~R.},
 title         = {Atmospheric NLTE-models for the spectroscopic analysis of blue stars with winds. II. Line-blanketed models},
 journal       = {A\&A},
 eprint        = {arXiv:astro-ph/0411398},
 year          = {2005},
 month         = {may},
 volume        = {435},
 firstpage     = {669},
 pages         = {669-698},
 doi           = {10.1051/0004-6361:20042365},
 adsurl        = {https://adsabs.net/abs/2005A&A...435..669P},
 comment       = {Comparaci{\'o}n entre FASTWIND (suyo), CMFGEN y WM-BASIC}}

@Article{Reed03,
 author        = {Reed, B.~C.},
 title         = {Catalog of Galactic OB Stars},
 journal       = {AJ},
 year          = {2003},
 month         = {may},
 volume        = {125},
 firstpage     = {2531},
 pages         = {2531-2533},
 adsurl        = {https://adsabs.net/abs/2003AJ....125.2531R}}

@Article{Renzetal19b,
 author        = {Renzo, M. and Zapartas, E. and de Mink, S.~E. and G{\"o}tberg, Y. and Justham, S. and Farmer, R.~J. and Izzard, R.~G. and Toonen, S. and Sana, H.},
 title         = {Massive runaway and walkaway stars. A study of the kinematical imprints of the physical processes governing the evolution and explosion of their binary progenitors},
 journal       = {A\&A},
 year          = 2019,
 month         = apr,
 volume        = {624},
 eid           = {A66},
 pages         = {A66},
 doi           = {10.1051/0004-6361/201833297},
 archivePrefix = {arXiv},
 eprint        = {arXiv:1804.09164},
 primaryClass  = {astro-ph.SR},
 adsurl        = {https://adsabs.net/abs/2019A&A...624A..66R}}

@Article{RivGetal11,
 author        = {Rivero Gonz{\'a}lez, J.~G. and Puls, J. and Najarro, F.},
 title         = {Nitrogen line spectroscopy of O-stars. I. Nitrogen III emission line formation revisited},
 journal       = {A\&A},
 archivePrefix = "arXiv",
 eprint        = {arXiv:1109.3595},
 primaryClass  = "astro-ph.SR",
 year          = 2011,
 month         = dec,
 volume        = 536,
 eid           = {A58},
 pages         = {A58},
 doi           = {10.1051/0004-6361/201117101},
 adsurl        = {https://adsabs.net/abs/2011A&A...536A..58R}}

@Inproceedings{Rivietal11b,
 author        = {Rivinius, T. and Stahl, O. and {\v S}tefl, S. and Baade, D. and Townsend, R. H. D. and Barrera, L.},
 title         = {The (B0+?)+O6 system FN CMa{\dagger}: a case for tidal-pulsational interaction?},
 booktitle     = {IAUS},
 year          = 2011,
 volume        = 272,
 month         = jul,
 pages         = {543-544},
 doi           = {10.1017/S1743921311011380},
 adsurl        = {https://adsabs.net/abs/2011IAUS..272..543R}}

@Article{Rosuetal25,
 author        = {Rosu, Sophie and Ma{\'\i}z Apell{\'a}niz, Jes{\'u}s and Sciarini, Luca and Gamen, Roberto C. and Molina-Calzada, Juan A. and Holgado, Gonzalo and Barb{\'a}, Rodolfo H.},
 title         = {One century data of {\ensuremath{\tau}} CMa: a (2+1)+1 system with a short-period overcontact binary and an eccentric intermediate orbit with probably no apsidal motion},
 year          = 2025,
 month         = oct,
 eid           = {arXiv:2510.16202},
 pages         = {arXiv:2510.16202},
 doi           = {10.48550/arXiv.2510.16202},
 archivePrefix = {arXiv},
 eprint        = {arXiv:2510.16202},
 primaryClass  = {astro-ph.SR},
 adsurl        = {https://ui.adsabs.harvard.edu/abs/2025arXiv251016202R}}

@Article{Santetal97,
 author        = {Santolaya-Rey, A.~E. and Puls, J. and Herrero, A.},
 title         = {Atmospheric NLTE-models for the spectroscopic analysis of luminous blue stars with winds.},
 journal       = {A\&A},
 year          = 1997,
 month         = jul,
 volume        = {323},
 pages         = {488-512},
 adsurl        = {https://ui.adsabs.harvard.edu/abs/1997A&A...323..488S}}

@Article{Schnetal19,
 author        = {Schneider, Fabian R.~N. and others},
 title         = {Stellar mergers as the origin of magnetic massive stars},
 journal       = {Nature},
 year          = 2019,
 month         = oct,
 volume        = {574},
 number        = {7777},
 pages         = {211-214},
 doi           = {10.1038/s41586-019-1621-5},
 archivePrefix = {arXiv},
 eprint        = {arXiv:1910.14058},
 primaryClass  = {astro-ph.SR},
 adsurl        = {https://adsabs.net/abs/2019Natur.574..211S}}

@Article{Sciaetal26,
 author        = {Sciarini, Luca and others},
 title         = {The IACOB project XIX. Revisiting massive-star evolution with empirical TAMS constraints: updated models, overshoot calibration, and the population of blue supergiants},
 journal       = {arXiv:2607.29650},
 year          = 2026,
 month         = jul,
 doi           = {10.48550/arXiv.2607.29650},
 archivePrefix = {arXiv},
 eprint        = {arXiv:2607.29650},
 primaryClass  = {astro-ph.SR},
 adsurl        = {https://ui.adsabs.harvard.edu/abs/2026arXiv260729650S}}

@Article{SimDHerr07,
 author        = {Sim{\'o}n-D{\'\i}az, S. and Herrero, A.},
 title         = {Fourier method of determining the rotational velocities in OB stars},
 journal       = {A\&A},
 year          = 2007,
 month         = jun,
 volume        = {468},
 number        = {3},
 pages         = {1063-1073},
 doi           = {10.1051/0004-6361:20066060},
 archivePrefix = {arXiv},
 eprint        = {astro-ph/0703216},
 primaryClass  = {astro-ph},
 adsurl        = {https://ui.adsabs.harvard.edu/abs/2007A&A...468.1063S}}

@Article{SimDHerr14,
 author        = {Sim{\'o}n-D{\'\i}az, S. and Herrero, A.},
 title         = {The IACOB project. I. Rotational velocities in northern Galactic O- and early B-type stars revisited. The impact of other sources of line-broadening},
 journal       = {A\&A},
 archivePrefix = "arXiv",
 eprint        = {arXiv:1311.3360},
 primaryClass  = "astro-ph.SR",
 year          = 2014,
 month         = feb,
 volume        = 562,
 eid           = {A135},
 pages         = {A135},
 doi           = {10.1051/0004-6361/201322758},
 adsurl        = {https://adsabs.net/abs/2014A&A...562A.135S}}

@Article{SimDetal11d,
 author        = {Sim{\'o}n-D{\'\i}az, S. and others},
 title         = {The IACOB project: A grid-based automatic tool for the quantitative spectroscopic analysis of O-stars},
 journal       = {Journal of Physics Conference Series},
 archivePrefix = "arXiv",
 eprint        = {arXiv:1111.1341},
 primaryClass  = "astro-ph.SR",
 year          = 2011,
 month         = dec,
 volume        = 328,
 number        = 1,
 pages         = {012021},
 doi           = {10.1088/1742-6596/328/1/012021},
 adsurl        = {https://adsabs.net/abs/2011JPhCS.328a2021S}}

@Article{SimDetal14,
 author        = {Sim{\'o}n-D{\'\i}az, S. and Herrero, A. and Sab{\'\i}n-Sanjuli{\'a}n, C. and Najarro, F. and Garcia, M. and Puls, J. and Castro, N. and Evans, C.~J.},
 title         = {The IACOB project. II. On the scatter of O-dwarf spectral type - effective temperature calibrations},
 journal       = {A\&A},
 archivePrefix = "arXiv",
 eprint        = {arXiv:1409.6653},
 primaryClass  = "astro-ph.SR",
 year          = 2014,
 month         = oct,
 volume        = 570,
 eid           = {L6},
 pages         = {L6},
 doi           = {10.1051/0004-6361/201424742},
 adsurl        = {https://adsabs.net/abs/2014A&A...570L...6S}}

@Article{SimDetal17,
 author        = {Sim{\'o}n-D{\'\i}az, S. and Godart, M. and Castro, N. and Herrero, A. and Aerts, C. and Puls, J. and Telting, J. and Grassitelli, L.},
 title         = {The IACOB project . III. New observational clues to understand macroturbulent broadening in massive O- and B-type stars},
 journal       = {A\&A},
 archivePrefix = "arXiv",
 eprint        = {arXiv:1608.05508},
 primaryClass  = "astro-ph.SR",
 year          = 2017,
 month         = jan,
 volume        = 597,
 eid           = {A22},
 pages         = {A22},
 doi           = {10.1051/0004-6361/201628541},
 adsurl        = {https://adsabs.net/abs/2017A&A...597A..22S}}

@Article{SimDetal24,
 author        = {Sim{\'o}n-D{\'\i}az, S. and others},
 title         = {The IACOB project: CVIII. Hunting for spectroscopic binaries in the O and B supergiant domain. The threat of pulsational variability},
 journal       = {arXiv e-prints},
 year          = 2024,
 month         = may,
 eid           = {arXiv:2405.11209},
 pages         = {arXiv:2405.11209},
 doi           = {10.48550/arXiv.2405.11209},
 archivePrefix = {arXiv},
 eprint        = {arXiv:2405.11209},
 primaryClass  = {astro-ph.SR},
 adsurl        = {https://ui.adsabs.harvard.edu/abs/2024arXiv240511209S}}

@Article{SimDetal26,
 author        = {Sim{\'o}n-D{\'\i}az, S. and others},
 title         = {The IACOB project: XVI. Surface helium abundances in Galactic O-type stars: Indications for identifying binary interaction products},
 journal       = {A\&A},
 year          = 2026,
 month         = jul,
 volume        = {711},
 eid           = {A151},
 pages         = {A151},
 doi           = {10.1051/0004-6361/202659020},
 archivePrefix = {arXiv},
 eprint        = {arXiv:2601.20698},
 primaryClass  = {astro-ph.SR},
 adsurl        = {https://ui.adsabs.harvard.edu/abs/2026A&A...711A.151S}}

@Article{Sotaetal11a,
 author        = {Sota, A. and Ma{\'\i}z Apell{\'a}niz, J. and Walborn, N.~R. and Alfaro, E.~J. and Barb{\'a}, R.~H. and Morrell, N.~I. and Gamen, R.~C. and Arias, J.~I.},
 title         = {The Galactic O-Star Spectroscopic Survey. I. Classification System and Bright Northern Stars in the Blue-violet at R \~{} 2500},
 journal       = {ApJS},
 archivePrefix = "arXiv",
 primaryClass  = "astro-ph.GA",
 year          = 2011,
 month         = apr,
 volume        = 193,
 pages         = {24},
 doi           = {10.1088/0067-0049/193/2/24},
 adsurl        = {https://adsabs.net/abs/2011ApJS..193...24S}}

@Article{Sotaetal14,
 author        = {Sota, A. and Ma{\'\i}z Apell{\'a}niz, J. and Morrell, N.~I. and Barb{\'a}, R.~H. and Walborn, N.~R. and Gamen, R.~C. and Arias, J.~I. and Alfaro, E.~J.},
 title         = {The Galactic O-Star Spectroscopic Survey (GOSSS). II. Bright Southern Stars},
 journal       = {ApJS},
 archivePrefix = "arXiv",
 eprint        = {arXiv:1312.6222},
 primaryClass  = "astro-ph.GA",
 year          = 2014,
 month         = mar,
 volume        = 211,
 eid           = {10},
 pages         = {10}, 
 doi           = {10.1088/0067-0049/211/1/10},
 adsurl        = {https://adsabs.net/abs/2014ApJS..211...10S}}

@Article{Tremetal26,
 author        = {Tremblay, Patrick and Beauchamp, Alain and Bergeron, Pierre and B{\'e}dard, Antoine},
 title         = {Improved Stark-broadened Profiles for Neutral Helium Lines Using Computer Simulations},
 journal       = {ApJ},
 year          = 2026,
 month         = apr,
 volume        = {1000},
 number        = {2},
 eid           = {253},
 pages         = {253},
 doi           = {10.3847/1538-4357/ae4e26},
 archivePrefix = {arXiv},
 eprint        = {arXiv:2603.04374},
 primaryClass  = {astro-ph.SR},
 adsurl        = {https://ui.adsabs.harvard.edu/abs/2026ApJ..1000..253T}}

@Article{Trigetal21,
 author        = {Trigueros P{\'a}ez, E. and Barb{\'a}, R.~H. and Negueruela, I. and Ma{\'\i}z Apell{\'a}niz, J. and Sim{\'o}n-D{\'\i}az, S. and Holgado, G.},
 title         = {MONOS: Multiplicity Of Northern O-type Spectroscopic systems. II. Orbit review and analysis for 35 single-lined spectroscopic binary systems and candidates},
 journal       = {A\&A},
 year          = 2021,
 month         = nov,
 volume        = {655},
 eid           = {A4},
 pages         = {A4},
 doi           = {10.1051/0004-6361/202141018},
 archivePrefix = {arXiv},
 eprint        = {arXiv:2106.08865},
 primaryClass  = {astro-ph.SR},
 adsurl        = {https://adsabs.net/abs/2021A&A...655A...4T}}

@Article{Valletal23,
 author        = {Vallenari, A. and others},
 title         = {Gaia Data Release 3. Summary of the content and survey properties},
 journal       = {A\&A},
 year          = 2023,
 month         = jun,
 volume        = {674},
 eid           = {A1},
 pages         = {A1},
 doi           = {10.1051/0004-6361/202243940},
 archivePrefix = {arXiv},
 eprint        = {arXiv:2208.00211},
 primaryClass  = {astro-ph.GA},
 adsurl        = {https://adsabs.net/abs/2023A&A...674A...1G}}

@Article{Walb71a,
 author        = {Walborn, N.~R.},
 title         = {Some Spectroscopic Characteristics of the OB Stars: an Investigation of the Space Distribution of Certain OB Stars and the Reference Frame of the Classification},
 journal       = {ApJS},
 year          = {1971},
 month         = {aug},
 volume        = {23},
 firstpage     = {257},
 pages         = {257-282},
 adsurl        = {https://adsabs.net/abs/1971ApJS...23..257W}}

@Article{Walb72,
 author        = {Walborn, N.~R.},
 title         = {Spectral classification of OB stars in both hemispheres and the absolute magnitude calibration},
 journal       = {AJ},
 year          = {1972},
 month         = {may},
 volume        = {77},
 firstpage     = {312},
 pages         = {312-318},
 adsurl        = {https://adsabs.net/abs/1972AJ.....77..312W}}

@Inproceedings{Walb01,
 author        = {Walborn, N.~R.},
 title         = {Selective Emission Lines in O-Type Spectra},
 booktitle     = {ASPCS Vol. 242},
 year          = {2001},
 firstpage     = {217},
 pages         = {217},
 adsurl        = {https://adsabs.net/abs/2001ASPC..242..217W}}

@Article{Walbetal11,
 author        = {Walborn, N.~R. and Ma{\'\i}z Apell{\'a}niz, J. and Sota, A. and Alfaro, E.~J. and Morrell, N.~I. and Barb{\'a}, R.~H. and Arias, J.~I. and Gamen, R.~C.},
 title         = {Further Results from the Galactic O-Star Spectroscopic Survey: Rapidly Rotating Late ON Giants},
 journal       = {AJ},
 archivePrefix = "arXiv",
 eprint        = {arXiv:1109.0515},
 primaryClass  = "astro-ph.SR",
 year          = 2011,
 month         = nov,
 volume        = 142,
 eid           = {150},
 pages         = {150},
 doi           = {10.1088/0004-6256/142/5/150},
 adsurl        = {https://adsabs.net/abs/2011AJ....142..150W}}

@Article{Walbetal16,
 author        = {Walborn, N.~R. and Morrell, N.~I. and Barb{\'a}, R.~H. and Sota, A.},
 title         = {New Luminous ON Spectra from the Galactic O-Star Spectroscopic Survey},
 journal       = {AJ},
 archivePrefix = "arXiv",
 eprint        = {arXiv:1603.04749},
 primaryClass  = "astro-ph.SR",
 year          = 2016,
 month         = apr,
 volume        = 151,
 eid           = {91},
 pages         = {91},
 doi           = {10.3847/0004-6256/151/4/91},
 adsurl        = {https://adsabs.net/abs/2016AJ....151...91W}}

@Inproceedings{WernRauc01,
  author        = {Werner, K. and Rauch, T.},
  title         = {Identification of the 4486, 4504{\AA} Emission Lines in O-type Spectra},
  booktitle     = {ASPCS Vol. 242},
  year          = 2001,
  pages         = {229},
  adsurl        = {https://adsabs.net/abs/2001ASPC..242..229W}}

@Article{Wolf63,
 author        = {Wolff, Richard J.},
 title         = {Two Unidentified Emission Lines in Of Stars},
 journal       = {PASP},
 year          = 1963,
 month         = dec,
 volume        = {75},
 number        = {447},
 pages         = {485},
 doi           = {10.1086/128011},
 adsurl        = {https://ui.adsabs.harvard.edu/abs/1963PASP...75..485W}}

%
%
%
%
%

\begin{appendix}

\setcounter{section}{1}

\begin{table*}[!t]
\caption{Sample information, spectral classifications, and number of epochs in the blue-violet range for the first UNWIND run.}
\renewcommand{\arraystretch}{1.2}
\centerline{
\begin{tabular}{lclllr@{}lrrrrrrrrrrrr}
\midrule
Name                  & $B$   & \mciii{Spectral}          & \mcii{$d$}              & tot & uni & MT  & OHP & OLS & NOT & TNG & OLS & VLT & SA  \\
                      & (mag) & \mciii{classification}    & \mcii{(pc)}             &     &     &     & 1.9 & 2.2 &     &     & 3.6 &     & LT  \\
\midrule
$\zeta$ Pup           & 2.0 & O4    & I    & (n)fp      &  488&$^{+139}_{-88}$    &  54 &  11 &   4 & --- &   7 &   3 & --- & --- &  40 & --- \\
$\zeta$ Oph           & 2.6 & ON9.5 & IV   & nn(e)      &  143&$^{+14}_{-11}$     & 373 &  64 &  32 & --- &  19 & 181 &   8 &  61 &  72 & --- \\
$\lambda$ Ori A       & 3.6 & O8    & III  & ((f))      &  369&$^{+93}_{-62}$     & 988 & 125 & 887 &   2 &   1 &  52 & --- & --- &  46 & --- \\
$\xi$ Per             & 4.1 & O7.5  & III  & (n)((f))p  &  413&$^{+110}_{-71}$    & 258 &  69 & 234 & --- & --- &  24 & --- & --- & --- & --- \\
$\alpha$ Cam          & 4.3 & O9    & Ia   &            & 2610&$^{+1180}_{-773}$  & 189 &  72 & 135 &   2 & --- &  52 & --- & --- & --- & --- \\
$\upsilon$ Ori        & 4.4 & O9.7  & V    &            &  406&$^{+81}_{-56}$     & 113 &  27 & 100 & --- &   2 &   7 &   2 & --- &   2 & --- \\
10 Lac                & 4.7 & O9    & V    &            &  449&$^{+90}_{-63}$     & 122 &  27 &  74 &   1 & --- &  44 &   3 & --- & --- & --- \\
$\mu$ Col             & 4.9 & O9.5  & V    &            &  594&$^{+109}_{-80}$    & 785 &  79 &   9 & --- &  10 &   2 &   2 &  99 & 663 & --- \\
68 Cyg                & 5.0 & O7.5  & III  & n((f))p    & 3044&$^{+7886}_{-1629}$ & 182 &  71 & 175 & --- & --- &   7 & --- & --- & --- & --- \\
$\mu$ Nor             & 5.0 & O9.7  & Iab  &            & 1108&$^{+162}_{-162}$   &  58 &  14 & --- & --- &  28 & --- & --- & --- &  30 & --- \\
19 Cep                & 5.2 & O9    & Ib   &            &  979&$^{+217}_{-148}$   & 150 &  44 & 124 & --- & --- &  24 &   2 & --- & --- & --- \\
HD \num{151804}       & 5.3 & O8    & Ia   & f          & 2009&$^{+306}_{-306}$   &  92 &  27 & --- & --- &  92 & --- & --- & --- & --- & --- \\
$\lambda$ Cep         & 5.3 & O6.5  & Iab  & (n)fp      &  919&$^{+252}_{-162}$   & 202 &  55 & 178 & --- & --- &  22 &   2 & --- & --- & --- \\
HD \num{135591}       & 5.4 & O8    & IV   &            &  884&$^{+234}_{-153}$   &  14 &   4 & --- & --- &   4 & --- & --- & --- &  10 & --- \\
HD \num{188209}       & 5.6 & O9.5  & Iab  &            & 1112&$^{+187}_{-187}$   & 127 &  46 & 107 &   2 & --- &  18 & --- & --- & --- & --- \\
HD \num{55879}        & 5.8 & O9.7  & III  &            &  937&$^{+195}_{-137}$   &  67 &  20 &   6 & --- &   6 &   4 &   1 & --- &  50 & --- \\
HD \num{152408}       & 5.9 & O8:   & Ia   & fpe        & 2264&$^{+388}_{-388}$   & 134 &  35 & --- & --- &  98 & --- & --- & --- &  36 & --- \\
1 Cam A               & 6.0 & O9.7  & II   & n          &  774&$^{+90}_{-73}$     &  22 &  20 &   8 & --- & --- &  14 & --- & --- & --- & --- \\
HD \num{155806} A,B   & 6.1 & O7.5  & V    & ((f))e     & 1169&$^{+587}_{-271}$   &  17 &  12 &   5 & --- &   7 & --- & --- &   3 &   2 & --- \\
63 Oph                & 6.2 & O8    & II   & ((f))      &  799&$^{+107}_{-107}$   &  38 &  32 &  19 & --- &   5 &  13 & --- & --- &   1 & --- \\
9 Sge                 & 6.2 & O7.5  & Iab  & f          & 1756&$^{+396}_{-264}$   & 101 &  40 &  75 & --- &   5 &  21 & --- & --- & --- & --- \\
AE Aur                & 6.2 & O9.5  & V    &            &  380&$^{+14}_{-13}$     & 128 &  42 &  24 &  22 & --- &  78 &   2 & --- &   2 & --- \\
HD \num{57682}        & 6.2 & O9.2  & IV   & (e)        & 1159&$^{+364}_{-221}$   &  30 &  20 &  18 & --- &   3 &   8 &   1 & --- & --- & --- \\
HD \num{207198}       & 6.3 & O8.5  & II   & ((f))      &  965&$^{+97}_{-81}$     &  61 &  25 &  45 & --- & --- &  16 & --- & --- & --- & --- \\
HD \num{47432}        & 6.4 & O9.7  & Iab  &            & 1694&$^{+387}_{-268}$   &  68 &  64 &  56 & --- &   2 &   9 &   1 & --- & --- & --- \\
HD \num{68450}        & 6.4 & O9.7  & II   &            & 1296&$^{+167}_{-133}$   &  17 &  12 &   3 & --- &  14 & --- & --- & --- & --- & --- \\
HD \num{152249}       & 6.6 & OC9   & Iab  &            & 1551&$^{+25}_{-24}$     & 222 &  47 & --- & --- & 207 & --- & --- &   5 &  10 & --- \\
HD \num{154368}       & 6.6 & O9.2  & Iab  &            & 1023&$^{+110}_{-90}$    &  21 &  11 & --- & --- &   5 & --- & --- & --- &  16 & --- \\
HD \num{202214} Aa,Ab & 6.6 & O9.5  & V    &            & 4553&$^{+2682}_{-2143}$ &  36 &  22 &  13 & --- & --- &  23 & --- & --- & --- & --- \\
BN Gem                & 6.7 & O8.5: & V    & (n)e       &  988&$^{+875}_{-362}$   &  45 &  43 &  33 & --- & --- &  11 &   1 & --- & --- & --- \\
HD \num{152424}       & 6.7 & OC9.2 & Ia   &            & 1640&$^{+356}_{-247}$   &  31 &  19 & --- & --- &  20 & --- & --- & --- &  11 & --- \\
HD \num{41161}        & 6.7 & O8    & V    & n          & 1639&$^{+600}_{-352}$   &  23 &  21 &  11 & --- & --- &  12 & --- & --- & --- & --- \\
PZ Gem                & 6.7 & O9.5: & III: & ne         &  872&$^{+81}_{-68}$     &  76 &  62 &  53 &   1 &   3 &  11 & --- & --- &   8 & --- \\
HD \num{175876}       & 6.8 & O7    & III  & n(f)p      & 2660&$^{+1093}_{-647}$  &  47 &  36 &  23 & --- &  11 &  13 & --- & --- & --- & --- \\
HD \num{209339} A     & 6.8 & O9.7  & IV   &            &  933&$^{+69}_{-60}$     &  30 &  27 &  20 & --- & --- &   9 &   1 & --- & --- & --- \\
HD \num{93521}        & 6.8 & ON9.5 & III  & nn(e)      & 1110&$^{+234}_{-179}$   & 242 &  96 & 170 & --- & --- &  69 &   3 & --- & --- & --- \\
HD \num{161807}       & 6.9 & ON9.5 & II   & nn(e)      & 1245&$^{+197}_{-148}$   &  19 &  15 &   8 & --- &   5 & --- & --- & --- & --- &   6 \\
HD \num{175754}       & 6.9 & O8    & II   & (n)((f))p  & 1748&$^{+288}_{-288}$   &  39 &  31 &  23 & --- &   8 &   8 & --- & --- & --- & --- \\
HD \num{34656} Aa     & 6.9 & O7.5  & II   & (f)        & 1307&$^{+948}_{-378}$   & 134 &  32 &  85 & --- & --- &  48 &   1 & --- & --- & --- \\
HD \num{46150}        & 6.9 & O4    & V    & ((f))z     & 1421&$^{+21}_{-20}$     &  61 &  36 &   5 & --- &   7 &  17 & --- &  29 &   3 & --- \\
HD \num{69106}        & 7.0 & O9.7  & II   & n          & 1440&$^{+222}_{-169}$   &  11 &  11 &   3 & --- &   8 & --- & --- & --- & --- & --- \\
\midrule
\multicolumn{17}{l}{Notes. tot: Total number of epochs. uni: Unique evening dates. The spectral classification of HD~\num{46150} corresponds to the primary (Aa) alone.} \\
\end{tabular}
}
\renewcommand{\arraystretch}{1.0}
\label{sample}
\end{table*}

\FloatBarrier

\section{Spectral classification plots}\label{appendixb}

\begin{figure*}
 \vspace{-2mm}
 \centerline{\includegraphics*[width=\linewidth]{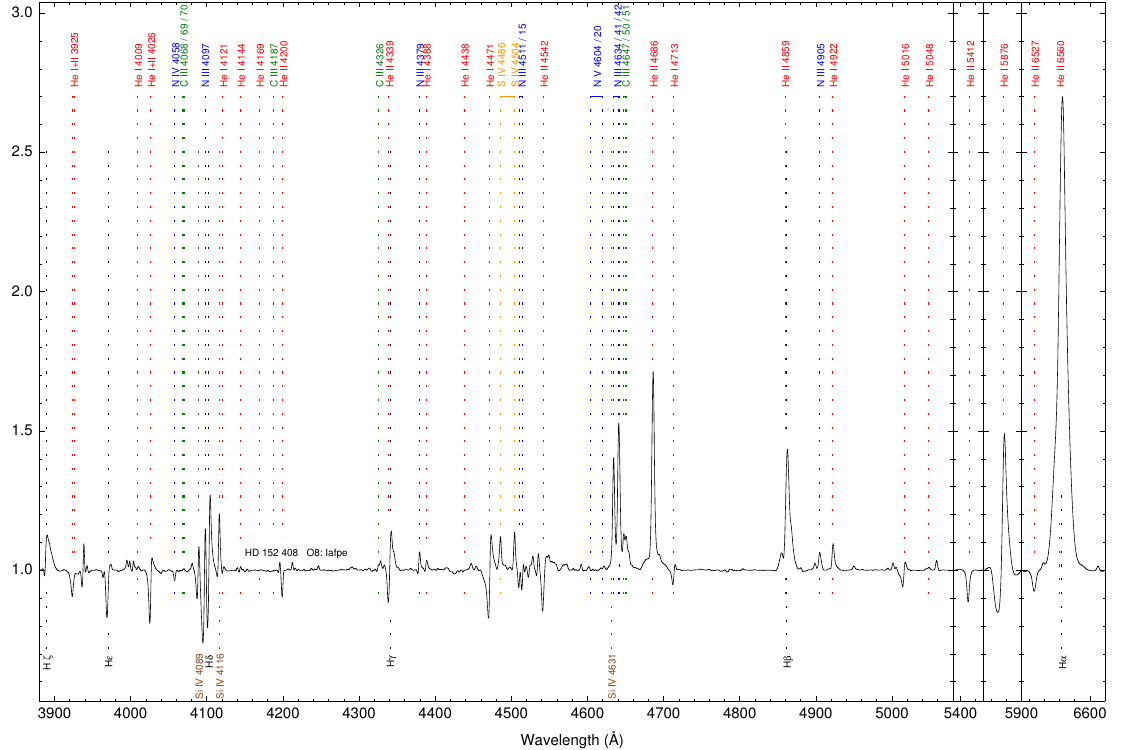}}
 \centerline{\includegraphics*[width=\linewidth]{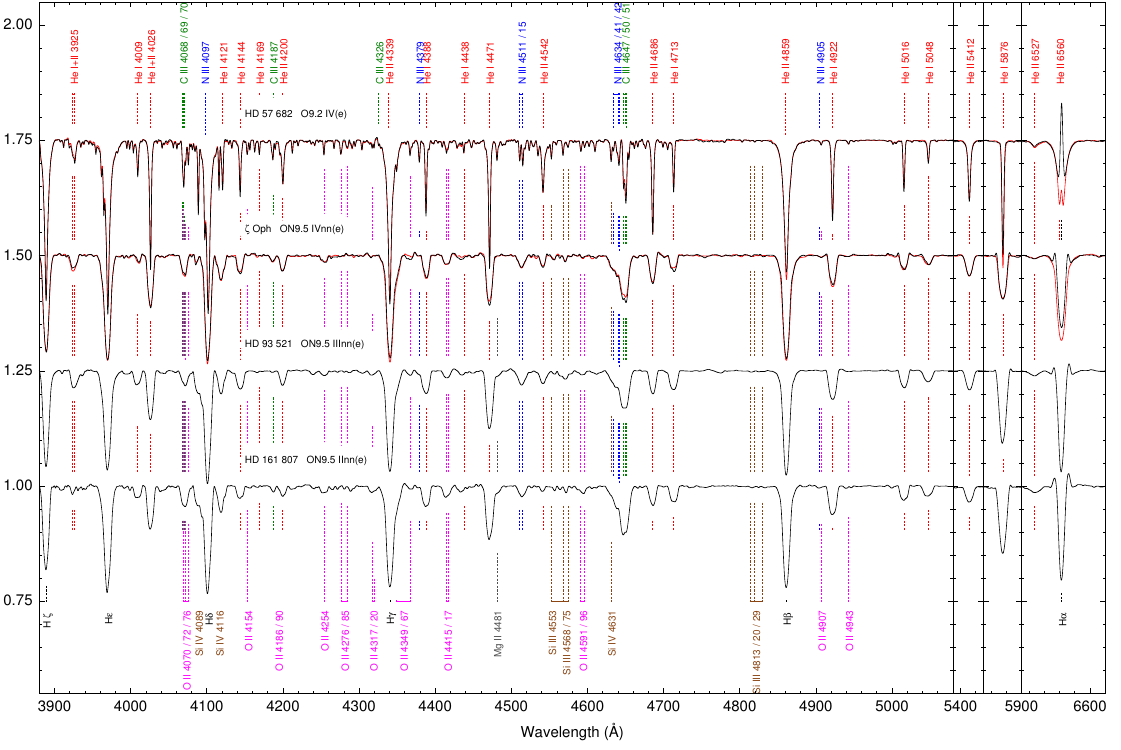}}
 \vspace{-2mm}
 \caption{Spectral classification plots ($R=2500$) for the O~Iafpe star (top) and the weak Oe stars (bottom). Two average spectra are given 
          for HD~\num{57682} and $\zeta$~Oph, one for the high emission state (black) and the other one for the low emission state (red).}
 \label{ALS_spectra_1}      
\end{figure*}

\begin{figure*}
 \centerline{\includegraphics*[width=\linewidth]{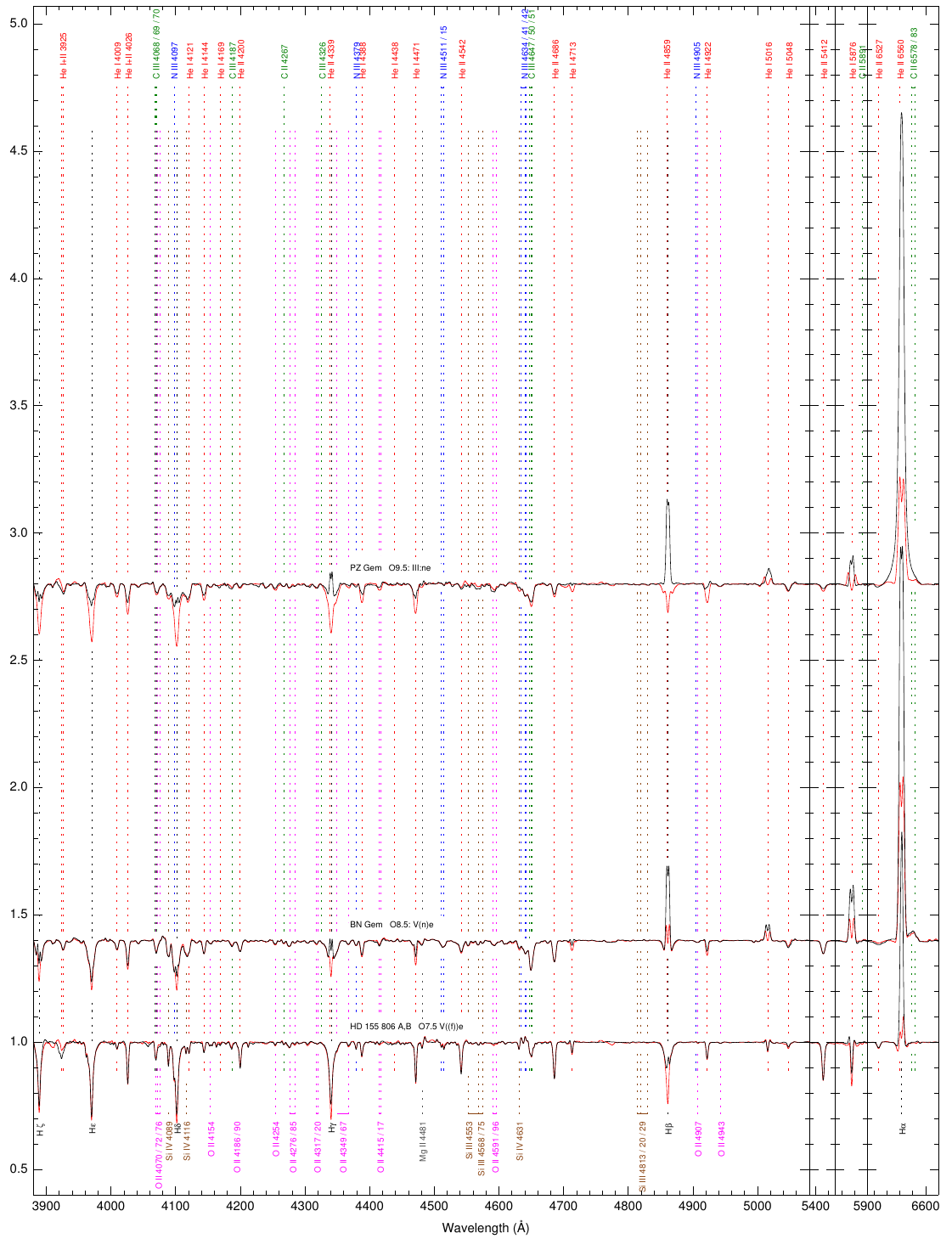}}
 \caption{Spectral classification plots ($R=2500$) for the strong Oe stars. Two average spectra are given for each star, one
          for the high emission state (black) and the other one for the low emission state (red).}
 \label{ALS_spectra_Oe_strong}      
\end{figure*}

\begin{figure*}
 \centerline{\includegraphics*[width=\linewidth]{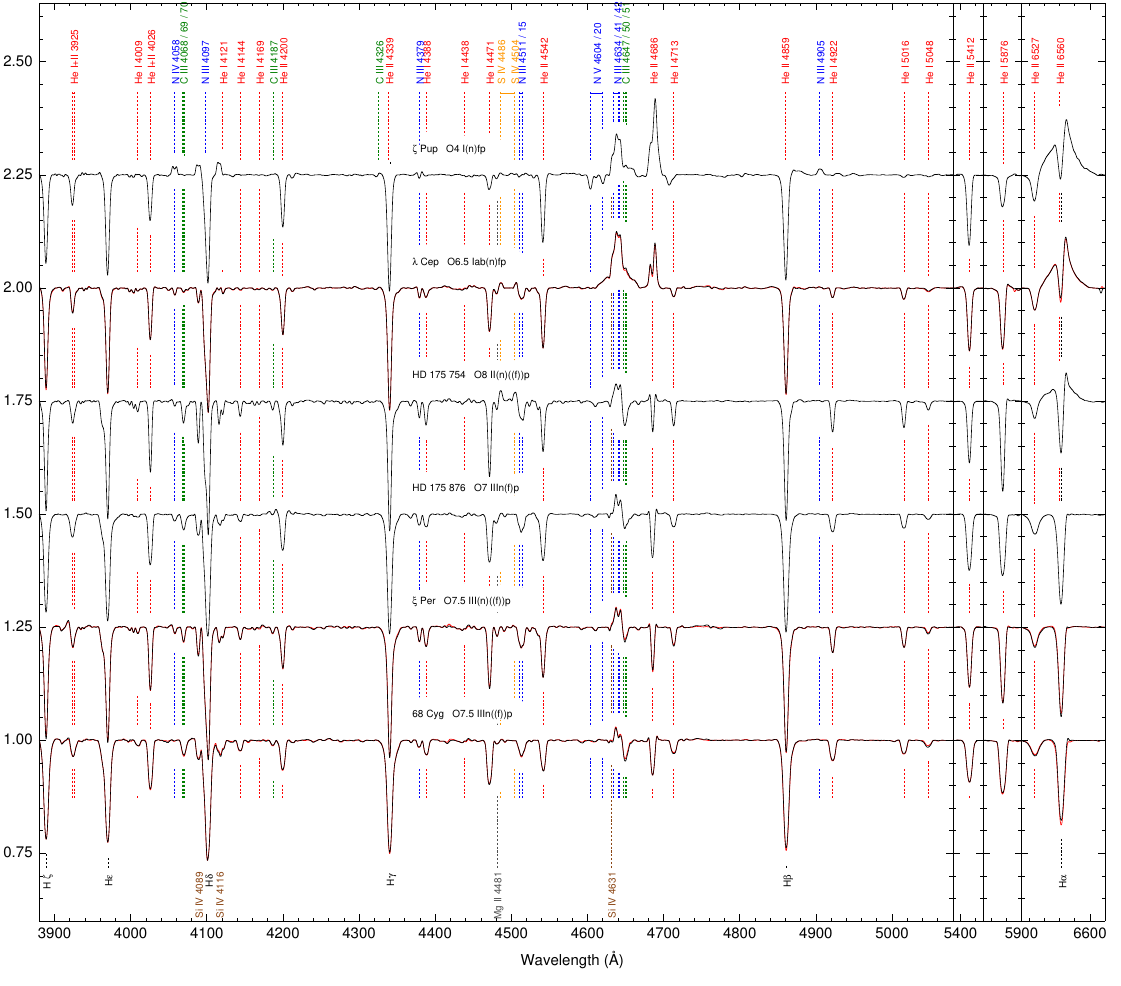}}
 \caption{Spectral classification plots ($R=2500$) for the Onfp stars. Two average spectra are given for $\lambda$~Cep, $\xi$~Per, and 
          68~Cyg, one for the high emission state (black) and the other one for the low emission state (red).}
 \label{ALS_spectra_Onfp}      
\end{figure*}

\begin{figure*}
 \centerline{\includegraphics*[width=\linewidth]{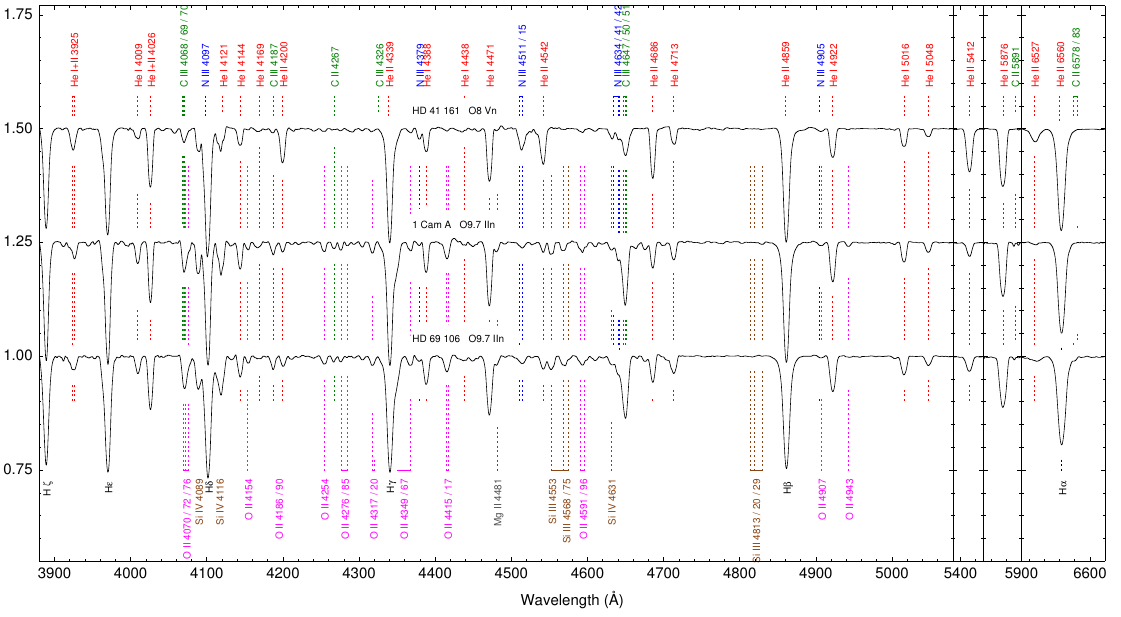}}
 \caption{Spectral classification plots ($R=2500$) for the normal fast rotators.}
 \label{ALS_spectra_On}      
\end{figure*}

\begin{figure*}
 \centerline{\includegraphics*[width=\linewidth]{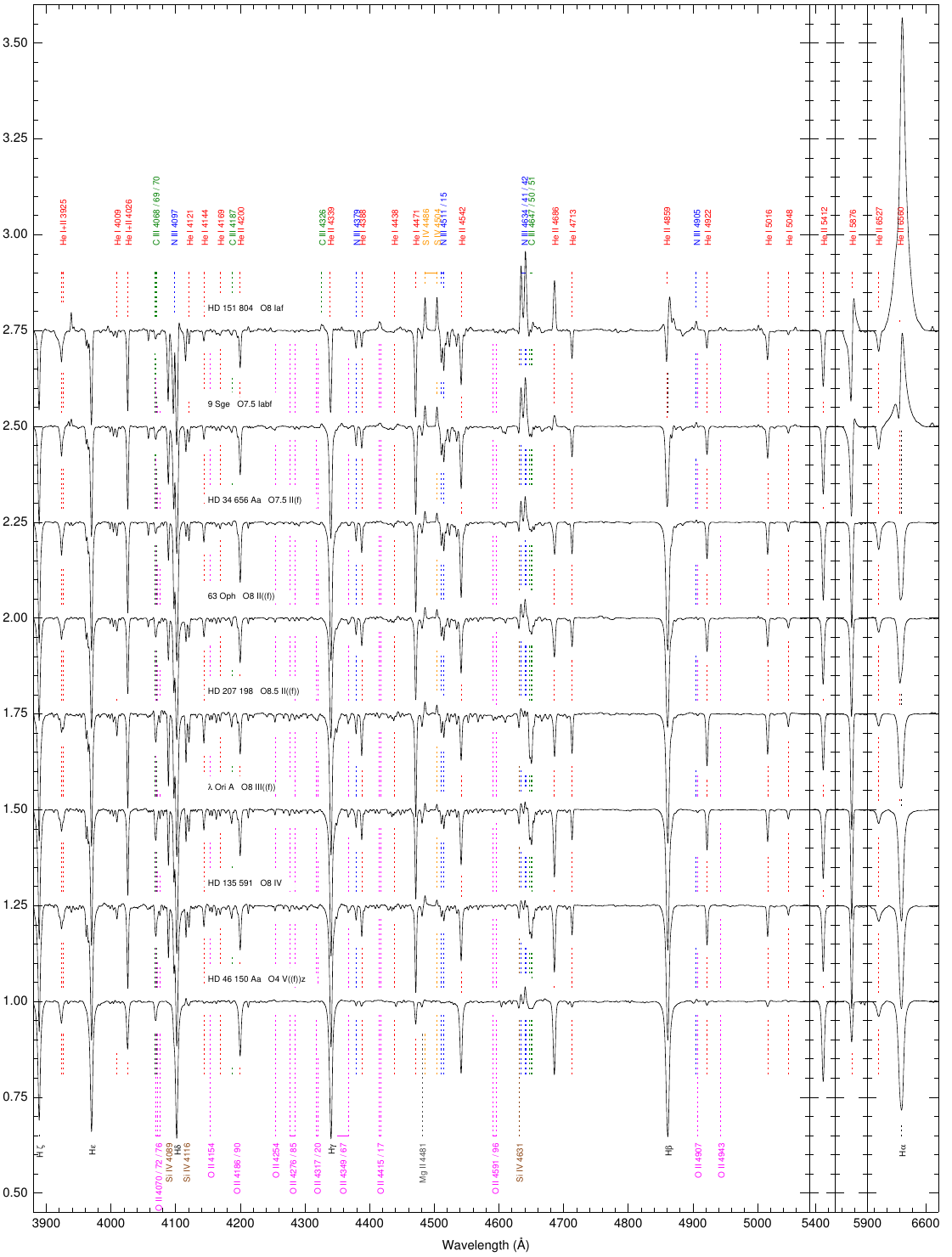}}
 \caption{Spectral classification plots ($R=2500$) for the normal O4-O8.5 slow rotators.}
 \label{ALS_spectra_Oslow1}      
\end{figure*}

\begin{figure*}
 \centerline{\includegraphics*[width=\linewidth]{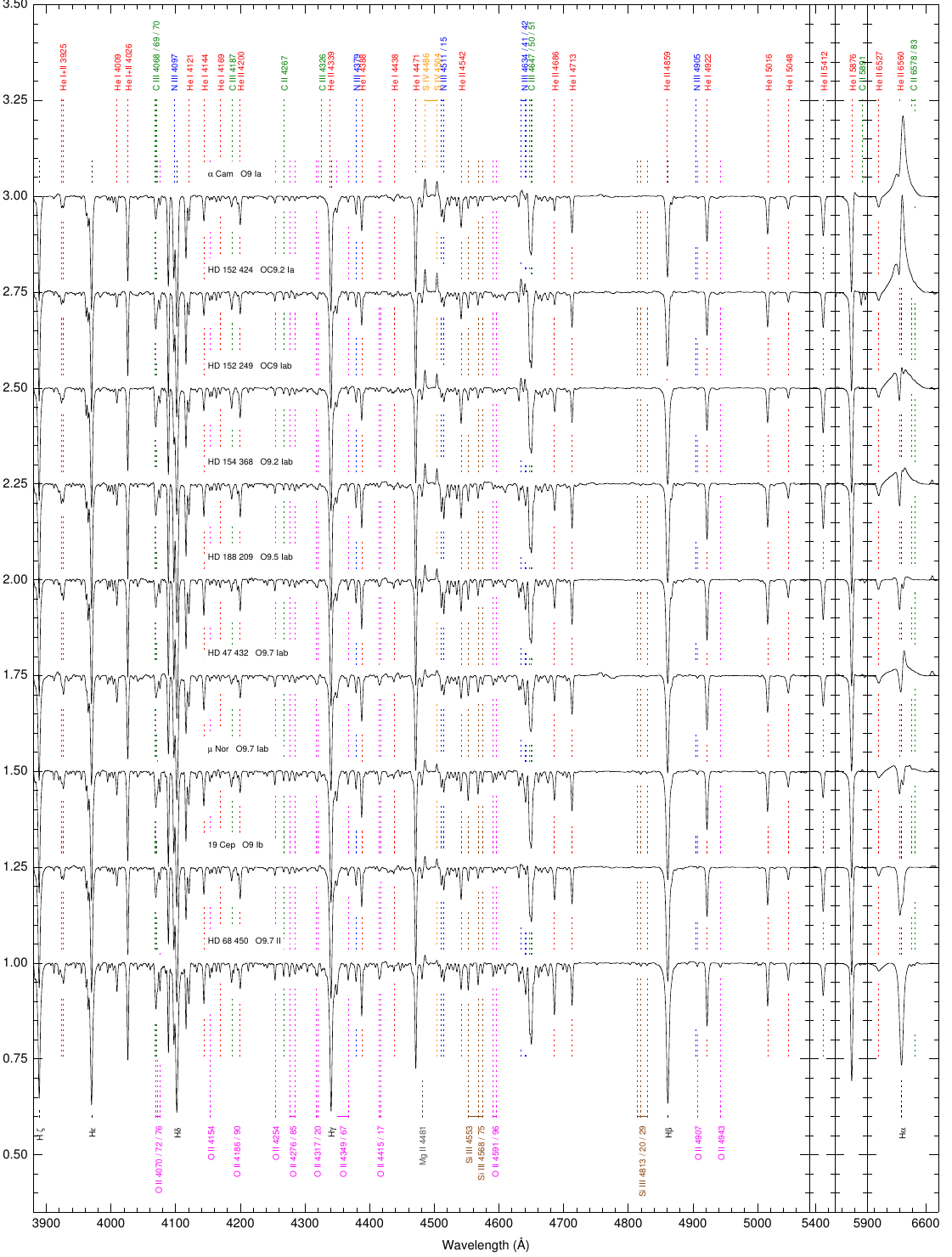}}
 \caption{Spectral classification plots ($R=2500$) for the normal O9-O9.7 slow rotators of LC II to Ia.}
 \label{ALS_spectra_Oslow2}      
\end{figure*}

\begin{figure*}
 \centerline{\includegraphics*[width=\linewidth]{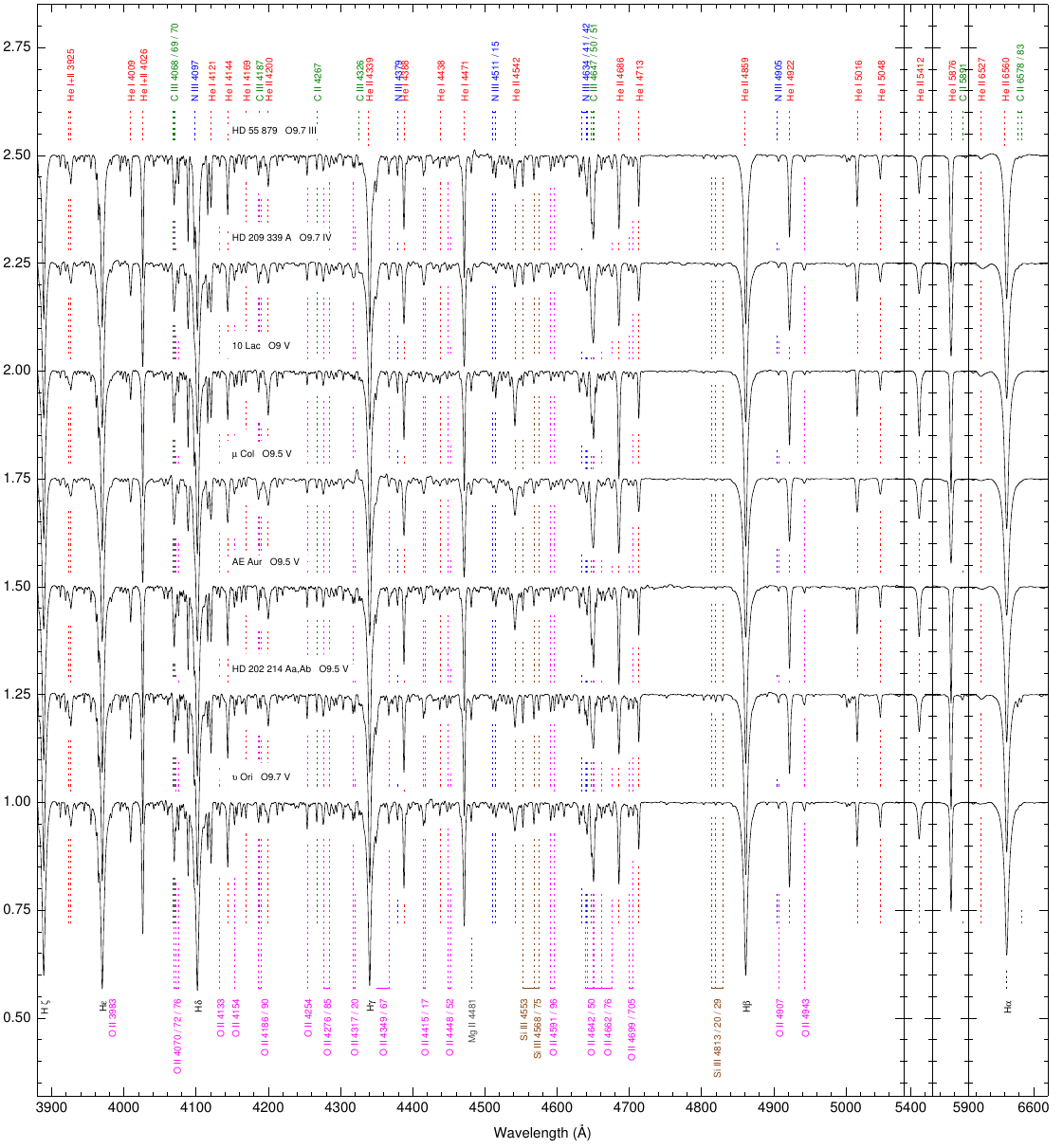}}
 \caption{Spectral classification plots ($R=2500$) for the normal O9-O9.7 slow rotators of LC V to III.}
 \label{ALS_spectra_Oslow3}      
\end{figure*}

$\,\!$\indent Figures~\ref{ALS_spectra_1} to \ref{ALS_spectra_Oslow3} show the spectral classification plots for the sample in this paper.

\section{The FASTWIND grid}\label{appendixc}

$\,\!$\indent We have generated a FASTWIND \citep{Santetal97,Pulsetal05,RivGetal11} stellar-atmosphere model (sub-)grid for this paper to analyse
the three types of RV shifts described in the main body: [a] contamination of Balmer lines by \ion{He}{ii} lines, [b] presence of forbidden 
components in some \ion{He}{i} lines and [c] wind infilling in H and He lines. Our grid is based on the complete FASTWIND v10.6.5 grid 
described in \citet{SimDetal26} and was designed to include models representative of the stars in our sample, considering two wind regimes: 
a nearly negligible wind ($\logQ=-14$ solar) and a maximum characteristic wind appropriate for each spectral subtype and luminosity class. As 
listed in Table~\ref{gridfastwind}, we adopted representative surface gravities for LCs I, III, and V ($\logg=3.4$, 3.7, and 4.0 cgs,
respectively) and the corresponding maximum characteristic wind strengths. For each combination of \logg\ and wind strength, we computed 
six effective temperatures covering the range spanned by our sample (limited to five for LC III and to four for LC I). Solar helium abundances 
were assumed throughout and the full grid was computed neglecting wind clumping, whose impact on the present analysis is expected to be negligible, 
except for \ion{He}{ii}~$\lambda$4685.71, for which the effect is expected to be limited \citep{Holgetal18}. We convolved the resulting spectra with 
two values of \vsini\ and an appropriate $v_{\rm macro}$ value in each case \citep{SimDetal17}. The grid was used for the analyses in 
subsections~\ref{subsection33},~\ref{subsection34},~and~\ref{subsection35}.

\begin{table}[!t]
 \caption{Parameter space covered by the FASTWIND grid of models.}
 \label{gridfastwind}
 \centerline{
   \begin{tabular}{lcc}
    \midrule
    Parameter & Units & Values \\
    \midrule
    \logg  & cgs   & 3.4\,$^a$, 3.7\,$^b$, 4.0\,$^c$                      \\
    \logQ  & solar & $-$14.0, $-$12.3\,$^a$, $-$13.0\,$^b$, $-$13.5\,$^c$ \\
    \Teff  & kK    & 30, 33, 36, 39, 42\,$^{b,c}$, 45\,$^c$               \\
    \vsini & km/s  & 20, 300                                              \\
    \midrule
    \multicolumn{3}{l}{$^a$ LC I; $^b$ LC III; $^c$ LC V.}
   \end{tabular}
 }
\end{table}

$\,\!$

\vfill

\eject

$\,\!$

\vfill

\eject

\section{Glossary}\label{appendixd}

$\,\!$\indent We provide a list of acronyms and terms used in this paper. 

\begin{itemize}
 \item ALS: Alma Luminous Star catalog \citep{Reed03,Pantetal21,Pantetal25b}.
 \item CAHA: Centro Astron\'omico Hispano-Alem\'an/Andaluz.
 \item CAHA3.5: CAHA 3.5~m telescope.
 \item DIB: Diffuse Interstellar Band.
 \item ESO: European Southern Observatory.
 \item EW: Equivalent Width (m\AA).
 \item FWHM: Full-Width at Half Maximum (\AA).
 \item $\gamma$: Systemic velocity.
 \item GLS: Galactic Luminous Star \citep{Pantetal25b}.
 \item GOSC: Galactic O-Star Catalog \citep{Maizetal04b}.
 \item GOSSS: Galactic O-Star Spectroscopic Survey \citep{Maizetal11}.
 \item $i$: (Orbital or rotational) inclination (deg).
 \item IMF: Initial Mass Function.   
 \item $K$: Semi-amplitude of the RV curve (km/s).
 \item LC: Luminosity Class.
 \item LiLiMaRlin: \textbf{Li}brary of \textbf{Li}braries of \textbf{Ma}ssive-Star 
  High-\textbf{R}eso\textbf{l}ut\textbf{i}o\textbf{n} Spectra \citep{Maizetal19a}.
 \item \logg: Logarithmic surface gravity (cgs).
 \item \logQ: Wind-strength parameter (solar, \citealt{Holgetal18}), with $Q=\dot{M}/(v_\infty R)^{1.5}$.
 \item $\dot{M}$: Mass loss rate (solar).
 \item MGB: Marxist Ghost Buster \citep{Maizetal12}.
 \item MONOS: Multiplicity Of Northern O-type Stars \citep{Maizetal19b}.
 \item MT: Mercator Telescope.
 \item \MTW: Multiplicity of Massive stars in the Milky Way (Paper I).
 \item NOT: Nordic Optical Telescope.
 \item OHP: Observatoire de Haute-Provence.
 \item OHP1.9: Observatoire de Haute-Provence 1.9~m telescope.
 \item OLS: Observatorio de La Silla.
 \item OLS2.2: Observatorio de La Silla 2.2~m telescope.
 \item OLS3.6: Observatorio de La Silla 3.6~m telescope.
 \item OP: Observatorio Paranal.
 \item ORM: Observatorio del Roque de los Muchachos.
 \item $R$: Stellar radius (solar).
 \item RV: Radial Velocity (km/s).
 \item SAAO: South African Astronomical Observatory.
 \item SALT: Southern African Large Telescope.
 \item SB: Spectroscopic Binary.
 \item SB1(E): Single-lined Spectroscopic (Eclipsing) Binary.
 \item SB2(E): Double-lined Spectroscopic (Eclipsing) Binary.
 \item \Teff: Effective temperature (K).
 \item TNG: Telescopio Nazionale Galileo.
 \item $v_\infty$: Wind terminal velocity (km/s).
 \item $v_{\rm macro}$: Macroturbulent velocity (km/s).
 \item $v_{\rm micro}$: Microturbulent velocity (km/s).
 \item \vsini: Projected equatorial rotational velocity (km/s).
 \item VLT: Very Large Telescope.         
 \item WDS: Washington Double Star catalog \citep{Masoetal01}.
\end{itemize}

\end{appendix}

\end{document}